\documentclass[aps,reprint,twocolumn,superscriptaddress,showpacs,prl,nofootinbib,floatfix]{revtex4-1}
\usepackage{amsmath}
 
 \usepackage{wasysym} 
 \usepackage{mathrsfs} 
 \usepackage{graphicx} 
 \usepackage{amsfonts} 
 \usepackage{amsbsy} 
 \usepackage{amscd} 
 \usepackage{amssymb}
\usepackage{bbm}  
 \usepackage{accents} 
 \usepackage{multirow} 
\usepackage{color}

\def\beq{\begin{equation}}
\def\eeq{\end{equation}}
\def\beqa{\begin{eqnarray}}
\def\eeqa{\end{eqnarray}}
 
\def\ben{\begin{enumerate}}
\def\een{\end{enumerate}}
\def\bit{\begin{itemize}}
\def\eit{\end{itemize}}
 
\begin{document}
 
\title{Lepton Number  Violation and `Diracness' of massive neutrinos\\ composed of Majorana states}

\author{Janusz Gluza}
\author{Tomasz Jeli\'nski}
\affiliation{Institute of Physics, University of Silesia, Uniwersytecka 4, 40-007 Katowice, Poland}
\author{Robert Szafron}
\affiliation{Department of Physics, University of Alberta, Edmonton, Alberta, Canada T6G 2G7}

\begin{abstract}
Majorana neutrinos naturally lead to a Lepton Number Violation (LNV). 
A superposition of Majorana states can mimic Dirac-type neutrinos, leading to its conservation (LNC). We demonstrate on the example of specific observables related to high and low energy processes how a strength of LNV correlates with neutrino parameters such as CP-phases, flavor mixings, mass ratios. We stress the coaction between low and high energy studies for putting phenomenological models to a quantitative test.
Secondly, we conclude that in order to fully study the role of heavy neutrinos in search for New Physics (NP) signals, a departure from trivial scenarios assuming degeneracy in mass and no flavor mixing or CP-phases becomes necessary for a proper physical analysis.
\end{abstract}

\maketitle 
\allowdisplaybreaks

\section{Introduction \label{intro} }

The neutrino oscillation phenomenon
established that at least two out of three known neutrinos are massive, though their masses are very tiny, at most at the electronvolt level,
$m_{\nu}\sim {\cal{O}}(1)$ eV.  It was a tremendous effort that led to this result as experimental studies of neutrino physics face the challenge of low event statistics for an anyhow scarce set of observables. It started around half a century ago with the pioneered Homestake experiment and the so-called solar neutrino problem  \cite{Bahcall:1976zz}, ending with the last year's Nobel award for Takaaki Kajita and Arthur B. McDonald (Super-Kamiokande and SNO Collaborations \cite{Fukuda:1998mi,Ahmad:2002jz}) "for the discovery of neutrino oscillations, which shows that neutrinos have mass".
 
Evidence of new phenomena involving neutrinos often stirs up a lot of attention in both the physics community and public media, however sometimes for no good reason c.f. the 17-keV neutrino (dubbed a Simpson's neutrino) signal in tritium decays \cite{Wietfeldt:1995ja,Franklin:1995pk}, the OPERA faster-than-light neutrino controversy \cite{Adam:2011faa,Antonello:2012hg}, or positive neutrinoless double beta decay signals \cite{KlapdorKleingrothaus:2006ff} (see also comments in \cite{Akhmedov:2014kxa}).
Certainly, we can expect more such situations in the future, as neutrino experiments explore  by definition {\it weak} effects and belong to the most challenging endeavours in physics.

Though the Large Hadron Collider (LHC) collides protons and deals  
predominantly with hadron effects, it is also sensitive to the
electroweak and New Physics (NP) effects. 
One of the exciting NP
options  is the teraelectronvolt heavy neutrino physics,  $M_{N}\sim {\cal{O}}(1)$  TeV.
Heavier particles hardly can be produced at a detectable level in collisions or observed indirectly in precise low energy experiments.

In theory,  heavy neutrino states are commonly embedded within  the
seesaw mass mechanism \cite{Minkowski:1977sc,Yanagida:1979as,GellMann:1980vs,Mohapatra:1979ia}. It
explains smallness of the known  neutrino masses  using
notion of Majorana states where lepton number violation (LNV) is present. Typical is the neutrinoless double beta decay $(\beta \beta)_{0 \nu}$, considered in this work,  
where a nuclear transition ends with two electrons in the final state
\cite{Schechter:1981cv,Rodejohann:2011mu,Pas:2015eia}.
Alternatively, in the inverse seesaw mechanism depending on choice of mass parameters either Majorana or Dirac
neutrinos can appear \cite{Mohapatra:1986aw,Mohapatra:1986bd}. 
The question whether neutrino states are of Majorana or Dirac type, or maybe their mixture, is a core problem in particle and astroparticle physics 
\cite{Mohapatra:1998rq,Giunti:2007ry,Drewes:2013gca}.  
Theories involving Dirac-type particles obey LNC,  in the inverse seesaw scenario, lepton number violation can vary
naturally and may be substantial or negligible. 
Apart from the LNV issue massive neutrinos regardless of their type can lead to appearance of the charged lepton
flavor violation (CLFV) \cite{deGouvea:2013zba}. Here a change of the charged lepton flavor  requires nontrivial neutrino mixing matrices. 
 
As even a single unambiguous LNV or CLFV event detection  would be a signal of NP, there are many efforts to upgrade or create new experimental setups.
Present bounds for low energy CLFV signals, such as nuclear $\mu$ to $e$ conversion will become more stringent at the so-called intensity frontier experiments \cite{Kuno:2013mha, Brown:2015cka}. 
The same is true for  $(\beta \beta)_{0 \nu}$ experiments, see e.g. \cite{Abgrall:2013rze}.

We show that limits on the low energy processes are essential for LHC searches and give deep insight into neutrino scenarios. {{In these studies crucial is an investigation of nontrivial parameters such as non-diagonal neutrino mixings, non-zero CP-phases or non-degenerated neutrino masses.}}

We consider a pending and relevant topic: how much can we simplify models in an experimental analysis? For instance, in searching for right-handed currents and heavy Majorana/Dirac neutrinos? This issue is important especially in a context of such searches by CMS and ATLAS collaborations.

We track the whole discussion in the context of the $pp \to lljj$ process. This process (coined
`golden' or `smoking-gun' process) is a good witness to NP due to its sensitivity to right-handed currents. They are not suppressed if heavy
neutrinos exist. We will consider such heavy {\it Majorana} neutrinos and observe how composed Majorana neutrinos mimic Dirac states, and how it affects LNV.  
Finally, $pp \to lljj$ connects three different high and
low energy experiments and is a perfect workhorse for a general
discussion of LNV effects and mutual constraints imposed on NP
model parameters. 
\section{Discussion and results \label{disc} }

Our discussion is based on the following Lagrangian
\begin{eqnarray}
\mathcal{L}&=& \frac{{g}}{\sqrt{2}}\sum\limits_{a=1}^3\overline{\nu}_a\gamma^{\mu}P_L(U_{\mathrm{PMNS}})_{aj}^\dagger l_jW_{1\mu}^{+}+\mathrm{h.c.}\label{lagr1}\\
&+&\frac{\widetilde{g}}{\sqrt{2}}\sum\limits_{a=1}^3\overline{N}_a\gamma^{\mu}P_R(K_{R})_{aj}l_jW_{2\mu}^{+} +\mathrm{h.c.},\label{lagr2}
\end{eqnarray}
where $N_a$ stands for massive  {\it Majorana} states (most naturally, in popular models based on left-right gauge symmetry there are three $M_a$ heavy neutrinos). 
The term~\eqref{lagr1} describes the SM physics of charged currents. It includes the celebrated neutrino mixing matrix 
$U_{\mathrm{PMNS}}$, responsible for neutrino oscillations phenomena. The term~\eqref{lagr2} is responsible for non-standard effects connected with heavy neutrinos $M_a$, their mixing matrix $K_R$ and right-handed currents mediated by additional charged gauge boson $W_2$. These are three types of modeled unknowns.
For models and phenomenology of heavy neutrinos without right-handed currents, see for instance \cite{Pilaftsis:1991ug,Bray:2007ru}.

\begin{figure}[h!]
\begin{center}
\includegraphics[scale=0.5]{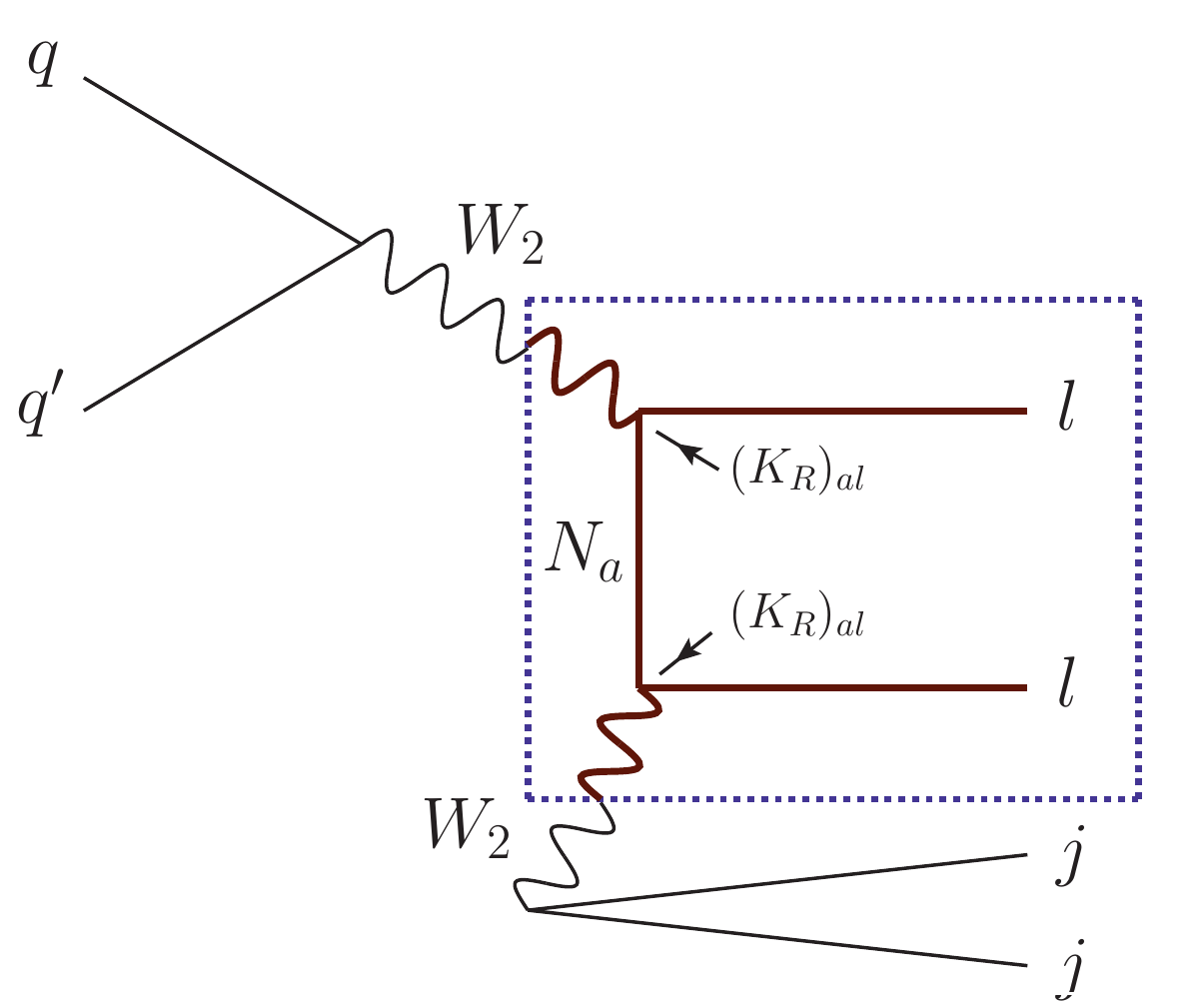}
\caption{Feynman diagram responsible for the `golden' $pp\to  lljj$
  signal. For  Majorana neutrinos two signals are possible with
  Same-Sign leptons $pp\to W_2^{\pm}\to l_i^{\pm} N_a\to l_i^{\pm}
  l_j^{\pm}jj$ and Opposite-Sign leptons $pp\to W_2^{\pm}\to l_i^{\pm}
  N_a\to l_i^{\pm} l_j^{\mp}jj$. In the internal frame two
  related LNV processes can be identified: $ll \to W_2 W_2$
  (possible in  the future lepton colliders) and  $W_2^- W_2^- \to
  e^- e^-$ (a part of the low energy neutrinoless nuclear double beta decay $(\beta \beta)_{0 \nu}$).
}\label{lljj}
\end{center}
\end{figure}

The main NP contribution to the $pp \to lljj$ process  is
sketched in Fig.~\ref{lljj}.
It is a prominent process that has been discussed long before LHC, at the dawn of  the Tevatron era \cite{Keung:1983uu}.
In this process same-sign (SS) leptons indicate LNV, which can be naturally explained by Majorana-type neutrinos. If only
excess is seen with opposite-sign (OS) leptons, the situation is more complicated.
According to Feynman rules \cite{Gluza:1991wj,Denner:1992me},  virtual Majorana particle can change the lepton number 
on the way from one vertex to the another, which may results in production of the same charged  dileptons.
However, as we will see, 
 since Majorana virtual states act in a coherent way, much depends on couplings and the neutrino mixing matrix elements $K_R$.   

In Fig.~\ref{lljj} yet another two processes $e^-e^- \to W_2^-W_2^-$ and
 $W_2^- W_2^- \to e^- e^-$ can be spotted. The first process can be a good option for LNV
 search at future electron colliders while the second one is a weak-interaction
 part of the neutrinoless double beta decay $(\beta \beta)_{0 \nu}$
 experiment.  All three processes have the same couplings, though they
 differ in channels and characteristic kinematics. That is why $pp \to lljj$ depends strongly on experimental bounds derived from  $(\beta \beta)_{0 \nu}$.

Look into  the details of the low energy $(\beta \beta)_{0 \nu}$ process, for other important low energy processes see the Appendix. 
In $(\beta \beta)_{0 \nu}$, the light and heavy neutrinos
contribution to the half-life reads 
\begin{eqnarray}
\frac{1}{T_{1/2}^{0\nu}} \ = \  {G_{}}{}\left|m^\nu+ m^N\right|^2 , 
\label{halft}
\end{eqnarray}
where $G_{}$ gathers all nuclear parameters. For light
neutrinos, the effective mass term $m^\nu$ is proportional to light neutrino masses $m_i$, namely
$m^\nu = \sum\limits_{i=1}^3 (U_{\mathrm{PMNS}})_{ei}^2 m_i$\;.
%
The heavy neutrinos exchange due to the interaction \eqref{lagr2} yields a term inversely proportional to the heavy neutrino masses $M_a$ ($|p^2| \sim 100$ $\rm {MeV}^2$)
\begin{eqnarray}
m^N \ = \left(\frac{\widetilde g}{g}\right)^4 \left(\frac{M_{W_1}}{M_{W_2}}\right)^4 |p|^2\sum_a\frac{{(K_R)}_{ea}^2}{M_a}.
\label{mNee}
\end{eqnarray} 
 
Two sorts of cancellations are possible: either between light and heavy contributions or inside each of them separately \cite{Hirsch:1996qw,Tello:2010am,Mitra:2011qr,Faessler:2011qw,Nemevsek:2012iq,Meroni:2012qf,Barry:2013xxa,Pascoli:2013fiz,Dev:2013oxa,Dev:2013vxa,Meroni:2014tba,Meroni:2015oya}.
 In $m^\nu$ and $m^N$ 
the mixing matrices $U_{\mathrm{PMNS}}$ and $K_R$ are squared, so without complex phases they always contribute positively.
Cancellations appear if nonzero CP phases are involved. 
$m^\nu$ depends on the absolute value of the lightest neutrino mass and mass hierarchies related to the  neutrino oscillation analysis \cite{Czakon:1999cd,Barger:1999na,Dev:2013vxa}. 
 Allowing whole range of CP-phases in $U_{\mathrm{PMNS}}$, results for $m^\nu$ can span 
from zero to the values of $T_{1/2}^{0\nu}$ allowed by $(\beta \beta)_{0 \nu}$ experiments \cite{Nemevsek:2012iq,Xing:2015zha,Dev:2013vxa,Awasthi:2015ota}.
 Interestingly, negligible $m^\nu$
 contributions to $(\beta \beta)_{0 \nu}$ are possible only in the
 case of normal light neutrinos mass hierarchy \cite{Nemevsek:2012iq,Czakon:1999cd,Barger:1999na,Dev:2013vxa}. 
 In inverted mass
 hierarchy $m^\nu$ cannot reach zero. In this case results and constrains on the NP parameters coming from $m^N$ presented below would relax due to possible CP effects in $U_{\mathrm{PMNS}}$ and $m^\nu-m^N$ cancellations. 
  
 To procede further and calculate $(\beta \beta)_{0 \nu}$ half-life, NP parameters $K_R$ and $M_i$ must be specified. 
   In what follows we will consider two physically interesting scenarios and their variants.
\begin{enumerate}
\item[(A)] ${(K_R)}_{aj}=\delta_{aj}, M_a=M_{W_2}/2$ and $M_{W_2}=2.2\,\mathrm{TeV}$.  
This is the simplest case of degenerated heavy neutrinos and without flavor mixings. It was used in 
CMS  $M_a - M_{W_2}$ exclusion analysis  in the context of the excess in $\sigma (pp \to eejj)$ at the invariant mass about $2.2$ TeV \cite{Khachatryan:2014dka};
\item[(B)] Masses as in (A), $M_a=M_{W_2}/2$ and $M_{W_2}=2.2\,\mathrm{TeV}$. $K_R$ includes the simplest two-variable parametrisation  mixing matrix (one nontrivial rotational mixing angle $\theta_{13}$ and one CP-phase $\phi_3$)
\beq\label{KRB13}
K_{R}=
\left(
\begin{array}{ccc}
\cos\theta_{13}&0&\sin\theta_{13}\\
0&1&0\\
-e^{i\phi_3}\sin\theta_{13}&0&e^{i\phi_3}\cos\theta_{13}
\end{array}
\right).
\eeq 
\item[(C)] As in (B), but with $M_{W_2}=3\,\mathrm{TeV}$.
\item[(D)] As in (B), but with $g_R/g_L=0.6$. 
\end{enumerate}
 
In case (A), there are no mixings. In case (B) they occur between $N_1$ and $N_3$, leading to possible electron-tau CLFV effects, see Eq.~(\ref{lagr2}) and the Appendix. $N_2$ does not mix here and remains of Majorana type. Its LNV effects could be then detected in some process, e.g. $\mu^- \mu^- \to W_{2}^- W_{2}^-$, analogously to the 
$e^- e^- \to W_2^- W_2^-$ case considered in the Appendix. 
In our study $K_R$ is always unitary and therfore leaving no room for additional light-heavy neutrino 
mixings that are tightly constrained or need special constructions \cite{Gluza:2002vs,Chen:2013foz}.
 
\begin{figure}[h!]
\begin{center}
\includegraphics[width=0.8\columnwidth]{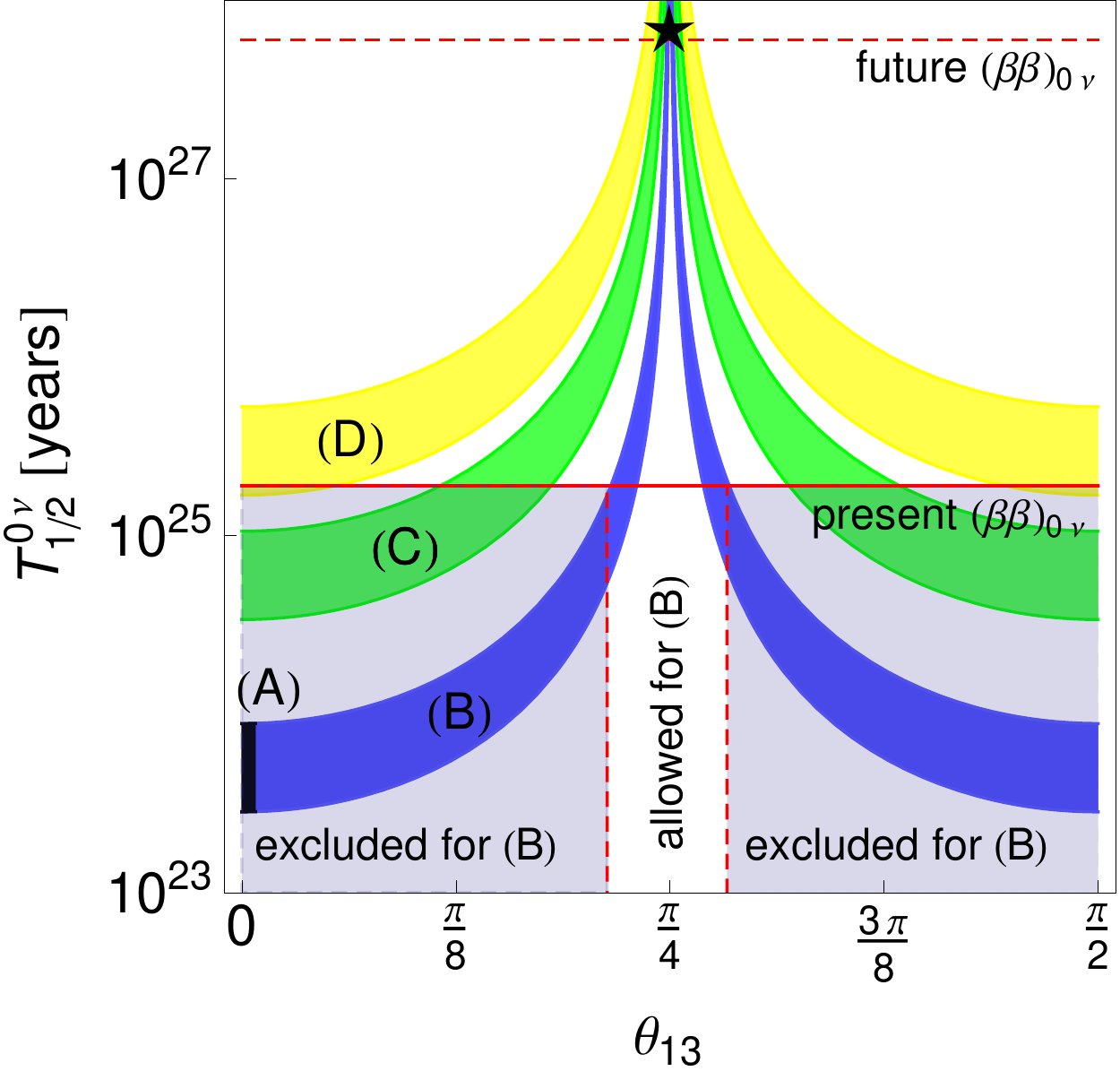}
\caption{\label{0nu}
`Christmas Tree'  exclusion plot for $m^N$ dominating $(\beta \beta)_{0 \nu}$ half-life $T_{1/2}^{0\nu}\left[^{76}\mathrm{Ge}\rightarrow ^{76}\mathrm{Se}\right]$ as a function of $\theta_{13}$. $\eta_{CP}(N_1)=-\eta_{CP}(N_3)=i$,
$\phi_3=\pi/2$ and $M_{1,2,3}=M_{W_2}/2=1.1\,\mathrm{TeV}$. A star on its top represents infinite half-life Dirac scenario (maximal $\theta_{13}$ mixing).  The bands correspond to different evaluations of the
nuclear matrix elements \cite{Meroni:2012qf}. Excluded region (below solid horizontal line) 
comes from \cite{Baudis:1999xd}. Dashed horizontal line represents expected future bound by the Majorana+GERDA experiment \cite{Abgrall:2013rze}.}
\end{center}
\end{figure}

 Fig.~\ref{0nu} shows predictions for $(\beta \beta)_{0 \nu}$ in scenarios (A)-(D) for $(\beta\beta)_{0\nu}$ when $m^N$ term  dominates. In (A) and (B) we take relatively small $M_{W_2}=2.2$ TeV. It was a value of $eejj$ invariant mass for which an excess in $\sigma (pp \to eejj)$ was reported  by the CMS during the LHC run 1 \cite{Khachatryan:2014dka}. However, it is in an excluded region for scenarios in which $g=\widetilde{g}$ and heavy neutrinos do not mix and are degenerate in mass \cite{Khachatryan:2014dka}. We compare this case with LNV low energy predictions for the same scenario and for the scenario in which heavy neutrino mixings and non-degenerate masses are allowed.  
In Fig.~\ref{0nu}, we consider (A)-(D) scenarios with degenerate neutrino masses and $N_1,N_3$ which have opposite CP parities, $\phi_3=\pi/2$. (CP parities of neutrinos are strictly connected with imaginary part of  neutrino mixing elements, see the Appendix and 
 \cite{Bilenky:1987ty,Kayser:1984ge,Kayser:1989iu}).

The results are interesting. Firstly, let us note, that the typical scenarios tested by CMS/ATLAS with no mixings and degenerate heavy neutrinos are in excluded region (A) in Fig.~\ref{0nu}.
Secondly, flavor mixing $\theta_{13}$ in case (B) is limited to the region close to $\pi/4$ for which LNC  is restored. It can be seen directly using the parametrisation shown in Eq.~\eqref{KRB13} for which Eq.~\eqref{lagr2} explicitly reads
\beq
\mathcal{L}_{\mathrm{RHC}}=\frac{\widetilde{g}}{\sqrt{2}} \overline{N}_{\mathrm{eff}}\gamma^{\mu}P_R  l_1W_{2\mu}^{+}+\mathrm{h.c.,}
\label{lagr3}
\eeq
where 
\begin{equation}\label{Neff} 
N_{\mathrm{eff}}={{\left(
 \cos \theta_{13} {N}_1 - e^{i \phi_3}\sin \theta_{13} {N}_3
  \right)}}.
\end{equation}
For $\theta_{13}=\pi/4, \phi_3=\pi/2$ we have
$N_{\mathrm{eff}}=\left({N}_1+i{N}_3\right)/\sqrt{2}$. That is a definition of the
Dirac neutrino composed of two Majorana degenerated states with
opposite CP parities, see the Appendix and \cite{Bilenky:1987ty}. In this case $N_1$ and $N_3$ Majorana neutrinos give
opposite contributions to $m^N$ of equal weight, $\sum\limits_a
{(K_R)}^2_{ea}=1^2+i^2=0$  (infinite half-life time). In our opinion this is
the easiest way to see how two Majorana neutrinos act as the Dirac state and effectively lead to LNC.  
  Thirdly, we can see from Fig.~\ref{0nu} that a scenario (D) with $g \neq 
  \widetilde{g}$ is not constrained by present  $(\beta\beta)_{0\nu}$ data. This scenario is more suitable as far as GUT unification of couplings is concerned \cite{Deppisch:2015cua,Dev:2015pga}.
  
\begin{figure}[h!]
\begin{center}
\includegraphics[width=0.72\columnwidth]{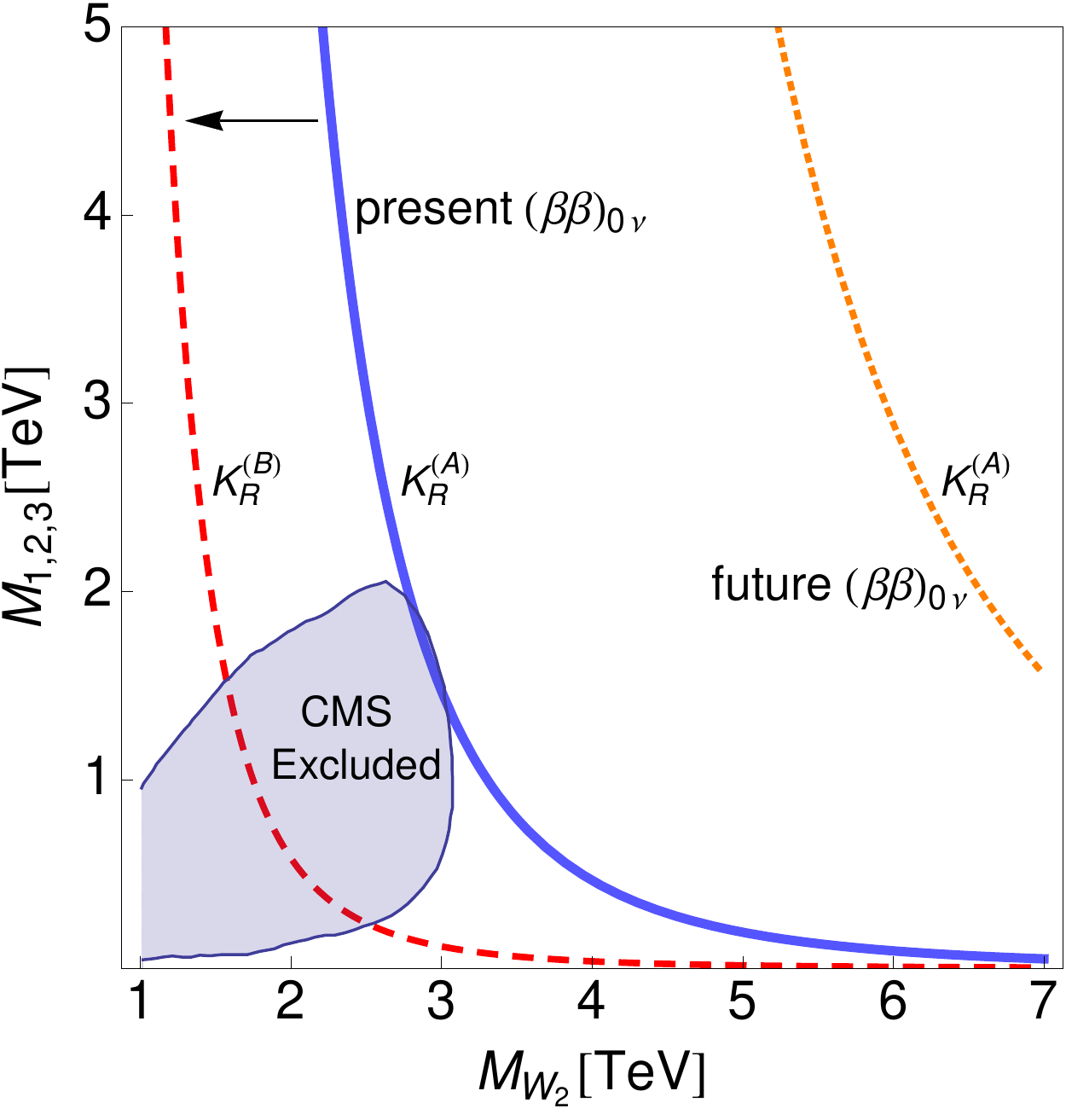}
\caption{\label{0nuCMS}
The CMS vs. $m^N$ dominant $(\beta\beta)_{0\nu}$ exclusion limits on masses of $W_2$ and $N_a$ in the case 
when $(K_R)_{aj}=\delta_{aj}$ as in the (A) scenario.
The shaded region is excluded by the CMS data related to $pp\to eejj$ at the LHC run 1
\cite{Khachatryan:2014dka}. 
Present $(\beta\beta)_{0\nu}$ experiments exclude the region under the blue solid curve. The dotted orange curve corresponds to a future bound on $T_{1/2}^{0\nu}$ 
\cite{Abgrall:2013rze}. For the comparison, when the mixing matrix $K_R$ is of the form \eqref{KRB13} 
with $\theta_{13}=0.9\times\pi/4$ and $\phi_3=\pi/2$, 
then only the region under the dashed red curve is excluded. There are no available LHC data exclusion analysis for such `almost' {}Dirac neutrinos.
}
\end{center}
\end{figure}

Assuming again that right-handed currents and $m^N$ dominate 
$(\beta\beta)_{0\nu}$, Fig.~\ref{0nuCMS} shows how strong low energy process can influence heavy mass parameters, comparing to the LHC run 1 studies. The CMS exclusion area for the (A) scenario is well within the exclusion region given by the present $(\beta\beta)_{0\nu}$ data.
For the case (B) when non-diagonal $K_R$ elements are assumed, the limits on $M_{W_2}$ and $M_a$ coming from $(\beta\beta)_{0\nu}$ are much weaker.  
For example, when $\theta_{13}=0.9\times\pi/4$ and $\phi_3=\pi/2$ which is a small distortion from the pure Dirac neutrino case (such states are also called pseudo-Dirac neutrinos) then the masses of $N_a$ and $W_2$ can be as low as $1$ and $2\,\mathrm{TeV}$ respectively, see Fig.~\ref{0nuCMS}.

 We expect that allowing wider range of mixing parameters in the CMS analysis of $pp\to eejj$ \cite{Khachatryan:2014dka} would analogously relax their bounds on $M_{W_2}$ and $M_a$. That is why it is desirable that the LHC collaborations include in heavy neutrino analysis more complicated but natural and less tuned mixing scenarios in future studies.

\begin{figure}[h!]
\begin{center}
\includegraphics[width=0.75\columnwidth]{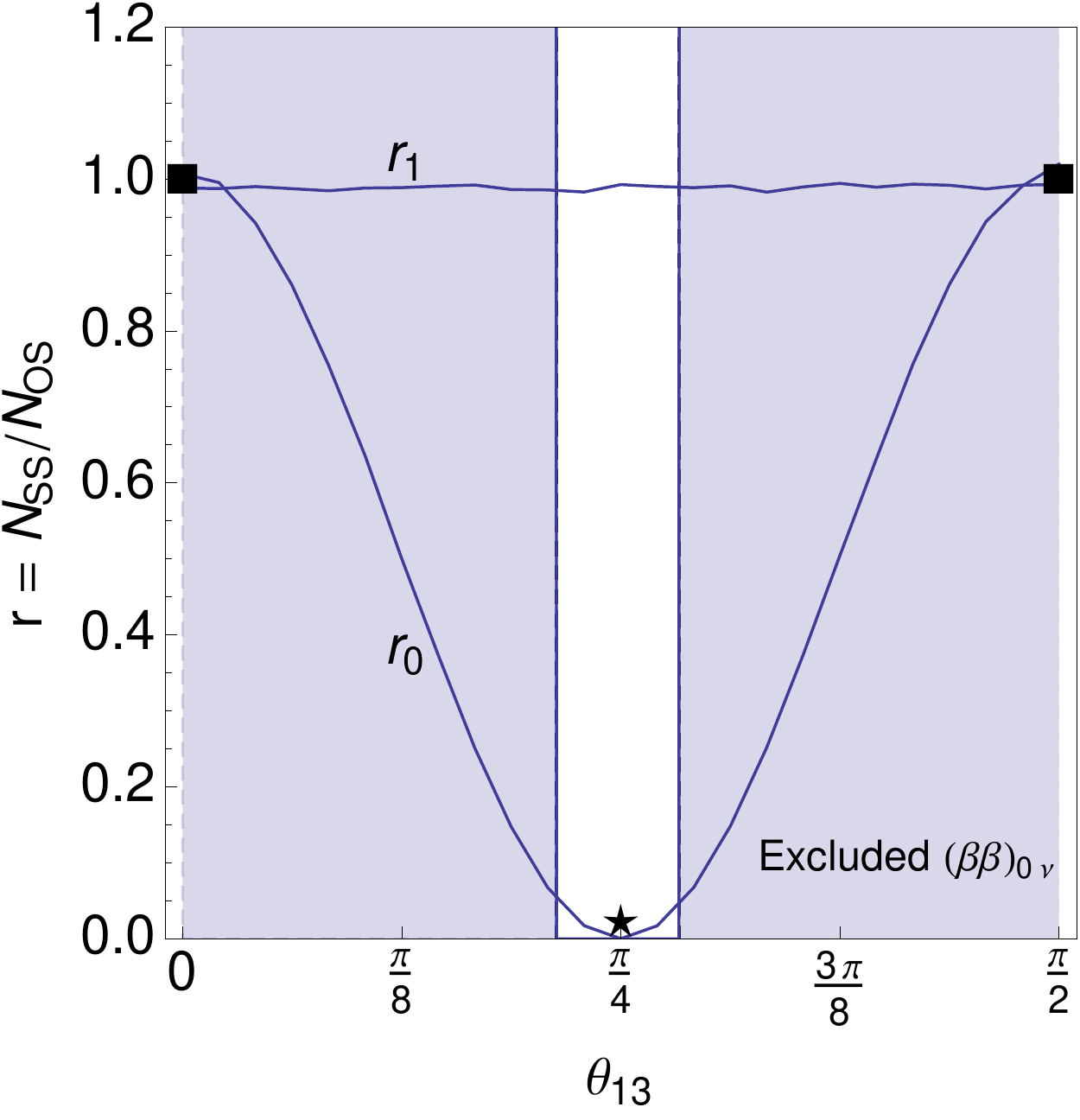}
\caption{Dependence of $r_0$ and $r_1$ on $\theta_{13}$ for case (B) for two
  neutrinos with opposite CP-parities, $\eta(N_1)=-\eta(N_3)=i$
  ($\phi_3=\pi/2$ in \eqref{Neff}). The subscript $0(1)$ of $r_{0(1)}$ means that there is
    $0(1)$  GeV  mass splitting between $N_1$ and $N_3$ states. 
A star represents a scenario with $\theta_{13}=\pi/4$ equivalent to one Dirac heavy neutrino. Black squares correspond to pure Majorana states (no interferences). The shaded region is already excluded by the $m^N$ dominant $(\beta\beta)_{0\nu}$ experimental data given in Fig.~\ref{0nu}.} \label{figr1}
\end{center}
\end{figure}

Let us come back to the LNV discussion in the LHC physics and  introduce a $r=N_{SS}/N_{OS}$ parameter that characterizes non-standard contributions to $\sigma(pp\to eejj)$,  see \cite{Gluza:2015goa,Dev:2015pga} for earlier studies of $r$ dependences in another contexts.
Here $N_{SS}(N_{OS})$ is the number of SS (OS) events, respectively.
$r=0$ corresponds to the LNC SM case.
The results are shown in Fig.~\ref{figr1} for the case (B) which includes
neutrino mixing and CP-phase parameters. Neutrinos $N_1$, $N_3$ contribute with different weights into vertices in Fig.~\ref{lljj}, see Eqs.~(\ref{lagr3}), (\ref{Neff}). If neutrinos are non-degenerate, additional weight factors appears. In this case lines in Fig.~\ref{figr1}
would not be symmetric over $\theta_{13} \in (0,\pi/2)$. Both neutrinos interfere leading to different values of $r$. For the Dirac case denoted by the star on the bottom ($r=0$), the mixing between two Majorana neutrino states should be maximal $\theta_{13}=\pi/4$, so the CP phase $\phi_3=\pi/2$. For two top black squares, there is no mixing among neutrino states and neutrino is purely of Majorana nature ($r=1$, maximal LNV). 
If $r$ is small (as dictated by $(\beta\beta)_{0\nu}$ bounds in Fig.~\ref{0nu}), the lepton number is slightly broken and we get a pseudo-Dirac neutrino.
The line $r_1$ in Fig.~\ref{figr1} shows that even small non-degeneracy of neutrino Majorana states at the GeV level spoils the interference effects (neutrino decay widths are $\sim$ MeV).  
Similar effect occurs when two neutrinos have the same CP-parities, $\eta(N_1)=\eta(N_3)$. Then the dependence on the mixing angle $\theta_{13}$ cancels out from physical observables like $r$. Let us also note that $r$ does not change when $\widetilde{g}$ deviates from $g$. It is a  consequence of the fact that both $N_{OS}$ and $N_{SS}$ scale as $\widetilde{g}^2$. 
Further technical details related to $r$ and its dependence on the mixing matrix $K_R$, decay widths and masses are discussed in the Appendix.
Based on results shown in Fig.~\ref{figr1}, different scenarios are summarized in Tab.~\ref{table1}.
\begin{table}[h!]
\begin{tabular}{|c|c|c|c|}
\hline \hline
$
\Delta M_{ab}/\max_{a,b}(\Gamma_a,\Gamma_b)$ &      
$r$ & $\Delta L$ violation   & nature \\
\hline
0 & 0 & 
0 & Dirac \\ 
 $\ll1$ & 
   small & moderate   & pseudo-Dirac  \\
  $\sim  1$ & 
   large   &  substantial & Majorana\\
  $\gg1$ & 
  1 & maximal & Majorana \\
\hline \hline
\end{tabular}
\caption{`Diracness' of neutrino states composed of Majorana massive states measured by the $r=N_{SS}/N_{OS}$ parameter  in $pp \to lljj$, see Fig.~\ref{figr1}. $\Delta M_{ab}=M_a-M_b$ and $\Gamma_{a,b}$ are Majorana neutrinos mass splittings and decay widths, respectively.} \label{table1}
\end{table}

Here we considered dependence of $r$ on physical parameters as masses of heavy neutrinos, decay widths, CP-parities. In \cite{Dev:2015pga} it has been shown that $r$ can be related in some models directly to the neutrino mass matrix entries. In a case of the inverse seesaw mechanism $r \simeq \mu_R^2/(\mu_R^2+4 M_N^2)$. See the Appendix for various mass matrix parametrizations and \cite{Dev:2015pga} for further details.

\section{Summary and outlook \label{sum} }

Assuming right-handed currents, we have shown on a couple of related processes how strength of LNV  varies with parameters of heavy Majorana neutrinos. 
In extreme cases Majorana neutrinos effectively do not violate lepton number. In all considered processes Majorana neutrinos were virtual, so LNV depends strongly on coherent effects connected with neutrino decay widths, CP-phases, non-degeneracy of heavy Majorana neutrino masses and their mixings.   
That is why  more refined LHC exclusion studies are highly
desirable. Such analysis could start from taking into account
CP-parities and a flavor mixing of two heavy neutrinos, as we sketched here. 
Furthermore, as there is a strong connection between high and low energy
experiments, progress in many different intensity frontier and neutrino oscillation experiments is highly welcome. For instance, determination of the light neutrino parameters, including normal or inverted neutrino mass hierarchies, affects  $(\beta \beta)_{0\nu}$ half-lifes predictions, which, in turn, influence heavy neutrino colliders physics. 
The reasoning can be equally well reversed: high energy collider processes give limits on the heavy sector of a given theory, which can in turn improve predictions for low energy signals. 
  
\section{Acknowledgements}
We would like to thank Marek Zra\l ek and Marek Gluza for useful comments. Work is supported by the Polish National Science Centre (NCN), Grant  No.~DEC-2013/11/B/ST2/04023. RS is supported by Natural Sciences and Engineering Research Council (NSERC) of Canada.
 
 \section{Appendix}
 
\subsection{1. Majorana neutrinos and the high energy processes. 
}
\label{appdir}

We give several remarks related to 
the  $pp\to W_2\to N_a l\to lljj$ process, useful for interpreting experimental data. We use the following notation:
\begin{eqnarray}
&&\sigma_{ij}^{\pm\pm}=\sigma(pp\to l_i^\pm l_j^\pm jj),\\
&&\sigma_{ij}^{\pm\mp}=\sigma(pp\to l_i^\pm l_j^\mp jj)
\end{eqnarray}
and collectivelly denote all these cross-sections by $\sigma_{ij}$.
Here LNV is present when final dileptons have the same charge. Sometimes lepton charge (lepton number) is defined for each lepton family separately $N_e,N_\mu,N_\tau$, (see e.g. \cite{Bilenky:1987ty}).
Then the  lepton number is violated, for instance, in the $\mu^\pm \to e^\pm \gamma$  process. However, instead LNV, it is often called Charged Lepton Flavor Violation (CLFV). So, for $i \neq j$,  the proces Eq.~(8)  breaks both lepton and flavor numbers (LNV, CLFV) while the process Eq.~(9) breaks a flavor number (CLFV).
In the SM possible CLFV effects are completely negligible due to smallness of active neutrino masses. 
(Commonly, we call light neutrinos {{\it active}} as opposed to  {{\it sterile}} or {{\it heavy}} neutrinos). Substantial effects may arise only if nonstandard heavy neutrinos exist.  
   
In the original paper \cite{Keung:1983uu} on the heavy Majorana neutrinos contribution to $pp \to lljj$, heavy neutrino mixings were assumed to be very small, suppressing CLFV  $\mu\to e\gamma$ process. Effectively   Majorana neutrinos with trivial diagonal $K_R$ were assumed. Moreover, due to $\Gamma_a/M_a\ll1$, a simple factorization of the process into $W_2$ production times branching ratios is possible. Then, of course, number of the same sign $eejj$ events is equal to the opposite sign $eejj$ events, a case $r=1$ in Fig.~4 in the main text. 
 However, in general the process depends on  mixing of Majorana states, decay widths, CP phases, Majorana heavy neutrino mass splittings, right-handed gauge boson mass and its gauge coupling $\widetilde{g}$. We found it worthwhile to revisit the case, especially 
as recently the $pp \to lljj$ process has been studied by CMS in \cite{Khachatryan:2014dka}. 
The CMS reported 13 electron-positron-jet-jet ($e^+e^-jj$) events which are above the Standard Model background, and  one event which breaks the lepton number was identified ($e^+e^+jj$), so definitely $r \ll1$.
The CMS report triggered a lot of theoretical activity
and  was a fruitful seed for new ideas in quest of New Physics (NP) at the LHC \cite{Deppisch:2014qpa,Heikinheimo:2014tba, Deppisch:2014zta,Aguilar-Saavedra:2014ola,Vasquez:2014mxa,Senjanovic:2014pva,Ng:2015hba,Dobrescu:2015qna,Brehmer:2015cia,Vasquez:2015una,Dev:2015pga,Coloma:2015una,Gluza:2015goa,
Das:2014jxa,Berger:2015qra,Dobrescu:2015yba,Hisano:2015gna,Krauss:2015nba,Dhuria:2015swa,Deppisch:2015cua,Bandyopadhyay:2015fka,Dev:2015kca,Deppisch:2015bbh,Deppisch:2015qwa,Banerjee:2015hoa,Dhuria:2015cfa,Awasthi:2015ota,Jelinski:2015ifw,Ko:2015uma,Leonardi:2015qna,Collins:2015wua,Dobrescu:2015jvn,Hati:2015awg,Aydemir:2015oob,Garcia-Cely:2015quu,Blumenhagen:2015fqn,Dasgupta:2015pbr,Shu:2015cxm,Hati:2016thk,Das:2016akd,Shu:2016exh}. 
For  earlier studies on $pp\to lljj$ see \cite{Datta:1993nm,Ferrari:2000sp,Kersten:2007vk,delAguila:2007qnc,delAguila:2008hw,Maiezza:2010ic,Tello:2010am,Nemevsek:2011hz,Chen:2011hc,Das:2012ii,Chakrabortty:2012pp,Nemevsek:2012iq,Han:2012vk,AguilarSaavedra:2012gf,Dev:2013oxa,Chen:2013foz,Dev:2013vba} and for more on heavy neutrino physics see for instance \cite{Giunti:2007ry,Mohapatra:1998rq,Kayser:1982br,Bilenky:1987ty} and  \cite{Dicus:1991fk, Pilaftsis:1991ug,Datta:1991mf,Datta:1992qw,Datta:1993nm,Gluza:1993gf,Gluza:1995js,Gluza:1997kg,Gluza:1997ts,delAguila:2005ssc,Han:2006ip,delAguila:2007qnc,Atre:2009rg,Ibarra:2010xw,Adhikari:2010yt,Chakrabortty:2012pp,Cely:2012bz,BhupalDev:2012zg,Das:2012ze,Dev:2013wba,Helo:2013esa,Alva:2014gxa,Antusch:2015mia,Banerjee:2015gca,Arganda:2015ija,Das:2016hof}.

Let us note that the ATLAS collaboration in the LHC run 1 \cite{Aad:2015xaa}
analyzed only possibility of Majorana neutrinos detection (search for SS dilepton final signals), so a possibility for OS  dilepton excess has been missed.
As similar situations can appear in future experiments,  it is therfore wise to know how far in interpretations of possible signals we can go with heavy neutrinos and right-handed currents physics, and how effectively parametrize cross sections. For initial discussion, see \cite{Gluza:2015goa}.
 
First, let us enumerate quantities which $\sigma_{ij}$ depend on. Relevant electroweak dimensionless parameters are: gauge couplings $\widetilde g$ (and $g$)
and the neutrino mixing matrix $K_R$.
On the other hand, the mass scales which are important for the process are: $M_{W_2}$, $M_a$ and $\sqrt{s}$. 

We shall discuss two setups in which it is possible to factor out the dependence on the mixing matrix $K_R$ from dependence on the above-mentioned scales.
In the following we focus on the scenario in which all heavy neutrinos are lighter than $W_2$. 
In such a case one can estimate the magnitude of decay widths of $W_2$ and $N_a$, $\Gamma_{W_2}$ and $\Gamma_a$ respectively: 
\begin{eqnarray}
\frac{\Gamma_{W_2}}{M_{W_2}}&=&\frac{g_R^2}{96\pi}\left(\sum_{a}F_W\left(x_a\right)+18\right)\sim10^{-2},\\
\frac{\Gamma_a}{M_a}&=&\frac{9g_R^4}{1024\pi^3}F\left(x_a\right)\sim10^{-5},
\end{eqnarray}
where $x_a=M_a^2/M_{W_2}^2$, while $F_W(x)$ and $F(x)$ are the following kinematical functions:
\begin{eqnarray}
F_W(x)&=&(2-3x+x^3)\theta(1-x),\\
F(x)&=&\frac{12}{x}\left[1-\frac{x}{2}-\frac{x^2}{6}+\frac{1-x}{x}\ln(1-x)\right].
\end{eqnarray}
In both cases $\Gamma/M\ll1$. Narrow Width Approximation (NWA) is valid  in the region in which $M_{W_2}>M_{N_{i}}$ and $M_{W_2}$ is neither close to $M_{N_i}$ nor to $\sqrt{s}$ (where the distance is measured in $\Gamma(W_2)$ units).
It turns out that when the mass splittings $M_a-M_b$, $a\neq b$ between heavy neutrinos are much different than $\Gamma_a$, then one can express $\sigma_{ij}$ as a product of two terms: one which depends only on mixing matrix $K_R$ and the second which depends on the remaining variables i.e. $\widetilde g$, $f(x,Q^2)$, $\sqrt{s}$, $M_{W_2}$ and $x_a=M_a^2/M_{W_2}^2$. We shall discuss two cases:
\begin{itemize}
\item[(a)] non-degenerate heavy neutrinos: \\
 $\min_{a\neq b}|M_a-M_b|\gg \max_a \Gamma_a$;
\item[(b)] degenerate heavy neutrinos:\\ $\max_{a\neq b}|M_a-M_b|\ll \min_a \Gamma_a$. 
\end{itemize}

\subsubsection{Case (a)} 
In this case  interferences between different diagrams are negligible. One can factor out dependence on the mixing matrix $K_R$ in the following way: 
\begin{eqnarray}
\sigma_{ij}^{\pm\pm}&=&\sum_a\widehat{\sigma}^{\pm\pm}_a|(K_R^\dag)_{ia}(K_R^*)_{aj}|^2+\mathrm{INT}, \label{bare1}\\
\sigma_{ij}^{\pm\mp}&=&\sum_a\widehat{\sigma}^{\pm\mp}_a|(K_R^\dag)_{ia}(K_R)_{aj}|^2+\mathrm{INT}, \label{bare2}
\end{eqnarray}
where $\mathrm{INT}$ stands for small interference terms, while $\widehat{\sigma}^{\pm\pm}_a$, $\widehat{\sigma}^{\pm\mp}_a$ are 'bare' cross-sections calculated for $(K_R)_{aj}=\delta_{aj}$. Obviously, they depend only on $\widetilde{g}$, $\sqrt{s}$, $M_{W_2}$, $x_a$ but not on $K_R$. Due to $\Gamma_a\ll M_a$ such `bare' cross-section can be approximated by:
\begin{eqnarray}
\widehat{\sigma}_a^{\pm\pm}&=&\sigma(pp\to W_2^\pm)\nonumber\\
&&\times\mathrm{BR}(W_2^\pm\to N_al_1^\pm)\mathrm{BR}(N_a\to l_1^\pm jj),\label{sighata1}\\
\widehat{\sigma}_a^{+-}&=&[\sigma(pp\to W_2^{+})+\sigma(pp\to W_2^{-})]\nonumber\\
&&\times\mathrm{BR}(W_2^{+}\to N_al_1^{+})\mathrm{BR}(N_a\to l_1^{-}jj).\label{sighata2}
\end{eqnarray}
By direct computation one can check that \eqref{sighata1} and \eqref{sighata2} lead to the following prediction of the ratio of the number of same-sign events ($N_{SS}$) to the number of opposite-sign events ($N_{OS}$): 
\begin{equation}
r_{ij}=\left(\frac{N_{SS}}{N_{OS}}\right)_{ij}=\frac{\sigma^{++}_{ij}+\sigma^{--}_{ij}}{\sigma^{+-}_{ij}}\approx1
\end{equation}
regardless on the form of $(K_R)_{aj}$. An example of such behaviour of $r_{ij}$ is presented in the Fig.~4, the main text. On that plot, the horizontal line corresponds to the value $r_{ee}$ in the scenario when splitting between masses of $N_1$ and $N_3$ is about $1\,\mathrm{GeV}$ while $\Gamma_a\sim1\,\mathrm{MeV}$.

\subsubsection{Case (b)}

In this case interferences are very important as they heavily influence the cross-section. We consider a setup in which all three heavy neutrinos have the same mass. Analogously to case (a), it is convenient to factor out dependence on the mixing matrix $K_R$:
\begin{eqnarray}
\sigma_{ij}^{\pm\pm}&=&\widehat{\sigma}^{\pm\pm}\left|\sum_a(K_R^\dag)_{ia}(K_R^*)_{aj}\right|^2, \label{bare3}\\
\sigma_{ij}^{\pm\mp}&=&\widehat{\sigma}^{\pm\mp}\left|\sum_a(K_R^\dag)_{ia}(K_R)_{aj}\right|^2. \label{bare4}
\end{eqnarray}
Let us note that when $K_R$ is unitary, then  $\sigma_{ij}^{\pm\mp}=\widehat{\sigma}^{\pm\mp}\delta_{ij}$. $\widehat{\sigma}$ are  cross-sections calculated for $(K_R)_{aj}=\delta_{aj}$. Due to $\Gamma_a\ll M_a$ the `bare' cross-sections $\widehat{\sigma}^{\pm\pm}$ and $\widehat{\sigma}^{\pm\mp}$ can be approximated by:
\begin{eqnarray}
\widehat{\sigma}^{\pm\pm}&=&\sigma(pp\to W_2^\pm)\nonumber\\
&&\times\mathrm{BR}(W_2^\pm\to N_1l_1^\pm)\mathrm{BR}(N_1\to l_1^\pm jj),\label{sighat1}\\
\widehat{\sigma}^{+-}&=&[\sigma(pp\to W_2^{+})+\sigma(pp\to W_2^{-})]\nonumber\\
&&\times\mathrm{BR}(W_2^{+}\to N_1l_1^{+})\mathrm{BR}(N_1\to l_1^{-}jj).\label{sighat2}
\end{eqnarray}
what leads to 
\begin{equation}
r_{ij}=
\frac{\sigma^{++}_{ij}+\sigma^{--}_{ij}}{\sigma^{+-}_{ij}}\approx\left|\sum_a(K_R^\dag)_{ia}(K_R^*)_{aj}\right|^2.
\end{equation}
When $K_R$ is unitary then one obtains $r_{ij}\leq1$.
\\

The results shown in Fig.~4 in the main text and vanishing of $r$-dependence for heavy Majorana mass splittings can be understood as follows. 
The LHC kinematics and masses of particles allows for the exchange of a neutrino  close to its mass shell. Then the total cross section is dominated by the exchange of the neutrino in the $s$-channel. Close to the mass pole, the neutrino propagator can be described by the relativistic Breit-Wigner distribution. If we consider mixing between neutrino states, amplitudes corresponding to different mass eigenstates have to be added coherently. The interference between amplitudes corresponding to different mass eigenstates can be destructive, leading to a suppression of the SS lepton production cross section. 

The size of the interference term can compete with the resonant contribution of the squared amplitudes only if the neutrino mass difference is of the order of the neutrinos widths. This kind of effects have been considered already in a model without right-handed currents in \cite{Bray:2007ru}. In our case the neutrino decay width is naturally very small, at the MeV level, and large interferences are possible, see the line $r_0$ in Fig.~4 in the main text. Otherwise the interference term contributes only to the continuum background and is negligible for the considered process (line $r_1$). Therefore, a suppression of the SS signal occurs only if the heavy Majorana neutrinos are (almost) degenerate.

In our numerical analysis scalar decay modes of heavy neutrinos are negligible (heavy neutrino masses are smaller than heavy scalar masses) and  a decay mode $N \to e^\pm W_1^\mp$ is dominating. For a discussion of possible decay modes scenarios of heavy neutrinos, see for instance \cite{Bambhaniya:2013wza} (minimal left-right model) and \cite{Pilaftsis:1991ug,Bray:2007ru} (models with heavy neutrino singlets).

\begin{figure}[t!]
\begin{center}
\includegraphics[width=0.75\columnwidth]{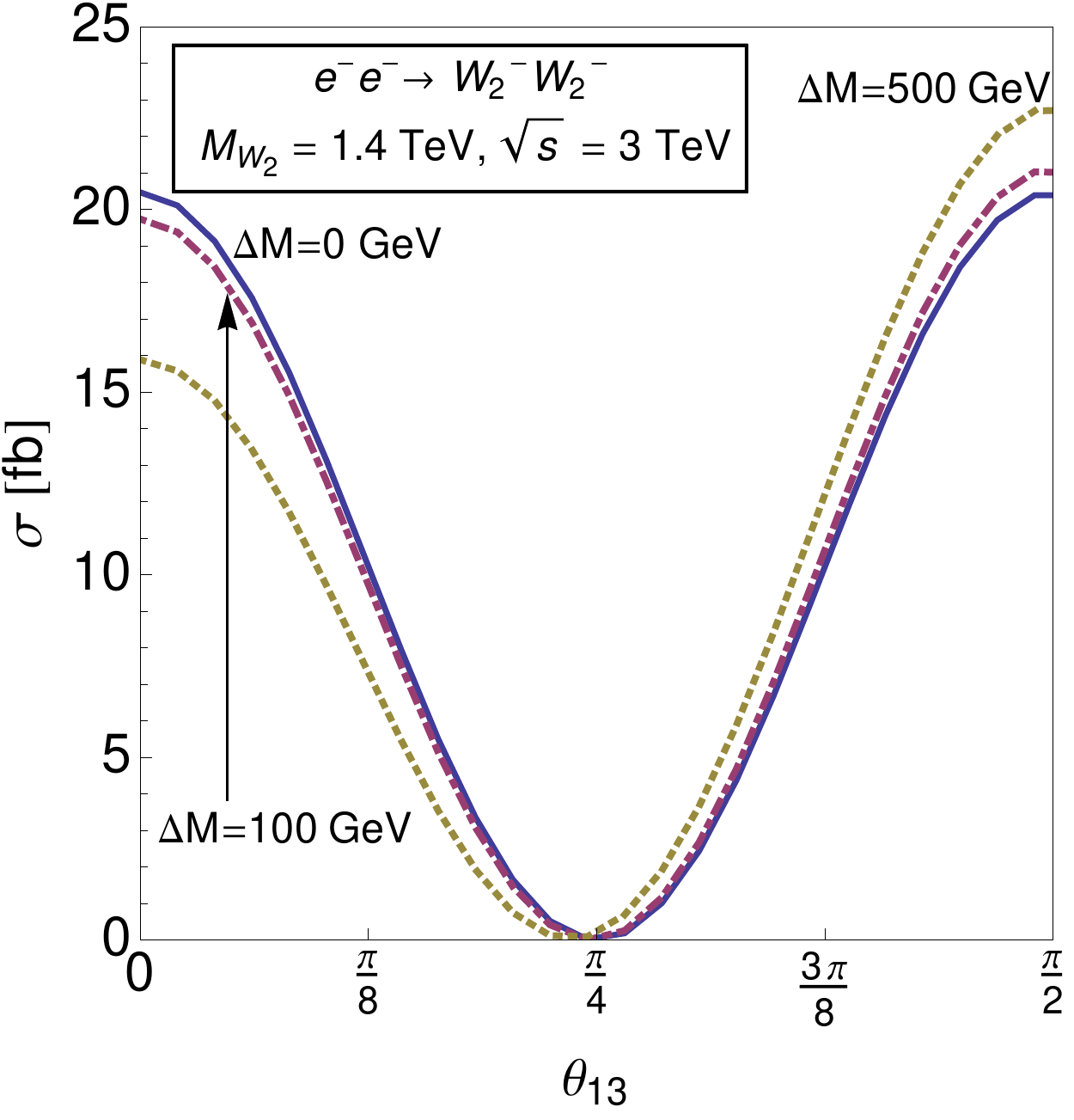}
\caption{\label{eeww}
Interference effects in the $e^-e^- \to W_2^-W_2^-$ LNV process for 
three different splittings $\Delta M=0$, $100$ and $500\,\mathrm{GeV}$ between masses of heavy neutrinos $N_{1,3}$. $M_{W_2}=1.4$ TeV, $\sqrt{s}=3$ TeV, $M_{1,3}=1\,\mathrm{TeV}\pm \Delta M/2$, 
$M_{2}=10\,\mathrm{TeV}$, $K_R$ is given as in Eq.~(6) in the main text, $\phi_3=\pi/2$.}
\end{center}
\end{figure}

We can also think about a situation where a process is $t$-channel dominated. In the $t$-channel, neutrino propagators are far from their poles. Here interferences can be substantial even for a very large Majorana neutrino mass splittings comparing to decay widths because the signal is
not dominated by the resonant contribution. Each neutrino propagator is a slowly varying function of
energy and mass. Large interference effects can effectively lead to LNC (destructive interference may totally suppress the cross section).
This can happen for another variant of the diagram shown in  a dashed box of Fig.~1 in the main text, namely $e^-e^- \to W_2^-W_2^-$ \cite{Belanger:1995nh,Gluza:1995ky,Gluza:1995ix,Asaka:2015oia,Antusch:2015mia,Banerjee:2015gca}. The numerical results 
are shown in Fig.~\ref{eeww}. The $\theta_{13}$ asymmetry for the cross section and a shift in a minimum is due to different weights in amplitudes of $N_1$ and $N_3$.

This example is of pedagogical value as $W_2$ taken to be $1.4$ TeV is much below present LHC direct limit, which is about $3$ TeV. However, note that this  strict LHC limit is based on non-degenerate and no-mixing heavy neutrino scenario (for consequences, see Fig.~3 and a discussion in the main text). 
The point is that relatively large cross section can be obtained only for on-shell $W_2$ pair production, and foreseen 
lepton colliders center of mass energies are atmost at the $3$ TeV level. Heavier off-shell $W_2$ pair production would require very high luminosities for a detection of this signal as the cross section drops down quickly below $\mathcal{O}$$(1)$ fb with increasing $W_2$ mass.

\subsection{2. Low energy CLFV processes}
\label{applfv}
 
In the Standard Model the  CLFV effects are negligible due to the small masses of
light, active neutrinos \cite{Lee:1977tib}. Similarly to the $pp\to lljj$ process, for substantial effects new heavy particles are needed.  
The best limits have been obtained so far in the muon-electron
sector, though nowadays also procesess involving the tau leptons start to play a role. Present and planned limits for the most important low energy CLFV processes are gathered  in Tab.~\ref{table2}.
 \begin{table}[h!]
 \noindent \begin{centering}
 \begin{tabular}{|c|c|c|}
 \hline 
 Process & Current Limit & Planned Limit\tabularnewline
 \hline 
 \hline 
 $\tau\rightarrow\mu\gamma$ & $4.4\times 10^{-8}$ & $3.0\times 10^{-9}$\tabularnewline
 \hline 
 $\tau\rightarrow e \gamma$ & 3.3$\times 10^{-8}$ & $3.0\times 10^{-9}$ \tabularnewline
 \hline 
 $\tau\rightarrow\mu\mu\mu$ & $2.1\times 10^{-8}$ & $1.0\times 10^{-9}$\tabularnewline
 \hline 
 $\tau\rightarrow eee$ & $2.7\times 10^{-8}$ & $1.0\times 10^{-9}$\tabularnewline
 \hline 
 $\mu\rightarrow e\gamma$ & $5.7\times 10^{-13}$ & $6.0\times 10^{-14}$\tabularnewline
 \hline 
 $\mu\rightarrow eee$ & $1.0\times 10^{-12}$ & $1.0\times 10^{-16}$\tabularnewline
 \hline 
 $\mu N\rightarrow eN$ & $7.0\times 10^{-13}$ & $1.0\times 10^{-17}$\tabularnewline
 \hline 
 \end{tabular}
 \par\end{centering}
\caption{Current and planned limits on 
the CFLV branching ratios
  \cite{Agashe:2014kda,Baldini:2013ke,Aushev:2010bq,
    Bartoszek:2014mya}. For the muon coherent conversion process $\mu N\rightarrow eN$ the limit
  is given as a ratio of  the conversion rate to
 the muon nuclear capture $\mu N\rightarrow \nu_\mu N'$ rate. 
}\label{table2}
\end{table}

Our choice of the mixing matrix, Eq.~6 in the main text, implies that a
mixing is present only between taus and electrons in the discussed model. Let us consider CLFV process $l\rightarrow l' \gamma$, in our case 
we are interested in an estimation of the nonzero $\tau
\rightarrow e \gamma $  branching ratio. 
The main
contribution to this process comes from the diagram containing virtual
heavy neutrinos.  The branching ratio for the general case $l\rightarrow
l'\gamma$ \cite{Bu:2008fx}, adopted to our case and including right-handed currents  is
\begin{eqnarray}
&\mathrm{BR}(l  \rightarrow  l'\gamma) \approx  \frac{3 }{8}\frac{\alpha}{\pi}\left(\frac{\widetilde{g}}{g}
                          \frac{M_{W_{1}}}{M_{W_{2}}}\right)^{4}
                               \nonumber &\\ 
 &\times  \left|  \sum\limits_{a}
  \left(K_{R}^{\dagger}\right)_{l'a}\left(K_{R}^{}\right)_{al}F\left(\frac{M_{a}^{2}}{M_{W_{2}}^{2}}
  \right)  \right|^{2},  &\label{eq:brllg}
\end{eqnarray}
where $M_{a}$ is the mass of the heavy neutrino and  

\begin{eqnarray}
F(x)&=&\frac{10-43x+78x^{2}-49x^{3}+4x^{4}+18x^{3}\ln
          x}{6\left(1-x\right)^{4}}. \nonumber \\
\end{eqnarray}
For unitary $K_R$ we can add any constant to the function $F(x)$
without affecting the branching ratio. It is convenient to define a
new function $\phi(x)$ such that $\phi(0)=0$. This can be obtained by
a redefinition
\begin{eqnarray}
F(x)\rightarrow \phi(x) = F(x)-F(0)= \nonumber
\\
-\frac{x \left[1 -6 x +3 x^2+2 x^3-6 x^2 \ln (x)\right]}{2 (1-x)^4}. 
\end{eqnarray}
In this instance we recover the classical result
\cite{Langacker:1988up}, which is valid only for the unitary mixing
matrix, but for any mass ratio of the neutrinos and vector bosons.
The function $\phi(x)$ is monotonically increasing and bounded in the physical domain, 
$0\leq\phi(x)< 1$ for $0 \leq x<\infty $. This allows us to derive an upper bound on the branching
ratio in the case of mixing between two generations 
\begin{eqnarray}
\mathrm{BR}(l  \rightarrow  l'\gamma) &<&  \frac{3 }{8}\frac{\alpha}{\pi}\left(\frac{\widetilde{g}}{g}
                          \frac{M_{W_{1}}}{M_{W_{2}}}\right)^{4}
                        \left(\sin\theta_{13}\cos\theta_{13}\right)^2
  \nonumber \\
&\leq & \frac{3 }{32}\frac{\alpha}{\pi}\left(\frac{\widetilde{g}}{g}
                          \frac{M_{W_{1}}}{M_{W_{2}}}\right)^{4}.
\end{eqnarray}

For completness let us also recall the formula  for
$M_{a}/M_{W_{2}}\ll1$. In this case Eq.~(\ref{eq:brllg})
can be futher simplified 
\begin{equation}
\mathrm{BR}(l  \rightarrow  l'\gamma)
=\frac{3\alpha}{32\pi}\left(\frac{\widetilde{g}}{g} \frac{M_{W_{1}}}{M_{W_{2}}}\right)^{4}\left(\sin\theta_{13}\cos\theta_{13}\frac{\Delta m_{12}^{2}}{M_{W_{2}}^{2}}\right)^{2}.\label{eq:mu2egamma}
\end{equation}

For $M_{W_{2}}=2.2$ TeV and $g=\widetilde{g}$ the branching ratio is
suppressed by
$(3\alpha/32\pi)\times\left(M_{W_1}/M_{W_2}\right)^{4}\sim4\times 10^{-10}$.
It gives a good estimation of the order of magnitude of the CLFV
effect.  To investigate it more carefully we chose the
maximal mixing $\theta_{13}=\pi/4$ and large mass difference between $N_1$
and $N_3$. (Note that for unitary $K_{R}$ the contribution does not depend  directly
on the absolute values of masses of neutrinos, but rather on their  difference, just as it is in the case of light neutrinos). 

\begin{figure}[h!]
\begin{center}
\includegraphics[width=\columnwidth]{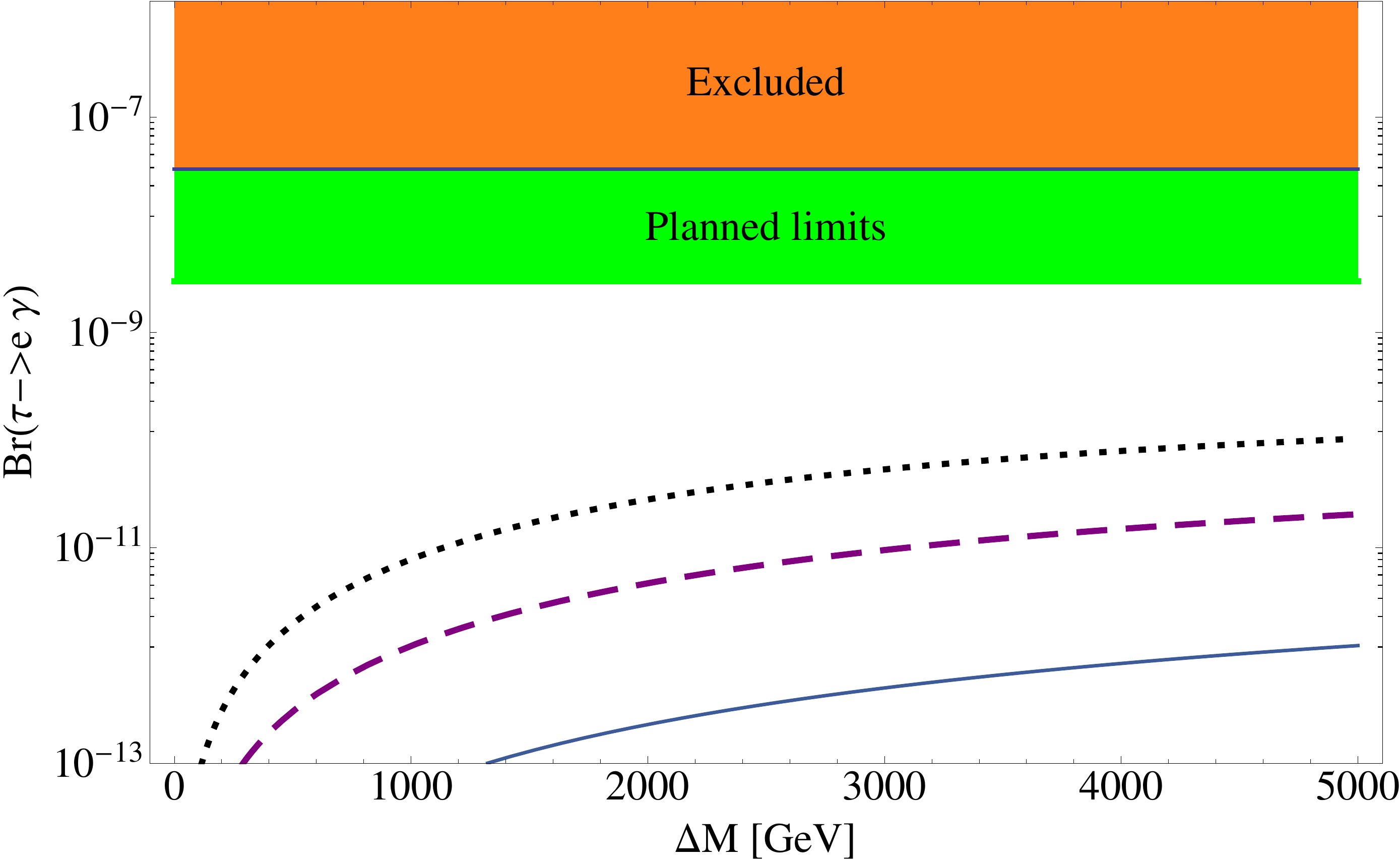}
\caption{
\label{meg2}
$\tau \rightarrow e \gamma$ branching ratio as a function of
mass splitting $\Delta M=M_3-M_1$. The mixing between the first
and the third generation is assumed to be maximal.  We chose
$M_{W_{2}}=2.2\,\mathrm{TeV}$  (dotted line),
$M_{W_{2}}=3\,\mathrm{TeV}$  (dashed-line), $M_{W_{2}}=5\,\mathrm{TeV}$
(solid line). $M_1=M_{W_{2}}/2$. 
  }
  \end{center}
  \end{figure}

In Fig.~\ref{meg2}  the branching ratio  $\tau
\rightarrow e \gamma $ is plotted for the maximal mixing between the first and the third generation.  We assume
$M_1=M_{W_{2}}/2$ and we vary $M_3$. Different lines
correspond to $M_{W_2}$ equal to $2.2$ TeV, 3 TeV and 5 TeV.

We can see that the $K_R$ parametrisation for effective
mixture of two heavy Majorana neutrinos given in Eq.~(6) in the main
text fullfils relevant CLFV limits. It is instructive to notice that
even if the planned limit would  improve further by an order of
magnitude, we would be able to probe only the large
mass splittings at the TeV level.

\subsection{3. Relations among heavy neutrino masses, CP-phases and mixing matrix elements. \label{appcp}} 

Complex couplings in interactions and complex elements of mass matrices may lead to CP-violating effects. Gauge transformations and unitary transformations on fields may reduce number of CP-phases \cite{Bilenky:1987ty,delAguila:1996ex}. To see which mass matrix leads to degenerate neutrino masses and to the mixing matrix $K_R$ discussed in the main text, enough is to consider two neutrino case ($N_2$ does not mix in Eq.~(6)). Studying this case in detail, also connection between CP-parities and mass eigenvalues can be seen.  

Let us start from a general $2\times2$ complex symmetric mass matrix (symmetry of the matrix  comes from Hermitian conjugated mass terms)
\begin{equation}
M = \left( 
\begin{matrix}
a e^{i \alpha} & b e^{i \beta} \\ b e^{i \beta} & c e^{i \gamma}
\end{matrix}
\right) .
\end{equation}
In general it can be diagonalized using bi-unitary matrices. However, taking $M^\dagger M$, Hermitian matrix emerges which can be diagonalized using single unitary matrix $V$
\begin{eqnarray}
&&V^\dagger (M^\dagger M) V = \mathrm{diag}(m_1^2, m_2^2),\label{dagg}
\\
&& \nonumber \\
&& m_{1,2}^2 = \frac{1}{2} \left[ a^2+c^2+2b^2
\pm \sqrt{(a^2-c^2)^2+\Omega} \right], \label{m1m2}\\ 
&& \nonumber \\
&&\Omega = 4 b^2[a^2+c^2+2 ac \cos(\alpha+\gamma-2 \beta)].
\end{eqnarray}
We can see that $m_1=m_2$ if $a=\pm c$ and $\alpha+\gamma-2 \beta=\pi$ (so element $b$ can take any value).

In general, squared eigenvalues in Eq.~\eqref{m1m2} are still complex, so we assume that for the same matrix $V$, 
\begin{equation}
V^T M V = \left( 
\begin{matrix}
m_1 e^{-2 i \phi_1} & 0 \\ 0 & m_2 e^{-2 i \phi_2} 
\end{matrix}
\right) 
\equiv 
\left( 
\begin{matrix}
m_1 \rho_1 & 0 \\ 0 & m_2 \rho_2
\end{matrix}
\right)  \label{transp}
\end{equation}
where $m_{1,2}\geq 0$, $m_{1,2} \in \mathbb{R}$ and $\rho_{1,2}=\pm 1$. Note that the Hermitian conjugation of $V$ in Eq.~\eqref{dagg} has been  replaced by transposition as $(M^\dagger M )^\ast = M^\dagger M$.

Absorbing phases into $V$, we get $U=V\mathrm{diag}(e^{i \phi_1},e^{i \phi_2})$.
Taking $V$ which leads to the squared eigenvalues in a following form
\begin{eqnarray}
V &=& \left( 
\begin{matrix}
\cos{\xi} & \sin{\xi} \\ -  e^{i \delta} \sin{\xi} &   e^{i \delta} \cos{\xi} \label{transp1} 
\end{matrix}
\right) , 
\end{eqnarray}
$\xi$ and $\delta$ are fixed through the relation Eq.~\eqref{dagg}
\begin{eqnarray}
\tan 2\xi &=& \frac{2 |ab  e^{i(\alpha-\beta)}+bc  e^{i(\beta-\gamma)}|}{c^2-a^2},  \\
&& \nonumber \\
e^{i \delta}&=& \frac{ab  e^{i(\alpha-\beta)}+bc  e^{i(\beta-\gamma)}}{|ab  e^{i(\alpha-\beta)}+bc  e^{i(\beta-\gamma)} |}.
\end{eqnarray}
We arrived in a general form of the unitary matrix $U$ which diagonalizes $M$ and gives positive eigenvalues [$s_\xi\equiv\sin{\xi}$, $c_\xi\equiv\cos\xi$]
\begin{eqnarray}
U &=& \left( 
\begin{matrix}
c_\xi e^{i \phi_1}& s_\xi e^{i \phi_2}\\ - s_\xi e^{i (\delta+\phi_1)} &  c_\xi e^{i (\delta+\phi_2)} \label{transp2}
\end{matrix}
\right),\\
&&\nonumber \\
e^{-2i \phi_1}&=& \frac{a c^2_\xi e^{i\alpha}-c s^2_\xi e^{i(\gamma+2\delta)}}{m_1 \cos{2 \xi}} , \\
e^{-2i \phi_2}&=& \frac{c c^2_\xi e^{i( \gamma +2\delta)}-a s^2_\xi e^{i\alpha}}{m_2 \cos{2 \xi}}.
\end{eqnarray}
Phases $\phi_{1,2}$ are fixed by solving the relation $U^T M U = \mathrm{diag}(m_1,m_2)$.

Let us assume real matrix $M$, $\alpha=\beta=\gamma=0$, $a>0$, $b>0$, $c<0$.
Such a matrix has one positive and one negative, real eigenvalue.
Assigning $\rho_1<0$, $\rho_2>0$, then $\phi_1=\pm \pi/2$, $\phi_2=0$
\begin{eqnarray}
U &=& \left( 
\begin{matrix}
i c_\xi & s_\xi \\ - i s_\xi  &  c_\xi  \label{ufin}
\end{matrix}
\right),
\end{eqnarray}
\\
which corresponds, up to transposition, to the mixing matrix $K_R$
 between $N_1$ and $N_3$ states in  Eq.~(6) in the main text (with $\phi_3=\pi/2$).
In addition,  for $a=-c$, $|m_1|=|m_2|$ (as already shown before).

We can see that the first column in Eq.~\eqref{ufin} is  just multiplied by $i$. It is connected with CP parities of neutrinos. To establish this
relation, the interaction term Eq.~(2) in the main text must be also studied. 
Let us first note that for degenerate $N_a$  states there is some 
symmetry in a mass term, as  $M \sim N^T N$, e.g. for our two-dimensional case we have
\begin{equation}
\left( 
\begin{matrix}
N_1 \\ N_2  
\end{matrix}
\right) = 
\left( 
\begin{matrix}
\cos \alpha & -\sin \alpha \\ \sin \alpha & \cos \alpha  
\end{matrix}
\right)
\left( 
\begin{matrix}
N_1' \\ N_2'  
\end{matrix}
\right) 
\end{equation} 

We demand that Lagrangian Eq.~(2) expressed through prime fields differs from its original by the phase factor $e^{\pm i \alpha}$, which can be absorbed later on, schematically
\begin{eqnarray}
\mathcal{L} &\sim& [\overline{N}_1 (U^T)_{1a} + \overline{N}_2 (U^T)_{2a}] l_{b} \\
&& \nonumber \\
&=& 
\overline{N}_1' \left[ \cos \alpha (U^T)_{1a} + \sin \alpha (U^T)_{2a}\right]  l_{b} \\
&& \nonumber \\
&+& \overline{N}_2' \left[ -\sin \alpha (U^T)_{1a} + \cos \alpha (U^T)_{2a}\right] l_{b}.
\end{eqnarray}
Now, if $(U^T)_{1a} = \pm i (U^T)_{2a}$,
\begin{eqnarray}
\mathcal{L} &\sim& 
e^{\pm i \alpha} \left( \frac{\overline{N}_1' \pm i \overline{N}_2'}{\sqrt{2}}\right) \sqrt{2}   (U^T)_{2a}  l_{b} \\
&& \nonumber \\
&\equiv&
\Phi_D (U^T)'_{2a},
\end{eqnarray}
 where $\Phi_D\equiv e^{\pm i \alpha} (\overline{N}_1' \pm i \overline{N}_2')/\sqrt{2}$ and $(U^T)'_{2a} \equiv \sqrt{2}   (U^T)_{2a}$.

As we can see, invariance of the Lagrangian for two Majorana neutrinos
with $\alpha$ symmetry implies a Dirac neutrino.  It remains to show
that these two Majorana neutrinos have opposite CP parities. 
Imaginary elements of the matrix Eq.~\eqref{transp2} will affect any terms
in the Lagrangian that are linear in the $N_1$. Those terms under CP symmetry
will change the sign, leading to apparent maximal CP
violation. However, we can also change the CP phase of the neutrino
field $N_2$. CP conservation can be restored if the field
transforms as $N_1\rightarrow -N_1$. More generally, if
$N_1\rightarrow \eta_{CP} N_1$, then $N_2\rightarrow -\eta_{CP} N_2$ \cite{Wolfenstein:1981rk}.  
 With this transformation properties our theory is CP conserving and, as
expected, both Majorana fields have opposite CP parities. 

It is a general
situation. For CP conserving case, neutrino $N_1$ with a negative mass eigenvalue has opposite CP parity to a neutrino $N_2$ with a positive mass eigenvalue, $-\eta_{CP}(N_1)=\eta_{CP}(N_2)=i$ \cite{Bilenky:1987ty}. In such a case, corresponding columns in the mixing matrix Eq.~\eqref{ufin} are purely real or complex.

It is worth mentioning that non-trivial mixing angles, as $\theta_{13}$ in $K_R$, Eq.~(6), have a physical meaning for degenerate neutrinos, only if CP parities of neutrinos are different {{(in other words, the mixing matrix can not be real)}}.  First, let us consider some process regardless of neutrino nature. In this case, the unitarity of
$K_R$ and degeneracy of neutrino masses makes the rate for this
process independent of the mixing angle. This is observed
for all LNC processes, like $\mu \rightarrow e \gamma$ or
$W_2^-\rightarrow e^- \overline{N}_a$.
On the other hand, if some
process is permitted only for Majorana neutrinos then the dependence on the
mixing angle can appear only if the Majorana phase is non-zero. This
Majorana process usually involves a charge conjugation operator and
its amplitude depends on elements of the matrix $K_R K_R^T$. 
Typical example is the double neutrinoless beta decay.
For neutrinos with different CP
parities this matrix is not  unity and can have a non trivial
dependence on the mixing angle. However, if  $K_R$ is real i.e. orthogonal, the neutrinos
have equal CP parities and can not be distinguished. 
For the corresponding $pp \to lljj$ cases, see Eqs.~(\ref{bare3}),(\ref{bare4}).

\subsection{4. Swinging between Dirac and Majorana states: from history to the last theoretical concepts  \label{appDM}} 

Majorana self-conjugated fields and operators are currently
being used in a wider context,  expanding to solid state physics (superconductors in the presence of vortices, fermionic lattice systems) or quantum statistics and computing \cite{wilczek,Elliott:2014iha,2001PhyU...44..131K}. In particle physics itself neutrinos as Majorana fermions are not the only possible option. There is a plethora of Majorana particles within supersymmetric models, such as a gluino or neutralinos. Some of these non-standard supersymmetric particles might hold the key to solving another mystery, which is the Dark Matter puzzle \cite{Bertone:2004pz}.

Until the end, let us remain with neutrino physics. Of course, the simplest description of neutrino states is possible in a massless limit. In this case Lorentz group implies two irreducible two-dimensional representations \cite{waerden1929} described by the two, two-dimensional Weyl equations \cite{Weyl:1929fm}. 
Van der Waerden--Weyl spinors built up Dirac states, which are already massive and restore P- and T- symmetries \cite{Dirac:1928hu,Dirac:1928ej}. 
Majorana in his seminal paper \cite{Majorana:1937vz} noted\footnote{The  idea has been known since 1933 and it took Fermi a while to convince Majorana to publish it.}    that a Dirac spinor $\Phi_{D}$ can be decomposed simple into two fields $\chi_1$, $\chi_2$ (for our purposes, this is a simplified version of quantized field equations which can be found in \cite{Majorana:1937vz})

\begin{eqnarray} 
\Phi_{D} &=& \frac{1}{\sqrt{2}} \left( \chi_1 + i  \chi_2 \right), \\
\chi_{1} &=& \frac{1}{\sqrt{2}} \left( \Phi_{D} +   \Phi_{D}^C \right), \label{c1} \\
\chi_2 &=& \frac{-i}{\sqrt{2}} \left( \Phi_{D} -   \Phi_{D}^C \right). \label{c2}
\end{eqnarray}
$C$  in Eqs.~(\ref{c1},\ref{c2}) is a charge operator defined with help of Dirac $\gamma$ matrices. Here we can refer for details to specialized textbooks  \cite{Bilenky:1987ty,Kayser:1989iu,Giunti:2007ry} or  modern QFT textbooks \cite{Maggiore:2005qv} where Majorana fields are discussed.
  
These new fields still obey Dirac equations, and are self-conjugated, 
  $\chi_{1,2}=\chi_{1,2}^c$. Due to the additional overall complex factor in Eq.~\eqref{c2}, it can be shown that $\chi_1$ and $\chi_2$ have opposite CP parities, similarly as was done in the previous section.
  The only technical difference with the Dirac equation is that the $\gamma^\mu$ matrix representation is purely complex for the Majorana case. 
  Majorana fields $\chi_{1,2}$ are just real and imaginary parts of the Dirac equation for a particle   and an antiparticle, respectively.
  In his work, Majorana wrote
  (translated from Italian by Luciano Maiani)
  "Even though it is perhaps not yet possible to ask experiments to decide between the new theory and a simple extension of the Dirac equations to neutral particles, one should keep in mind
  that the new theory introduces a smaller number of hypothetical entities, in this yet unexplored field."
    It is thus said that Majorana fields are more economical, i.e. Majorana fields are more concise in terms of parameters.
    Indeed, only two degrees of freedom are needed (either a left- or a right-handed Weyl spinor) to describe massive fermions. In the Majorana case the right-handed Weyl spinor can be expressed through a left-handed one, and vice versa. 
  
Majorana $\chi_{1,2}$ fields do not carry any $U(1)$ charge $Q_M$, as 
\begin{eqnarray}
j_i^\mu&=& Q_M \bar{\chi}_i \gamma^\mu \chi_i = Q_M \bar{\chi}_i^c \gamma^\mu \chi_i^c\nonumber\\ 
&=&-Q_M \bar{\chi}_i \gamma^\mu \chi_i=0\;.
\end{eqnarray}
They are strictly neutral and no global $U(1)$ symmetry and conserved number such as
the lepton number $L$ can be assigned to them. This feature is
directly related to the presence of a Majorana mass term in the
Lagrangian. 
Let us take for example  a mass term $\nu_L^c \nu_R$ that includes
right-handed spinors.  We have $\nu_L^c \nu_R = -C \nu_R^T \nu_R$ that is not invariant under any global 
$U(1)$ transformation. In particular, the lepton number $L$ is
defined as $\nu_R \to e^{i L} \nu_R$ and the Majorana mass term
breaks $L$ by two units.

In the limit of vanishing  mass Dirac and Majorana neutrinos are both equivalent to Weyl fermions and the lepton conservation is restored
\cite{Kayser:1989iu,Giunti:2007ry,
Maggiore:2005qv}. 
It is also known that in the Standard Model (meaning only left-handed
interactions and neutrino masses in the eV range), active Dirac and Majorana neutrinos are practically not
distinguishable, as in almost all processes involving neutrinos,
their masses are much smaller than the energies involved and the
helicity plays a role of a conserved charge (non-standard or wrong helicities are suppressed by the neutrinos' mass over energy ratios). This is known as Dirac-Majorana practical confusion theorem \cite{Kayser:1982br}.
Let us note that an interesting option to tell apart Majorana from Dirac neutrinos lighter than the $W^\pm$ gauge boson mass has been proposed recently in \cite{Dib:2015oka}. The idea is to study the muon spectrum in the $W^+$ decays. 

How to detect lepton number violation? We need to look for some rare processes. 
It has been noticed  \cite{Racah:1937qq} that Majorana neutrinos can  lead to emission of two electrons in the following chain of nuclear reactions
$$\left.
\begin{array}{l}
(A,Z)\to (A,Z+1)+e^-+\nu \\
 \nu + (A',Z') \to (A',Z'+1) + e^-
 \end{array} 
 \right\} \to 
 \begin{array}{l}
 (A,Z+1)+\\
 (A',Z'+1)+\\ 2 e^-
 \end{array},
$$
which obviously breaks the total lepton number by two units.

In 1939 Furry proposed another reaction \cite{Furry:1939qr} which
would undoubtfully reveal Majorana nature of neutrinos
\begin{equation}
(A,Z) \to (A,Z+2) + 2e^-. \label{bb}
\end{equation}
For further historical remarks, see e.g. \cite{Zralek1997sa}.

\begin{figure}[h!]
\begin{center}
\includegraphics[width=0.9\columnwidth]{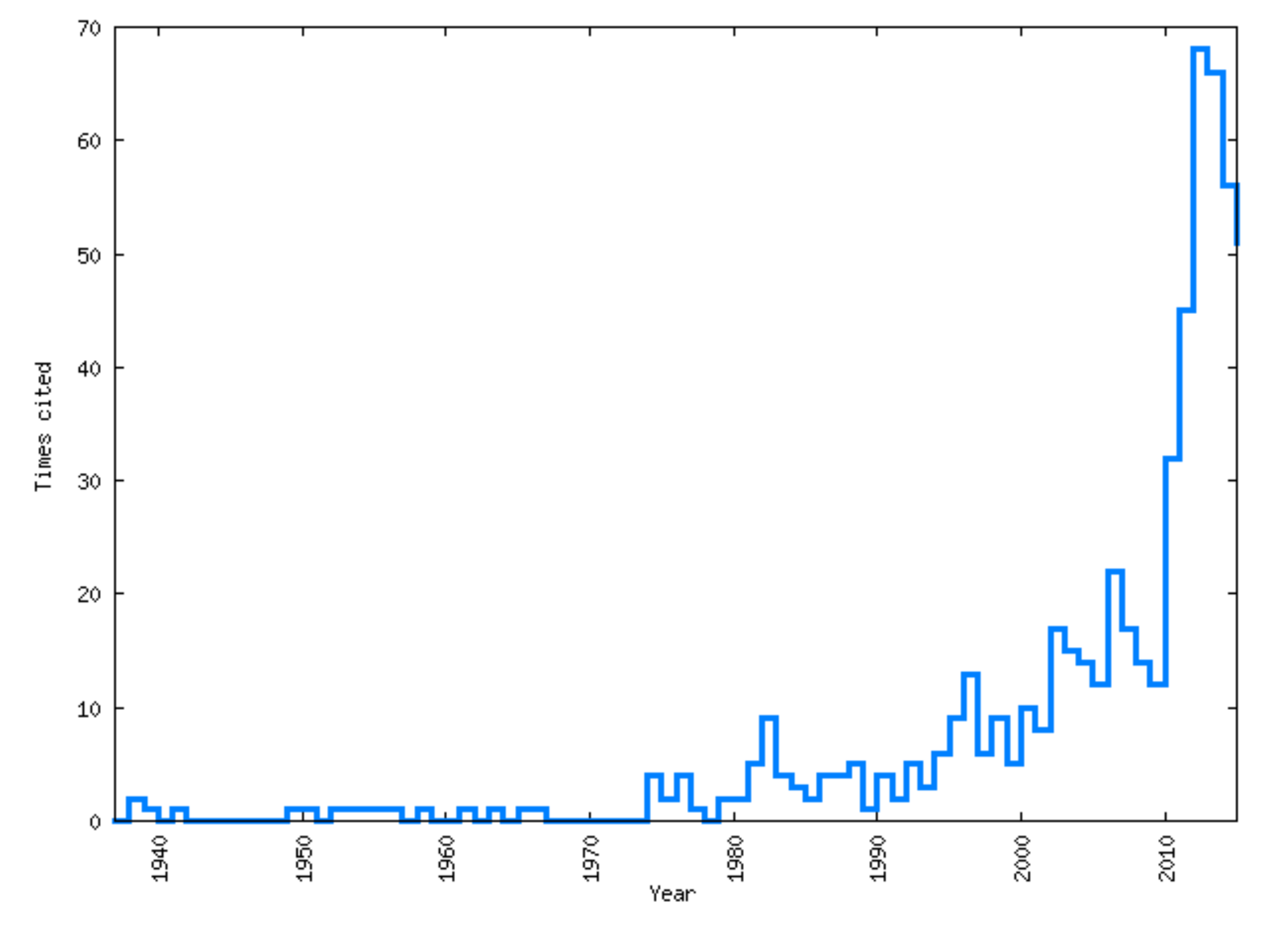}
\caption{\label{cit}
The INSPIRE citations record of the Majorana original paper \cite{Majorana:1937vz}. 
The plot taken from \cite{httpmaj}.
  }
  \end{center}
  \end{figure}

The search for the reaction Eq.~\eqref{bb} is  nowadays in many present and planned experiments \cite{Agostini:2013mzu,Gando:2012zm,Albert:2014awa,Alfonso:2015wka,Majorovits:2015vka,Abgrall:2013rze,Albert:2014afa,Wang:2015raa}. It is also  
explored in theoretical and phenomenological studies
 \cite{Czakon:1999cd,Barger:1999na,Czakon:2001uh,Prezeau:2003xn,Xing:2015zha,Minakata:2014jba,Rodejohann:2012xd,Rodejohann:2011mu,Mohapatra:2005wg,Chakrabortty:2012mh,Nemevsek:2012iq,Dev:2013vxa,Mahajan:2013ixa,Mahajan:2014nca,Dev:2014xea,Ge:2015bfa,Gonzalez:2015ady,Horoi:2015gdv,Awasthi:2015ota,Bambhaniya:2015ipg}. As seen in Fig.~\ref{cit}, according to Inspire literature database, the Majorana original article is cited about six hundred times so far, mostly interest started with the New Millenium, after neutrino oscillations have been confirmed (Particle Data Group included neutrino oscillation parameters for the first time in its 2000 review).

The composition of neutrino states origins from the neutrino mass matrix.  
There are many mechanisms for neutrino mass generation. Typically, these are radiative mass generations or tree level (effective) 
constructions, in both cases non-standard fields, interactions or symmetries are 
necessarily involved. 
 
Ranging neutrino masses from zero to $M_i \geq 10^9$ GeV, mass mechanisms introduce different neutrino states \cite{Drewes:2013gca}. Apart from Dirac or Majorana types, there are pseudo-Dirac (or quasi-Dirac) \cite{Akhmedov:2014kxa}, schizophrenic \cite{Allahverdi:2010us}, or vanilla  \cite{Dev:2013vba} neutrinos, to call some of them. 
Popular seesaw mechanisms give a possibility for a dynamical explanation why known active neutrino states are so light. They appear to be of  Majorana type (recently dynamical explanation for Dirac light neutrinos has been proposed \cite{Valle:2016kyz}).
 Seesaw type-I models have been worked out in  \cite{Minkowski:1977sc,GellMann:1980vs,Yanagida:1980xy,Mohapatra:1980yp},
  type-II in \cite{Magg:1980ut}, 
  type-III in \cite{Foot:1988aq}. A hybrid mechanism is also possible
  \cite{Franco:2015pva}.
For original inverse seesaw, see \cite{Mohapatra:1986aw,Mohapatra:1986bd} and its generalizations, see \cite{Gavela:2009cd,Dev:2012sg,Dev:2012bd,Awasthi:2013ff,Dev:2015pga}.

Here we recap the classical type-I and the more universal but in the same time more complex in construction the inverse seesaw mechanism. 
For the seesaw type-I mass matrix and SM  leptonic (L) and 
scalar $\widetilde{\phi}$ fields we have

\begin{eqnarray}
\mathcal{L}_{\text{Y}} &=& -Y_{ij} \, \overline{L'_{iL}} N'_{jR} \, \widetilde{\phi} +
\text{h.c.}\, \\
\mathcal{L}_\text{M} &=& -\frac{1}{2} M_{ij} \overline{N'_{iL}} N'_{jR} + \mathrm{h.c.} \,,\\
\mathcal{L}_{\text{Y}}+\mathcal{L}_\text{M}=  & = & - \frac{1}{2} \,
\left(\bar \nu'_L \; \bar N'_L \right)
\left( \! \begin{array}{cc}
0 & \frac{v}{\sqrt 2} Y \\ \frac{v}{\sqrt 2} Y^T & M
\end{array} \! \right) \,
\left( \!\! \begin{array}{c} \nu'_R \\ N'_R \end{array} \!\! \right).
\nonumber \\ &&
\end{eqnarray}

If indices $i, j$ run from 1 to 3, we have in general $6 \times 6$ mass matrix.
For the above, the neutrino mass matrix can be identified with Dirac $M_D$ and Majorana $M_R$ mass terms
\begin{equation}
M_\nu = \left( \begin{array}{cc}  0  & M_D 
\\ M^T_D & M_R {{}{ (v_R)}} \end{array} \right). \label{massm}
\end{equation}
\\
With $  M_D \ll M_R$
\begin{eqnarray}
m_N & \sim & M_R \\
m_{\rm light} & \sim & M_D^2/M_R.
\end{eqnarray}
For $M_D \sim {\cal{O}}(1)\;\rm GeV$ and demanding light neutrino masses of the order of $0.1$ eV, without artificial fine-tunings, we get the heavy Majorana mass scale
${}{M_R \sim 10^{15}\; \rm GeV}$.

As typically $M_D [{\cal{O}}(\mathrm{MeV})] \ll M_R [{\cal{O}}(\mathrm{TeV})]$, most of possible connections between heavy $N$ and light $\nu$ neutrino sectors are cut away, and we can explore heavy sector effects exclusively, as done in Eq.~(6) in the main text.
In this case the neutrino mixing matrix takes the following form  
\begin{eqnarray}\label{V}
U\approx\left(
\begin{array}{cc}
U_{\mathrm{PMNS}}
& 
0\\
0 & K_{R}^\dag
\end{array}\right),
\end{eqnarray}
where, $K_{R}$ is an unitary $3\times3$ matrix defined by $M_R=K_{R}^T\mathrm{diag}(M_1,M_2,M_3)K_{R}$, $M_a>0$.  

A more universal neutrino mass construction is connected with
the so-called inverse or linear seesaw mechanism.
Classically pseudo-Dirac neutrino has been introduced for light neutrinos demanding $m_D\gg m_R$ in {Eq.~\eqref{massm}}. In such a case neutrinos can mix maximally leading to the almost degenerate mass states  with opposite CP phases \cite{Bilenky:1987ty}.
 
In the inverse seesaw neutrino ranges from pure Majorana case, through pseudo-Dirac to pure Dirac scenario. Also relatively large light-heavy neutrino mixings can be obtained here \cite{Mohapatra:1986aw,Chen:2013foz,Deppisch:2015cua}.  It means that in the main text we explore mainly the type-I seesaw scenarios, Eq.~\eqref{V}.
 
In the original inverse seesaw proposal, the lepton number violation is small, 
being directly proportional to the light neutrino masses.
 
The generalized inverse seesaw  neutrino mass matrix in the extended flavor basis $\{\nu^
C,N,S^C\}$ is given by
\begin{eqnarray}
{\cal M} \ = \ \left(\begin{array}{ccc} 0 & M_D&0\\M^{\sf T}_D & \mu_R & M_N^{\sf T}\\
0 & M_N & \mu_S\end{array}\right) \; 
\label{nu2}
\end{eqnarray}
with two eigenvalues, which are
\begin{equation}
M_{N_{1,2}} \ \simeq \ \frac{1}{2}\left[\mu_R\pm \sqrt{\mu_R^2+4M_N^2}\right].
\end{equation}
For $\mu_R \ll M_N$, $N_{1,2}$  -  pseudo-Dirac pair emerges. For   $\mu_R \gg M_N$, $N_1$ -  purely Majorana with $M_1=\mu_R$ is 
realized.
 
Thus, for intermediate values of $\mu_R$, we can have scenarios with a varying degree of lepton number breaking \cite{Chen:2011hc,Han:2006ip,delAguila:2007em,Mohapatra:1986aw,Aad:2015xaa}.
 
In the inverse seesaw case, since there are two pairs of SM singlet fermions, they can always form Dirac pairs in the limit of small LNV. However, in the Type-I seesaw with three Majorana neutrinos, we can get one Dirac and one un-paired Majorana neutrino, as in the case of Eq.~(\ref{KRB13}).
 
In the context of considered models and phenomenological studies, let us focus on the $r$ parameter in the $pp\to lljj$ process. The following question naturally emerges: is it possible to establish if we have a pure Dirac state or a Majorana composition?
In the setup discussed in this work, the lepton number conservation is realized ($r=0$) only when two heavy Majorana neutrinos are degenerate and have opposite parities, or equivalently, heavy neutrino is of the Dirac type. But, there is no possibility to measure $r=0$ exactly. The best what can be done is to derive more and more precise bounds on $r$. Inturn, such bounds on the lepton number violation provide limits on the mass splittings and mixing parametes of heavy Majorana neutrinos. 

 One should also keep in mind that in more complicated scenarios $r=0$ can be realized also when masses of heavy neutrinos are not degenerate. For example, effective operators which violate the lepton number may be present in the model. Then contributions to $pp\to lljj$ coming from heavy neutrinos and those coming from additional sector of the theory may interfere and lead to $r=0$.  It is clear that such configuration would need severe fine-tuning of model parameters.
      
\providecommand{\href}[2]{#2}

\bibliographystyle{apsrev4-1}

\begin{thebibliography}{191}%
\makeatletter
\providecommand \@ifxundefined [1]{%
 \@ifx{#1\undefined}
}%
\providecommand \@ifnum [1]{%
 \ifnum #1\expandafter \@firstoftwo
 \else \expandafter \@secondoftwo
 \fi
}%
\providecommand \@ifx [1]{%
 \ifx #1\expandafter \@firstoftwo
 \else \expandafter \@secondoftwo
 \fi
}%
\providecommand \natexlab [1]{#1}%
\providecommand \enquote  [1]{``#1''}%
\providecommand \bibnamefont  [1]{#1}%
\providecommand \bibfnamefont [1]{#1}%
\providecommand \citenamefont [1]{#1}%
\providecommand \href@noop [0]{\@secondoftwo}%
\providecommand \href [0]{\begingroup \@sanitize@url \@href}%
\providecommand \@href[1]{\@@startlink{#1}\@@href}%
\providecommand \@@href[1]{\endgroup#1\@@endlink}%
\providecommand \@sanitize@url [0]{\catcode `\\12\catcode `\$12\catcode
  `\&12\catcode `\#12\catcode `\^12\catcode `\_12\catcode `\%12\relax}%
\providecommand \@@startlink[1]{}%
\providecommand \@@endlink[0]{}%
\providecommand \url  [0]{\begingroup\@sanitize@url \@url }%
\providecommand \@url [1]{\endgroup\@href {#1}{\urlprefix }}%
\providecommand \urlprefix  [0]{URL }%
\providecommand \Eprint [0]{\href }%
\providecommand \doibase [0]{http://dx.doi.org/}%
\providecommand \selectlanguage [0]{\@gobble}%
\providecommand \bibinfo  [0]{\@secondoftwo}%
\providecommand \bibfield  [0]{\@secondoftwo}%
\providecommand \translation [1]{[#1]}%
\providecommand \BibitemOpen [0]{}%
\providecommand \bibitemStop [0]{}%
\providecommand \bibitemNoStop [0]{.\EOS\space}%
\providecommand \EOS [0]{\spacefactor3000\relax}%
\providecommand \BibitemShut  [1]{\csname bibitem#1\endcsname}%
\let\auto@bib@innerbib\@empty
\bibitem [{\citenamefont {Bahcall}\ and\ \citenamefont
  {Davis}(1976)}]{Bahcall:1976zz}%
  \BibitemOpen
  \bibfield  {author} {\bibinfo {author} {\bibfnamefont {J.~N.}\ \bibnamefont
  {Bahcall}}\ and\ \bibinfo {author} {\bibfnamefont {R.}~\bibnamefont
  {Davis}},\ }\href {\doibase 10.1126/science.191.4224.264} {\bibfield
  {journal} {\bibinfo  {journal} {Science}\ }\textbf {\bibinfo {volume}
  {191}},\ \bibinfo {pages} {264} (\bibinfo {year} {1976})}\BibitemShut
  {NoStop}%
\bibitem [{\citenamefont {Fukuda}\ \emph {et~al.}(1998)\citenamefont {Fukuda}
  \emph {et~al.}}]{Fukuda:1998mi}%
  \BibitemOpen
  \bibfield  {author} {\bibinfo {author} {\bibfnamefont {Y.}~\bibnamefont
  {Fukuda}} \emph {et~al.} (\bibinfo {collaboration} {Super-Kamiokande}),\
  }\href {\doibase 10.1103/PhysRevLett.81.1562} {\bibfield  {journal} {\bibinfo
   {journal} {Phys. Rev. Lett.}\ }\textbf {\bibinfo {volume} {81}},\ \bibinfo
  {pages} {1562} (\bibinfo {year} {1998})},\ \Eprint
  {http://arxiv.org/abs/hep-ex/9807003} {arXiv:hep-ex/9807003 [hep-ex]}
  \BibitemShut {NoStop}%
\bibitem [{\citenamefont {Ahmad}\ \emph {et~al.}(2002)\citenamefont {Ahmad}
  \emph {et~al.}}]{Ahmad:2002jz}%
  \BibitemOpen
  \bibfield  {author} {\bibinfo {author} {\bibfnamefont {Q.~R.}\ \bibnamefont
  {Ahmad}} \emph {et~al.} (\bibinfo {collaboration} {SNO}),\ }\href {\doibase
  10.1103/PhysRevLett.89.011301} {\bibfield  {journal} {\bibinfo  {journal}
  {Phys. Rev. Lett.}\ }\textbf {\bibinfo {volume} {89}},\ \bibinfo {pages}
  {011301} (\bibinfo {year} {2002})},\ \Eprint
  {http://arxiv.org/abs/nucl-ex/0204008} {arXiv:nucl-ex/0204008 [nucl-ex]}
  \BibitemShut {NoStop}%
\bibitem [{\citenamefont {Wietfeldt}\ and\ \citenamefont
  {Norman}(1996)}]{Wietfeldt:1995ja}%
  \BibitemOpen
  \bibfield  {author} {\bibinfo {author} {\bibfnamefont {F.~E.}\ \bibnamefont
  {Wietfeldt}}\ and\ \bibinfo {author} {\bibfnamefont {E.~B.}\ \bibnamefont
  {Norman}},\ }\href {\doibase 10.1016/0370-1573(95)00082-8} {\bibfield
  {journal} {\bibinfo  {journal} {Phys. Rept.}\ }\textbf {\bibinfo {volume}
  {273}},\ \bibinfo {pages} {149} (\bibinfo {year} {1996})}\BibitemShut
  {NoStop}%
\bibitem [{\citenamefont {Franklin}(1995)}]{Franklin:1995pk}%
  \BibitemOpen
  \bibfield  {author} {\bibinfo {author} {\bibfnamefont {A.}~\bibnamefont
  {Franklin}},\ }\href {\doibase 10.1103/RevModPhys.67.457} {\bibfield
  {journal} {\bibinfo  {journal} {Rev. Mod. Phys.}\ }\textbf {\bibinfo {volume}
  {67}},\ \bibinfo {pages} {457} (\bibinfo {year} {1995})}\BibitemShut
  {NoStop}%
\bibitem [{\citenamefont {Adam}\ \emph {et~al.}(2012)\citenamefont {Adam} \emph
  {et~al.}}]{Adam:2011faa}%
  \BibitemOpen
  \bibfield  {author} {\bibinfo {author} {\bibfnamefont {T.}~\bibnamefont
  {Adam}} \emph {et~al.} (\bibinfo {collaboration} {OPERA}),\ }\href {\doibase
  10.1007/JHEP10(2012)093} {\bibfield  {journal} {\bibinfo  {journal} {JHEP}\
  }\textbf {\bibinfo {volume} {10}},\ \bibinfo {pages} {093} (\bibinfo {year}
  {2012})},\ \Eprint {http://arxiv.org/abs/1109.4897} {arXiv:1109.4897
  [hep-ex]} \BibitemShut {NoStop}%
\bibitem [{\citenamefont {Antonello}\ \emph {et~al.}(2012)\citenamefont
  {Antonello} \emph {et~al.}}]{Antonello:2012hg}%
  \BibitemOpen
  \bibfield  {author} {\bibinfo {author} {\bibfnamefont {M.}~\bibnamefont
  {Antonello}} \emph {et~al.} (\bibinfo {collaboration} {ICARUS}),\ }\href
  {\doibase 10.1016/j.physletb.2012.05.033} {\bibfield  {journal} {\bibinfo
  {journal} {Phys. Lett.}\ }\textbf {\bibinfo {volume} {B713}},\ \bibinfo
  {pages} {17} (\bibinfo {year} {2012})},\ \Eprint
  {http://arxiv.org/abs/1203.3433} {arXiv:1203.3433 [hep-ex]} \BibitemShut
  {NoStop}%
\bibitem [{\citenamefont {Klapdor-Kleingrothaus}\ and\ \citenamefont
  {Krivosheina}(2006)}]{KlapdorKleingrothaus:2006ff}%
  \BibitemOpen
  \bibfield  {author} {\bibinfo {author} {\bibfnamefont {H.~V.}\ \bibnamefont
  {Klapdor-Kleingrothaus}}\ and\ \bibinfo {author} {\bibfnamefont {I.~V.}\
  \bibnamefont {Krivosheina}},\ }\href {\doibase 10.1142/S0217732306020937}
  {\bibfield  {journal} {\bibinfo  {journal} {Mod. Phys. Lett.}\ }\textbf
  {\bibinfo {volume} {A21}},\ \bibinfo {pages} {1547} (\bibinfo {year}
  {2006})}\BibitemShut {NoStop}%
\bibitem [{\citenamefont {Akhmedov}()}]{Akhmedov:2014kxa}%
  \BibitemOpen
  \bibfield  {author} {\bibinfo {author} {\bibfnamefont {E.}~\bibnamefont
  {Akhmedov}},\ }\href@noop {} {\ }\Eprint {http://arxiv.org/abs/1412.3320}
  {arXiv:1412.3320 [hep-ph]} \BibitemShut {NoStop}%
\bibitem [{\citenamefont {Minkowski}(1977)}]{Minkowski:1977sc}%
  \BibitemOpen
  \bibfield  {author} {\bibinfo {author} {\bibfnamefont {P.}~\bibnamefont
  {Minkowski}},\ }\href {\doibase 10.1016/0370-2693(77)90435-X} {\bibfield
  {journal} {\bibinfo  {journal} {Phys. Lett.}\ }\textbf {\bibinfo {volume}
  {B67}},\ \bibinfo {pages} {421} (\bibinfo {year} {1977})}\BibitemShut
  {NoStop}%
\bibitem [{\citenamefont {Yanagida}(1979)}]{Yanagida:1979as}%
  \BibitemOpen
  \bibfield  {author} {\bibinfo {author} {\bibfnamefont {T.}~\bibnamefont
  {Yanagida}},\ }\href@noop {} {\bibfield  {journal} {\bibinfo  {journal}
  {Conf. Proc.}\ }\textbf {\bibinfo {volume} {C7902131}},\ \bibinfo {pages}
  {95} (\bibinfo {year} {1979})}\BibitemShut {NoStop}%
\bibitem [{\citenamefont {Gell-Mann}\ \emph {et~al.}(1979)\citenamefont
  {Gell-Mann}, \citenamefont {Ramond},\ and\ \citenamefont
  {Slansky}}]{GellMann:1980vs}%
  \BibitemOpen
  \bibfield  {author} {\bibinfo {author} {\bibfnamefont {M.}~\bibnamefont
  {Gell-Mann}}, \bibinfo {author} {\bibfnamefont {P.}~\bibnamefont {Ramond}}, \
  and\ \bibinfo {author} {\bibfnamefont {R.}~\bibnamefont {Slansky}},\
  }\href@noop {} {\bibfield  {journal} {\bibinfo  {journal} {Conf. Proc.}\
  }\textbf {\bibinfo {volume} {C790927}},\ \bibinfo {pages} {315} (\bibinfo
  {year} {1979})},\ \Eprint {http://arxiv.org/abs/1306.4669} {arXiv:1306.4669
  [hep-th]} \BibitemShut {NoStop}%
\bibitem [{\citenamefont {Mohapatra}\ and\ \citenamefont
  {Senjanovic}(1980)}]{Mohapatra:1979ia}%
  \BibitemOpen
  \bibfield  {author} {\bibinfo {author} {\bibfnamefont {R.~N.}\ \bibnamefont
  {Mohapatra}}\ and\ \bibinfo {author} {\bibfnamefont {G.}~\bibnamefont
  {Senjanovic}},\ }\href {\doibase 10.1103/PhysRevLett.44.912} {\bibfield
  {journal} {\bibinfo  {journal} {Phys.Rev.Lett.}\ }\textbf {\bibinfo {volume}
  {44}},\ \bibinfo {pages} {912} (\bibinfo {year} {1980})}\BibitemShut
  {NoStop}%
\bibitem [{\citenamefont {Schechter}\ and\ \citenamefont
  {Valle}(1982)}]{Schechter:1981cv}%
  \BibitemOpen
  \bibfield  {author} {\bibinfo {author} {\bibfnamefont {J.}~\bibnamefont
  {Schechter}}\ and\ \bibinfo {author} {\bibfnamefont {J.~W.~F.}\ \bibnamefont
  {Valle}},\ }\href {\doibase 10.1103/PhysRevD.25.774} {\bibfield  {journal}
  {\bibinfo  {journal} {Phys. Rev.}\ }\textbf {\bibinfo {volume} {D25}},\
  \bibinfo {pages} {774} (\bibinfo {year} {1982})}\BibitemShut {NoStop}%
\bibitem [{\citenamefont {Rodejohann}(2011)}]{Rodejohann:2011mu}%
  \BibitemOpen
  \bibfield  {author} {\bibinfo {author} {\bibfnamefont {W.}~\bibnamefont
  {Rodejohann}},\ }\href {\doibase 10.1142/S0218301311020186} {\bibfield
  {journal} {\bibinfo  {journal} {Int. J. Mod. Phys.}\ }\textbf {\bibinfo
  {volume} {E20}},\ \bibinfo {pages} {1833} (\bibinfo {year} {2011})},\ \Eprint
  {http://arxiv.org/abs/1106.1334} {arXiv:1106.1334 [hep-ph]} \BibitemShut
  {NoStop}%
\bibitem [{\citenamefont {Pas}\ and\ \citenamefont
  {Rodejohann}(2015)}]{Pas:2015eia}%
  \BibitemOpen
  \bibfield  {author} {\bibinfo {author} {\bibfnamefont {H.}~\bibnamefont
  {Pas}}\ and\ \bibinfo {author} {\bibfnamefont {W.}~\bibnamefont
  {Rodejohann}},\ }\href {\doibase 10.1088/1367-2630/17/11/115010} {\bibfield
  {journal} {\bibinfo  {journal} {New J. Phys.}\ }\textbf {\bibinfo {volume}
  {17}},\ \bibinfo {pages} {115010} (\bibinfo {year} {2015})},\ \Eprint
  {http://arxiv.org/abs/1507.00170} {arXiv:1507.00170 [hep-ph]} \BibitemShut
  {NoStop}%
\bibitem [{\citenamefont {Mohapatra}(1986)}]{Mohapatra:1986aw}%
  \BibitemOpen
  \bibfield  {author} {\bibinfo {author} {\bibfnamefont {R.~N.}\ \bibnamefont
  {Mohapatra}},\ }\href {\doibase 10.1103/PhysRevLett.56.561} {\bibfield
  {journal} {\bibinfo  {journal} {Phys. Rev. Lett.}\ }\textbf {\bibinfo
  {volume} {56}},\ \bibinfo {pages} {561} (\bibinfo {year} {1986})}\BibitemShut
  {NoStop}%
\bibitem [{\citenamefont {Mohapatra}\ and\ \citenamefont
  {Valle}(1986)}]{Mohapatra:1986bd}%
  \BibitemOpen
  \bibfield  {author} {\bibinfo {author} {\bibfnamefont {R.~N.}\ \bibnamefont
  {Mohapatra}}\ and\ \bibinfo {author} {\bibfnamefont {J.~W.~F.}\ \bibnamefont
  {Valle}},\ }\href {\doibase 10.1103/PhysRevD.34.1642} {\bibfield  {journal}
  {\bibinfo  {journal} {Phys. Rev.}\ }\textbf {\bibinfo {volume} {D34}},\
  \bibinfo {pages} {1642} (\bibinfo {year} {1986})}\BibitemShut {NoStop}%
\bibitem [{\citenamefont {Mohapatra}\ and\ \citenamefont
  {Pal}(1998)}]{Mohapatra:1998rq}%
  \BibitemOpen
  \bibfield  {author} {\bibinfo {author} {\bibfnamefont {R.~N.}\ \bibnamefont
  {Mohapatra}}\ and\ \bibinfo {author} {\bibfnamefont {P.~B.}\ \bibnamefont
  {Pal}},\ }\href@noop {} {\bibfield  {journal} {\bibinfo  {journal} {World
  Sci. Lect. Notes Phys.}\ }\textbf {\bibinfo {volume} {60}},\ \bibinfo {pages}
  {1} (\bibinfo {year} {1998})},\ \bibinfo {note} {[World Sci. Lect. Notes
  Phys.72,1(2004)]}\BibitemShut {NoStop}%
\bibitem [{\citenamefont {Giunti}\ and\ \citenamefont
  {Kim}(2007)}]{Giunti:2007ry}%
  \BibitemOpen
  \bibfield  {author} {\bibinfo {author} {\bibfnamefont {C.}~\bibnamefont
  {Giunti}}\ and\ \bibinfo {author} {\bibfnamefont {C.~W.}\ \bibnamefont
  {Kim}},\ }\href@noop {} {\emph {\bibinfo {title} {{Fundamentals of Neutrino
  Physics and Astrophysics}}}}\ (\bibinfo {year} {Oxford, UK: Univ. Pr.,
  2007})\BibitemShut {NoStop}%
\bibitem [{\citenamefont {Drewes}(2013)}]{Drewes:2013gca}%
  \BibitemOpen
  \bibfield  {author} {\bibinfo {author} {\bibfnamefont {M.}~\bibnamefont
  {Drewes}},\ }\href {\doibase 10.1142/S0218301313300191} {\bibfield  {journal}
  {\bibinfo  {journal} {Int. J. Mod. Phys.}\ }\textbf {\bibinfo {volume}
  {E22}},\ \bibinfo {pages} {1330019} (\bibinfo {year} {2013})},\ \Eprint
  {http://arxiv.org/abs/1303.6912} {arXiv:1303.6912 [hep-ph]} \BibitemShut
  {NoStop}%
\bibitem [{\citenamefont {de~Gouvea}\ and\ \citenamefont
  {Vogel}(2013)}]{deGouvea:2013zba}%
  \BibitemOpen
  \bibfield  {author} {\bibinfo {author} {\bibfnamefont {A.}~\bibnamefont
  {de~Gouvea}}\ and\ \bibinfo {author} {\bibfnamefont {P.}~\bibnamefont
  {Vogel}},\ }\href {\doibase 10.1016/j.ppnp.2013.03.006} {\bibfield  {journal}
  {\bibinfo  {journal} {Prog. Part. Nucl. Phys.}\ }\textbf {\bibinfo {volume}
  {71}},\ \bibinfo {pages} {75} (\bibinfo {year} {2013})},\ \Eprint
  {http://arxiv.org/abs/1303.4097} {arXiv:1303.4097 [hep-ph]} \BibitemShut
  {NoStop}%
\bibitem [{\citenamefont {Kuno}(2013)}]{Kuno:2013mha}%
  \BibitemOpen
  \bibfield  {author} {\bibinfo {author} {\bibfnamefont {Y.}~\bibnamefont
  {Kuno}} (\bibinfo {collaboration} {COMET}),\ }\href {\doibase
  10.1093/ptep/pts089} {\bibfield  {journal} {\bibinfo  {journal} {PTEP}\
  }\textbf {\bibinfo {volume} {2013}},\ \bibinfo {pages} {022C01} (\bibinfo
  {year} {2013})}\BibitemShut {NoStop}%
\bibitem [{\citenamefont {Brown}(2015)}]{Brown:2015cka}%
  \BibitemOpen
  \bibfield  {author} {\bibinfo {author} {\bibfnamefont {D.}~\bibnamefont
  {Brown}} (\bibinfo {collaboration} {Mu2e}),\ }\href {\doibase
  10.1016/j.nuclphysbps.2015.02.032} {\bibfield  {journal} {\bibinfo  {journal}
  {Nucl. Part. Phys. Proc.}\ }\textbf {\bibinfo {volume} {260}},\ \bibinfo
  {pages} {151} (\bibinfo {year} {2015})}\BibitemShut {NoStop}%
\bibitem [{\citenamefont {Abgrall}\ \emph {et~al.}(2014)\citenamefont {Abgrall}
  \emph {et~al.}}]{Abgrall:2013rze}%
  \BibitemOpen
  \bibfield  {author} {\bibinfo {author} {\bibfnamefont {N.}~\bibnamefont
  {Abgrall}} \emph {et~al.} (\bibinfo {collaboration} {Majorana}),\ }\href
  {\doibase 10.1155/2014/365432} {\bibfield  {journal} {\bibinfo  {journal}
  {Adv. High Energy Phys.}\ }\textbf {\bibinfo {volume} {2014}},\ \bibinfo
  {pages} {365432} (\bibinfo {year} {2014})},\ \Eprint
  {http://arxiv.org/abs/1308.1633} {arXiv:1308.1633 [physics.ins-det]}
  \BibitemShut {NoStop}%
\bibitem [{\citenamefont {Pilaftsis}(1992)}]{Pilaftsis:1991ug}%
  \BibitemOpen
  \bibfield  {author} {\bibinfo {author} {\bibfnamefont {A.}~\bibnamefont
  {Pilaftsis}},\ }\href {\doibase 10.1007/BF01482590} {\bibfield  {journal}
  {\bibinfo  {journal} {Z. Phys.}\ }\textbf {\bibinfo {volume} {C55}},\
  \bibinfo {pages} {275} (\bibinfo {year} {1992})},\ \Eprint
  {http://arxiv.org/abs/hep-ph/9901206} {arXiv:hep-ph/9901206 [hep-ph]}
  \BibitemShut {NoStop}%
\bibitem [{\citenamefont {Bray}\ \emph {et~al.}(2007)\citenamefont {Bray},
  \citenamefont {Lee},\ and\ \citenamefont {Pilaftsis}}]{Bray:2007ru}%
  \BibitemOpen
  \bibfield  {author} {\bibinfo {author} {\bibfnamefont {S.}~\bibnamefont
  {Bray}}, \bibinfo {author} {\bibfnamefont {J.~S.}\ \bibnamefont {Lee}}, \
  and\ \bibinfo {author} {\bibfnamefont {A.}~\bibnamefont {Pilaftsis}},\ }\href
  {\doibase 10.1016/j.nuclphysb.2007.07.002} {\bibfield  {journal} {\bibinfo
  {journal} {Nucl. Phys.}\ }\textbf {\bibinfo {volume} {B786}},\ \bibinfo
  {pages} {95} (\bibinfo {year} {2007})},\ \Eprint
  {http://arxiv.org/abs/hep-ph/0702294} {arXiv:hep-ph/0702294 [HEP-PH]}
  \BibitemShut {NoStop}%
\bibitem [{\citenamefont {Keung}\ and\ \citenamefont
  {Senjanovic}(1983)}]{Keung:1983uu}%
  \BibitemOpen
  \bibfield  {author} {\bibinfo {author} {\bibfnamefont {W.-Y.}\ \bibnamefont
  {Keung}}\ and\ \bibinfo {author} {\bibfnamefont {G.}~\bibnamefont
  {Senjanovic}},\ }\href {\doibase 10.1103/PhysRevLett.50.1427} {\bibfield
  {journal} {\bibinfo  {journal} {Phys.Rev.Lett.}\ }\textbf {\bibinfo {volume}
  {50}},\ \bibinfo {pages} {1427} (\bibinfo {year} {1983})}\BibitemShut
  {NoStop}%
\bibitem [{\citenamefont {Gluza}\ and\ \citenamefont
  {Zra{\l}ek}(1992)}]{Gluza:1991wj}%
  \BibitemOpen
  \bibfield  {author} {\bibinfo {author} {\bibfnamefont {J.}~\bibnamefont
  {Gluza}}\ and\ \bibinfo {author} {\bibfnamefont {M.}~\bibnamefont
  {Zra{\l}ek}},\ }\href {\doibase 10.1103/PhysRevD.45.1693} {\bibfield
  {journal} {\bibinfo  {journal} {Phys. Rev.}\ }\textbf {\bibinfo {volume}
  {D45}},\ \bibinfo {pages} {1693} (\bibinfo {year} {1992})}\BibitemShut
  {NoStop}%
\bibitem [{\citenamefont {Denner}\ \emph {et~al.}(1992)\citenamefont {Denner},
  \citenamefont {Eck}, \citenamefont {Hahn},\ and\ \citenamefont
  {Kublbeck}}]{Denner:1992me}%
  \BibitemOpen
  \bibfield  {author} {\bibinfo {author} {\bibfnamefont {A.}~\bibnamefont
  {Denner}}, \bibinfo {author} {\bibfnamefont {H.}~\bibnamefont {Eck}},
  \bibinfo {author} {\bibfnamefont {O.}~\bibnamefont {Hahn}}, \ and\ \bibinfo
  {author} {\bibfnamefont {J.}~\bibnamefont {Kublbeck}},\ }\href {\doibase
  10.1016/0370-2693(92)91045-B} {\bibfield  {journal} {\bibinfo  {journal}
  {Phys. Lett.}\ }\textbf {\bibinfo {volume} {B291}},\ \bibinfo {pages} {278}
  (\bibinfo {year} {1992})}\BibitemShut {NoStop}%
\bibitem [{\citenamefont {Hirsch}\ \emph {et~al.}(1996)\citenamefont {Hirsch},
  \citenamefont {Klapdor-Kleingrothaus},\ and\ \citenamefont
  {Panella}}]{Hirsch:1996qw}%
  \BibitemOpen
  \bibfield  {author} {\bibinfo {author} {\bibfnamefont {M.}~\bibnamefont
  {Hirsch}}, \bibinfo {author} {\bibfnamefont {H.~V.}\ \bibnamefont
  {Klapdor-Kleingrothaus}}, \ and\ \bibinfo {author} {\bibfnamefont
  {O.}~\bibnamefont {Panella}},\ }\href {\doibase 10.1016/0370-2693(96)00185-2}
  {\bibfield  {journal} {\bibinfo  {journal} {Phys. Lett.}\ }\textbf {\bibinfo
  {volume} {B374}},\ \bibinfo {pages} {7} (\bibinfo {year} {1996})},\ \Eprint
  {http://arxiv.org/abs/hep-ph/9602306} {arXiv:hep-ph/9602306 [hep-ph]}
  \BibitemShut {NoStop}%
\bibitem [{\citenamefont {Tello}\ \emph {et~al.}(2011)\citenamefont {Tello},
  \citenamefont {Nemevsek}, \citenamefont {Nesti}, \citenamefont {Senjanovic},\
  and\ \citenamefont {Vissani}}]{Tello:2010am}%
  \BibitemOpen
  \bibfield  {author} {\bibinfo {author} {\bibfnamefont {V.}~\bibnamefont
  {Tello}}, \bibinfo {author} {\bibfnamefont {M.}~\bibnamefont {Nemevsek}},
  \bibinfo {author} {\bibfnamefont {F.}~\bibnamefont {Nesti}}, \bibinfo
  {author} {\bibfnamefont {G.}~\bibnamefont {Senjanovic}}, \ and\ \bibinfo
  {author} {\bibfnamefont {F.}~\bibnamefont {Vissani}},\ }\href {\doibase
  10.1103/PhysRevLett.106.151801} {\bibfield  {journal} {\bibinfo  {journal}
  {Phys.Rev.Lett.}\ }\textbf {\bibinfo {volume} {106}},\ \bibinfo {pages}
  {151801} (\bibinfo {year} {2011})},\ \Eprint {http://arxiv.org/abs/1011.3522}
  {arXiv:1011.3522 [hep-ph]} \BibitemShut {NoStop}%
\bibitem [{\citenamefont {Mitra}\ \emph {et~al.}(2012)\citenamefont {Mitra},
  \citenamefont {Senjanovic},\ and\ \citenamefont {Vissani}}]{Mitra:2011qr}%
  \BibitemOpen
  \bibfield  {author} {\bibinfo {author} {\bibfnamefont {M.}~\bibnamefont
  {Mitra}}, \bibinfo {author} {\bibfnamefont {G.}~\bibnamefont {Senjanovic}}, \
  and\ \bibinfo {author} {\bibfnamefont {F.}~\bibnamefont {Vissani}},\ }\href
  {\doibase 10.1016/j.nuclphysb.2011.10.035} {\bibfield  {journal} {\bibinfo
  {journal} {Nucl. Phys.}\ }\textbf {\bibinfo {volume} {B856}},\ \bibinfo
  {pages} {26} (\bibinfo {year} {2012})},\ \Eprint
  {http://arxiv.org/abs/1108.0004} {arXiv:1108.0004 [hep-ph]} \BibitemShut
  {NoStop}%
\bibitem [{\citenamefont {Faessler}\ \emph {et~al.}(2011)\citenamefont
  {Faessler}, \citenamefont {Meroni}, \citenamefont {Petcov}, \citenamefont
  {Simkovic},\ and\ \citenamefont {Vergados}}]{Faessler:2011qw}%
  \BibitemOpen
  \bibfield  {author} {\bibinfo {author} {\bibfnamefont {A.}~\bibnamefont
  {Faessler}}, \bibinfo {author} {\bibfnamefont {A.}~\bibnamefont {Meroni}},
  \bibinfo {author} {\bibfnamefont {S.~T.}\ \bibnamefont {Petcov}}, \bibinfo
  {author} {\bibfnamefont {F.}~\bibnamefont {Simkovic}}, \ and\ \bibinfo
  {author} {\bibfnamefont {J.}~\bibnamefont {Vergados}},\ }\href {\doibase
  10.1103/PhysRevD.83.113003} {\bibfield  {journal} {\bibinfo  {journal} {Phys.
  Rev.}\ }\textbf {\bibinfo {volume} {D83}},\ \bibinfo {pages} {113003}
  (\bibinfo {year} {2011})},\ \Eprint {http://arxiv.org/abs/1103.2434}
  {arXiv:1103.2434 [hep-ph]} \BibitemShut {NoStop}%
\bibitem [{\citenamefont {Nemevsek}\ \emph {et~al.}(2013)\citenamefont
  {Nemevsek}, \citenamefont {Senjanovic},\ and\ \citenamefont
  {Tello}}]{Nemevsek:2012iq}%
  \BibitemOpen
  \bibfield  {author} {\bibinfo {author} {\bibfnamefont {M.}~\bibnamefont
  {Nemevsek}}, \bibinfo {author} {\bibfnamefont {G.}~\bibnamefont
  {Senjanovic}}, \ and\ \bibinfo {author} {\bibfnamefont {V.}~\bibnamefont
  {Tello}},\ }\href {\doibase 10.1103/PhysRevLett.110.151802} {\bibfield
  {journal} {\bibinfo  {journal} {Phys. Rev. Lett.}\ }\textbf {\bibinfo
  {volume} {110}},\ \bibinfo {pages} {151802} (\bibinfo {year} {2013})},\
  \Eprint {http://arxiv.org/abs/1211.2837} {arXiv:1211.2837 [hep-ph]}
  \BibitemShut {NoStop}%
\bibitem [{\citenamefont {Meroni}\ \emph {et~al.}(2013)\citenamefont {Meroni},
  \citenamefont {Petcov},\ and\ \citenamefont {Simkovic}}]{Meroni:2012qf}%
  \BibitemOpen
  \bibfield  {author} {\bibinfo {author} {\bibfnamefont {A.}~\bibnamefont
  {Meroni}}, \bibinfo {author} {\bibfnamefont {S.~T.}\ \bibnamefont {Petcov}},
  \ and\ \bibinfo {author} {\bibfnamefont {F.}~\bibnamefont {Simkovic}},\
  }\href {\doibase 10.1007/JHEP02(2013)025} {\bibfield  {journal} {\bibinfo
  {journal} {JHEP}\ }\textbf {\bibinfo {volume} {02}},\ \bibinfo {pages} {025}
  (\bibinfo {year} {2013})},\ \Eprint {http://arxiv.org/abs/1212.1331}
  {arXiv:1212.1331 [hep-ph]} \BibitemShut {NoStop}%
\bibitem [{\citenamefont {Barry}\ and\ \citenamefont
  {Rodejohann}(2013)}]{Barry:2013xxa}%
  \BibitemOpen
  \bibfield  {author} {\bibinfo {author} {\bibfnamefont {J.}~\bibnamefont
  {Barry}}\ and\ \bibinfo {author} {\bibfnamefont {W.}~\bibnamefont
  {Rodejohann}},\ }\href {\doibase 10.1007/JHEP09(2013)153} {\bibfield
  {journal} {\bibinfo  {journal} {JHEP}\ }\textbf {\bibinfo {volume} {09}},\
  \bibinfo {pages} {153} (\bibinfo {year} {2013})},\ \Eprint
  {http://arxiv.org/abs/1303.6324} {arXiv:1303.6324 [hep-ph]} \BibitemShut
  {NoStop}%
\bibitem [{\citenamefont {Pascoli}\ \emph {et~al.}(2014)\citenamefont
  {Pascoli}, \citenamefont {Mitra},\ and\ \citenamefont
  {Wong}}]{Pascoli:2013fiz}%
  \BibitemOpen
  \bibfield  {author} {\bibinfo {author} {\bibfnamefont {S.}~\bibnamefont
  {Pascoli}}, \bibinfo {author} {\bibfnamefont {M.}~\bibnamefont {Mitra}}, \
  and\ \bibinfo {author} {\bibfnamefont {S.}~\bibnamefont {Wong}},\ }\href
  {\doibase 10.1103/PhysRevD.90.093005} {\bibfield  {journal} {\bibinfo
  {journal} {Phys. Rev.}\ }\textbf {\bibinfo {volume} {D90}},\ \bibinfo {pages}
  {093005} (\bibinfo {year} {2014})},\ \Eprint {http://arxiv.org/abs/1310.6218}
  {arXiv:1310.6218 [hep-ph]} \BibitemShut {NoStop}%
\bibitem [{\citenamefont {Lee}\ \emph {et~al.}(2013)\citenamefont {Lee},
  \citenamefont {Bhupal~Dev},\ and\ \citenamefont {Mohapatra}}]{Dev:2013oxa}%
  \BibitemOpen
  \bibfield  {author} {\bibinfo {author} {\bibfnamefont {C.-H.}\ \bibnamefont
  {Lee}}, \bibinfo {author} {\bibfnamefont {P.~S.}\ \bibnamefont {Bhupal~Dev}},
  \ and\ \bibinfo {author} {\bibfnamefont {R.~N.}\ \bibnamefont {Mohapatra}},\
  }\href {\doibase 10.1103/PhysRevD.88.093010} {\bibfield  {journal} {\bibinfo
  {journal} {Phys. Rev.}\ }\textbf {\bibinfo {volume} {D88}},\ \bibinfo {pages}
  {093010} (\bibinfo {year} {2013})},\ \Eprint {http://arxiv.org/abs/1309.0774}
  {arXiv:1309.0774 [hep-ph]} \BibitemShut {NoStop}%
\bibitem [{\citenamefont {Bhupal~Dev}\ \emph {et~al.}(2013)\citenamefont
  {Bhupal~Dev}, \citenamefont {Goswami}, \citenamefont {Mitra},\ and\
  \citenamefont {Rodejohann}}]{Dev:2013vxa}%
  \BibitemOpen
  \bibfield  {author} {\bibinfo {author} {\bibfnamefont {P.~S.}\ \bibnamefont
  {Bhupal~Dev}}, \bibinfo {author} {\bibfnamefont {S.}~\bibnamefont {Goswami}},
  \bibinfo {author} {\bibfnamefont {M.}~\bibnamefont {Mitra}}, \ and\ \bibinfo
  {author} {\bibfnamefont {W.}~\bibnamefont {Rodejohann}},\ }\href {\doibase
  10.1103/PhysRevD.88.091301} {\bibfield  {journal} {\bibinfo  {journal} {Phys.
  Rev.}\ }\textbf {\bibinfo {volume} {D88}},\ \bibinfo {pages} {091301}
  (\bibinfo {year} {2013})},\ \Eprint {http://arxiv.org/abs/1305.0056}
  {arXiv:1305.0056 [hep-ph]} \BibitemShut {NoStop}%
\bibitem [{\citenamefont {Meroni}\ and\ \citenamefont
  {Peinado}(2014)}]{Meroni:2014tba}%
  \BibitemOpen
  \bibfield  {author} {\bibinfo {author} {\bibfnamefont {A.}~\bibnamefont
  {Meroni}}\ and\ \bibinfo {author} {\bibfnamefont {E.}~\bibnamefont
  {Peinado}},\ }\href {\doibase 10.1103/PhysRevD.90.053002} {\bibfield
  {journal} {\bibinfo  {journal} {Phys. Rev.}\ }\textbf {\bibinfo {volume}
  {D90}},\ \bibinfo {pages} {053002} (\bibinfo {year} {2014})},\ \Eprint
  {http://arxiv.org/abs/1406.3990} {arXiv:1406.3990 [hep-ph]} \BibitemShut
  {NoStop}%
\bibitem [{\citenamefont {Meroni}(2015)}]{Meroni:2015oya}%
  \BibitemOpen
  \bibfield  {author} {\bibinfo {author} {\bibfnamefont {A.}~\bibnamefont
  {Meroni}},\ }\href {\doibase 10.1140/epjp/i2015-15232-0} {\bibfield
  {journal} {\bibinfo  {journal} {Eur. Phys. J. Plus}\ }\textbf {\bibinfo
  {volume} {130}},\ \bibinfo {pages} {232} (\bibinfo {year} {2015})},\ \Eprint
  {http://arxiv.org/abs/1510.05523} {arXiv:1510.05523 [hep-ph]} \BibitemShut
  {NoStop}%
\bibitem [{\citenamefont {Czakon}\ \emph {et~al.}(1999)\citenamefont {Czakon},
  \citenamefont {Gluza},\ and\ \citenamefont {Zra{\l}ek}}]{Czakon:1999cd}%
  \BibitemOpen
  \bibfield  {author} {\bibinfo {author} {\bibfnamefont {M.}~\bibnamefont
  {Czakon}}, \bibinfo {author} {\bibfnamefont {J.}~\bibnamefont {Gluza}}, \
  and\ \bibinfo {author} {\bibfnamefont {M.}~\bibnamefont {Zra{\l}ek}},\ }\href
  {\doibase 10.1016/S0370-2693(99)01008-4} {\bibfield  {journal} {\bibinfo
  {journal} {Phys. Lett.}\ }\textbf {\bibinfo {volume} {B465}},\ \bibinfo
  {pages} {211} (\bibinfo {year} {1999})},\ \Eprint
  {http://arxiv.org/abs/hep-ph/9906381} {arXiv:hep-ph/9906381 [hep-ph]}
  \BibitemShut {NoStop}%
\bibitem [{\citenamefont {Barger}\ and\ \citenamefont
  {Whisnant}(1999)}]{Barger:1999na}%
  \BibitemOpen
  \bibfield  {author} {\bibinfo {author} {\bibfnamefont {V.~D.}\ \bibnamefont
  {Barger}}\ and\ \bibinfo {author} {\bibfnamefont {K.}~\bibnamefont
  {Whisnant}},\ }\href {\doibase 10.1016/S0370-2693(99)00514-6} {\bibfield
  {journal} {\bibinfo  {journal} {Phys. Lett.}\ }\textbf {\bibinfo {volume}
  {B456}},\ \bibinfo {pages} {194} (\bibinfo {year} {1999})},\ \Eprint
  {http://arxiv.org/abs/hep-ph/9904281} {arXiv:hep-ph/9904281 [hep-ph]}
  \BibitemShut {NoStop}%
\bibitem [{\citenamefont {Xing}\ \emph {et~al.}(2015)\citenamefont {Xing},
  \citenamefont {Zhao},\ and\ \citenamefont {Zhou}}]{Xing:2015zha}%
  \BibitemOpen
  \bibfield  {author} {\bibinfo {author} {\bibfnamefont {Z.-z.}\ \bibnamefont
  {Xing}}, \bibinfo {author} {\bibfnamefont {Z.-h.}\ \bibnamefont {Zhao}}, \
  and\ \bibinfo {author} {\bibfnamefont {Y.-L.}\ \bibnamefont {Zhou}},\ }\href
  {\doibase 10.1140/epjc/s10052-015-3656-6} {\bibfield  {journal} {\bibinfo
  {journal} {Eur. Phys. J.}\ }\textbf {\bibinfo {volume} {C75}},\ \bibinfo
  {pages} {423} (\bibinfo {year} {2015})},\ \Eprint
  {http://arxiv.org/abs/1504.05820} {arXiv:1504.05820 [hep-ph]} \BibitemShut
  {NoStop}%
\bibitem [{\citenamefont {Awasthi}\ \emph {et~al.}(2016)\citenamefont
  {Awasthi}, \citenamefont {Dev},\ and\ \citenamefont
  {Mitra}}]{Awasthi:2015ota}%
  \BibitemOpen
  \bibfield  {author} {\bibinfo {author} {\bibfnamefont {R.~L.}\ \bibnamefont
  {Awasthi}}, \bibinfo {author} {\bibfnamefont {P.~S.~B.}\ \bibnamefont {Dev}},
  \ and\ \bibinfo {author} {\bibfnamefont {M.}~\bibnamefont {Mitra}},\ }\href
  {\doibase 10.1103/PhysRevD.93.011701} {\bibfield  {journal} {\bibinfo
  {journal} {Phys. Rev.}\ }\textbf {\bibinfo {volume} {D93}},\ \bibinfo {pages}
  {011701} (\bibinfo {year} {2016})},\ \Eprint
  {http://arxiv.org/abs/1509.05387} {arXiv:1509.05387 [hep-ph]} \BibitemShut
  {NoStop}%
\bibitem [{\citenamefont {Khachatryan}\ \emph {et~al.}(2014)\citenamefont
  {Khachatryan} \emph {et~al.}}]{Khachatryan:2014dka}%
  \BibitemOpen
  \bibfield  {author} {\bibinfo {author} {\bibfnamefont {V.}~\bibnamefont
  {Khachatryan}} \emph {et~al.} (\bibinfo {collaboration} {CMS}),\ }\href
  {\doibase 10.1140/epjc/s10052-014-3149-z} {\bibfield  {journal} {\bibinfo
  {journal} {Eur. Phys. J.}\ }\textbf {\bibinfo {volume} {C74}},\ \bibinfo
  {pages} {3149} (\bibinfo {year} {2014})},\ \Eprint
  {http://arxiv.org/abs/1407.3683} {arXiv:1407.3683 [hep-ex]} \BibitemShut
  {NoStop}%
\bibitem [{\citenamefont {Gluza}(2002)}]{Gluza:2002vs}%
  \BibitemOpen
  \bibfield  {author} {\bibinfo {author} {\bibfnamefont {J.}~\bibnamefont
  {Gluza}},\ }\href@noop {} {\bibfield  {journal} {\bibinfo  {journal} {Acta
  Phys.Polon.}\ }\textbf {\bibinfo {volume} {B33}},\ \bibinfo {pages} {1735}
  (\bibinfo {year} {2002})},\ \Eprint {http://arxiv.org/abs/hep-ph/0201002}
  {arXiv:hep-ph/0201002 [hep-ph]} \BibitemShut {NoStop}%
\bibitem [{\citenamefont {Chen}\ \emph {et~al.}(2013)\citenamefont {Chen},
  \citenamefont {Dev},\ and\ \citenamefont {Mohapatra}}]{Chen:2013foz}%
  \BibitemOpen
  \bibfield  {author} {\bibinfo {author} {\bibfnamefont {C.-Y.}\ \bibnamefont
  {Chen}}, \bibinfo {author} {\bibfnamefont {P.~S.~B.}\ \bibnamefont {Dev}}, \
  and\ \bibinfo {author} {\bibfnamefont {R.}~\bibnamefont {Mohapatra}},\ }\href
  {\doibase 10.1103/PhysRevD.88.033014} {\bibfield  {journal} {\bibinfo
  {journal} {Phys.Rev.}\ }\textbf {\bibinfo {volume} {D88}},\ \bibinfo {pages}
  {033014} (\bibinfo {year} {2013})},\ \Eprint {http://arxiv.org/abs/1306.2342}
  {arXiv:1306.2342 [hep-ph]} \BibitemShut {NoStop}%
\bibitem [{\citenamefont {Bilenky}\ and\ \citenamefont
  {Petcov}(1987)}]{Bilenky:1987ty}%
  \BibitemOpen
  \bibfield  {author} {\bibinfo {author} {\bibfnamefont {S.~M.}\ \bibnamefont
  {Bilenky}}\ and\ \bibinfo {author} {\bibfnamefont {S.~T.}\ \bibnamefont
  {Petcov}},\ }\href {\doibase 10.1103/RevModPhys.59.671} {\bibfield  {journal}
  {\bibinfo  {journal} {Rev. Mod. Phys.}\ }\textbf {\bibinfo {volume} {59}},\
  \bibinfo {pages} {671} (\bibinfo {year} {1987})},\ \bibinfo {note} {[Erratum:
  Rev. Mod. Phys.60,575(1988)]}\BibitemShut {NoStop}%
\bibitem [{\citenamefont {Kayser}(1984)}]{Kayser:1984ge}%
  \BibitemOpen
  \bibfield  {author} {\bibinfo {author} {\bibfnamefont {B.}~\bibnamefont
  {Kayser}},\ }\href {\doibase 10.1103/PhysRevD.30.1023} {\bibfield  {journal}
  {\bibinfo  {journal} {Phys. Rev.}\ }\textbf {\bibinfo {volume} {D30}},\
  \bibinfo {pages} {1023} (\bibinfo {year} {1984})}\BibitemShut {NoStop}%
\bibitem [{\citenamefont {Kayser}\ \emph {et~al.}(1989)\citenamefont {Kayser},
  \citenamefont {Gibrat-Debu},\ and\ \citenamefont {Perrier}}]{Kayser:1989iu}%
  \BibitemOpen
  \bibfield  {author} {\bibinfo {author} {\bibfnamefont {B.}~\bibnamefont
  {Kayser}}, \bibinfo {author} {\bibfnamefont {F.}~\bibnamefont {Gibrat-Debu}},
  \ and\ \bibinfo {author} {\bibfnamefont {F.}~\bibnamefont {Perrier}},\
  }\href@noop {} {\bibfield  {journal} {\bibinfo  {journal} {World Sci. Lect.
  Notes Phys.}\ }\textbf {\bibinfo {volume} {25}},\ \bibinfo {pages} {1}
  (\bibinfo {year} {1989})}\BibitemShut {NoStop}%
\bibitem [{\citenamefont {Deppisch}\ \emph {et~al.}(2016)\citenamefont
  {Deppisch}, \citenamefont {Graf}, \citenamefont {Kulkarni}, \citenamefont
  {Patra}, \citenamefont {Rodejohann}, \citenamefont {Sahu},\ and\
  \citenamefont {Sarkar}}]{Deppisch:2015cua}%
  \BibitemOpen
  \bibfield  {author} {\bibinfo {author} {\bibfnamefont {F.~F.}\ \bibnamefont
  {Deppisch}}, \bibinfo {author} {\bibfnamefont {L.}~\bibnamefont {Graf}},
  \bibinfo {author} {\bibfnamefont {S.}~\bibnamefont {Kulkarni}}, \bibinfo
  {author} {\bibfnamefont {S.}~\bibnamefont {Patra}}, \bibinfo {author}
  {\bibfnamefont {W.}~\bibnamefont {Rodejohann}}, \bibinfo {author}
  {\bibfnamefont {N.}~\bibnamefont {Sahu}}, \ and\ \bibinfo {author}
  {\bibfnamefont {U.}~\bibnamefont {Sarkar}},\ }\href {\doibase
  10.1103/PhysRevD.93.013011} {\bibfield  {journal} {\bibinfo  {journal} {Phys.
  Rev.}\ }\textbf {\bibinfo {volume} {D93}},\ \bibinfo {pages} {013011}
  (\bibinfo {year} {2016})},\ \Eprint {http://arxiv.org/abs/1508.05940}
  {arXiv:1508.05940 [hep-ph]} \BibitemShut {NoStop}%
\bibitem [{\citenamefont {Bhupal~Dev}\ and\ \citenamefont
  {Mohapatra}(2015)}]{Dev:2015pga}%
  \BibitemOpen
  \bibfield  {author} {\bibinfo {author} {\bibfnamefont {P.~S.}\ \bibnamefont
  {Bhupal~Dev}}\ and\ \bibinfo {author} {\bibfnamefont {R.~N.}\ \bibnamefont
  {Mohapatra}},\ }\href {\doibase 10.1103/PhysRevLett.115.181803} {\bibfield
  {journal} {\bibinfo  {journal} {Phys. Rev. Lett.}\ }\textbf {\bibinfo
  {volume} {115}},\ \bibinfo {pages} {181803} (\bibinfo {year} {2015})},\
  \Eprint {http://arxiv.org/abs/1508.02277} {arXiv:1508.02277 [hep-ph]}
  \BibitemShut {NoStop}%
\bibitem [{\citenamefont {Baudis}\ \emph {et~al.}(1999)\citenamefont {Baudis}
  \emph {et~al.}}]{Baudis:1999xd}%
  \BibitemOpen
  \bibfield  {author} {\bibinfo {author} {\bibfnamefont {L.}~\bibnamefont
  {Baudis}} \emph {et~al.},\ }\href {\doibase 10.1103/PhysRevLett.83.41}
  {\bibfield  {journal} {\bibinfo  {journal} {Phys. Rev. Lett.}\ }\textbf
  {\bibinfo {volume} {83}},\ \bibinfo {pages} {41} (\bibinfo {year} {1999})},\
  \Eprint {http://arxiv.org/abs/hep-ex/9902014} {arXiv:hep-ex/9902014 [hep-ex]}
  \BibitemShut {NoStop}%
\bibitem [{\citenamefont {Gluza}\ and\ \citenamefont
  {Jeli\'nski}(2015)}]{Gluza:2015goa}%
  \BibitemOpen
  \bibfield  {author} {\bibinfo {author} {\bibfnamefont {J.}~\bibnamefont
  {Gluza}}\ and\ \bibinfo {author} {\bibfnamefont {T.}~\bibnamefont
  {Jeli\'nski}},\ }\href {\doibase 10.1016/j.physletb.2015.06.077} {\bibfield
  {journal} {\bibinfo  {journal} {Phys. Lett.}\ }\textbf {\bibinfo {volume}
  {B748}},\ \bibinfo {pages} {125} (\bibinfo {year} {2015})},\ \Eprint
  {http://arxiv.org/abs/1504.05568} {arXiv:1504.05568 [hep-ph]} \BibitemShut
  {NoStop}%
\bibitem [{\citenamefont {Deppisch}\ \emph {et~al.}(2014)\citenamefont
  {Deppisch}, \citenamefont {Gonzalo}, \citenamefont {Patra}, \citenamefont
  {Sahu},\ and\ \citenamefont {Sarkar}}]{Deppisch:2014qpa}%
  \BibitemOpen
  \bibfield  {author} {\bibinfo {author} {\bibfnamefont {F.~F.}\ \bibnamefont
  {Deppisch}}, \bibinfo {author} {\bibfnamefont {T.~E.}\ \bibnamefont
  {Gonzalo}}, \bibinfo {author} {\bibfnamefont {S.}~\bibnamefont {Patra}},
  \bibinfo {author} {\bibfnamefont {N.}~\bibnamefont {Sahu}}, \ and\ \bibinfo
  {author} {\bibfnamefont {U.}~\bibnamefont {Sarkar}},\ }\href {\doibase
  10.1103/PhysRevD.90.053014} {\bibfield  {journal} {\bibinfo  {journal} {Phys.
  Rev.}\ }\textbf {\bibinfo {volume} {D90}},\ \bibinfo {pages} {053014}
  (\bibinfo {year} {2014})},\ \Eprint {http://arxiv.org/abs/1407.5384}
  {arXiv:1407.5384 [hep-ph]} \BibitemShut {NoStop}%
\bibitem [{\citenamefont {Heikinheimo}\ \emph {et~al.}(2014)\citenamefont
  {Heikinheimo}, \citenamefont {Raidal},\ and\ \citenamefont
  {Spethmann}}]{Heikinheimo:2014tba}%
  \BibitemOpen
  \bibfield  {author} {\bibinfo {author} {\bibfnamefont {M.}~\bibnamefont
  {Heikinheimo}}, \bibinfo {author} {\bibfnamefont {M.}~\bibnamefont {Raidal}},
  \ and\ \bibinfo {author} {\bibfnamefont {C.}~\bibnamefont {Spethmann}},\
  }\href {\doibase 10.1140/epjc/s10052-014-3107-9} {\bibfield  {journal}
  {\bibinfo  {journal} {Eur. Phys. J.}\ }\textbf {\bibinfo {volume} {C74}},\
  \bibinfo {pages} {3107} (\bibinfo {year} {2014})},\ \Eprint
  {http://arxiv.org/abs/1407.6908} {arXiv:1407.6908 [hep-ph]} \BibitemShut
  {NoStop}%
\bibitem [{\citenamefont {Deppisch}\ \emph
  {et~al.}(2015{\natexlab{a}})\citenamefont {Deppisch}, \citenamefont
  {Gonzalo}, \citenamefont {Patra}, \citenamefont {Sahu},\ and\ \citenamefont
  {Sarkar}}]{Deppisch:2014zta}%
  \BibitemOpen
  \bibfield  {author} {\bibinfo {author} {\bibfnamefont {F.~F.}\ \bibnamefont
  {Deppisch}}, \bibinfo {author} {\bibfnamefont {T.~E.}\ \bibnamefont
  {Gonzalo}}, \bibinfo {author} {\bibfnamefont {S.}~\bibnamefont {Patra}},
  \bibinfo {author} {\bibfnamefont {N.}~\bibnamefont {Sahu}}, \ and\ \bibinfo
  {author} {\bibfnamefont {U.}~\bibnamefont {Sarkar}},\ }\href {\doibase
  10.1103/PhysRevD.91.015018} {\bibfield  {journal} {\bibinfo  {journal} {Phys.
  Rev.}\ }\textbf {\bibinfo {volume} {D91}},\ \bibinfo {pages} {015018}
  (\bibinfo {year} {2015}{\natexlab{a}})},\ \Eprint
  {http://arxiv.org/abs/1410.6427} {arXiv:1410.6427 [hep-ph]} \BibitemShut
  {NoStop}%
\bibitem [{\citenamefont {Aguilar-Saavedra}\ and\ \citenamefont
  {Joaquim}(2014)}]{Aguilar-Saavedra:2014ola}%
  \BibitemOpen
  \bibfield  {author} {\bibinfo {author} {\bibfnamefont {J.~A.}\ \bibnamefont
  {Aguilar-Saavedra}}\ and\ \bibinfo {author} {\bibfnamefont {F.~R.}\
  \bibnamefont {Joaquim}},\ }\href {\doibase 10.1103/PhysRevD.90.115010}
  {\bibfield  {journal} {\bibinfo  {journal} {Phys. Rev.}\ }\textbf {\bibinfo
  {volume} {D90}},\ \bibinfo {pages} {115010} (\bibinfo {year} {2014})},\
  \Eprint {http://arxiv.org/abs/1408.2456} {arXiv:1408.2456 [hep-ph]}
  \BibitemShut {NoStop}%
\bibitem [{\citenamefont {Vasquez}(2014)}]{Vasquez:2014mxa}%
  \BibitemOpen
  \bibfield  {author} {\bibinfo {author} {\bibfnamefont {J.~C.}\ \bibnamefont
  {Vasquez}},\ }\href@noop {} {\  (\bibinfo {year} {2014})},\ \Eprint
  {http://arxiv.org/abs/1411.5824} {arXiv:1411.5824 [hep-ph]} \BibitemShut
  {NoStop}%
\bibitem [{\citenamefont {Senjanovic}\ and\ \citenamefont
  {Tello}(2015)}]{Senjanovic:2014pva}%
  \BibitemOpen
  \bibfield  {author} {\bibinfo {author} {\bibfnamefont {G.}~\bibnamefont
  {Senjanovic}}\ and\ \bibinfo {author} {\bibfnamefont {V.}~\bibnamefont
  {Tello}},\ }\href {\doibase 10.1103/PhysRevLett.114.071801} {\bibfield
  {journal} {\bibinfo  {journal} {Phys. Rev. Lett.}\ }\textbf {\bibinfo
  {volume} {114}},\ \bibinfo {pages} {071801} (\bibinfo {year} {2015})},\
  \Eprint {http://arxiv.org/abs/1408.3835} {arXiv:1408.3835 [hep-ph]}
  \BibitemShut {NoStop}%
\bibitem [{\citenamefont {Ng}\ \emph {et~al.}(2015)\citenamefont {Ng},
  \citenamefont {de~la Puente},\ and\ \citenamefont {Pan}}]{Ng:2015hba}%
  \BibitemOpen
  \bibfield  {author} {\bibinfo {author} {\bibfnamefont {J.~N.}\ \bibnamefont
  {Ng}}, \bibinfo {author} {\bibfnamefont {A.}~\bibnamefont {de~la Puente}}, \
  and\ \bibinfo {author} {\bibfnamefont {B.~W.-P.}\ \bibnamefont {Pan}},\
  }\href {\doibase 10.1007/JHEP12(2015)172} {\bibfield  {journal} {\bibinfo
  {journal} {JHEP}\ }\textbf {\bibinfo {volume} {12}},\ \bibinfo {pages} {172}
  (\bibinfo {year} {2015})},\ \Eprint {http://arxiv.org/abs/1505.01934}
  {arXiv:1505.01934 [hep-ph]} \BibitemShut {NoStop}%
\bibitem [{\citenamefont {Dobrescu}\ and\ \citenamefont
  {Liu}(2015{\natexlab{a}})}]{Dobrescu:2015qna}%
  \BibitemOpen
  \bibfield  {author} {\bibinfo {author} {\bibfnamefont {B.~A.}\ \bibnamefont
  {Dobrescu}}\ and\ \bibinfo {author} {\bibfnamefont {Z.}~\bibnamefont {Liu}},\
  }\href {\doibase 10.1103/PhysRevLett.115.211802} {\bibfield  {journal}
  {\bibinfo  {journal} {Phys. Rev. Lett.}\ }\textbf {\bibinfo {volume} {115}},\
  \bibinfo {pages} {211802} (\bibinfo {year} {2015}{\natexlab{a}})},\ \Eprint
  {http://arxiv.org/abs/1506.06736} {arXiv:1506.06736 [hep-ph]} \BibitemShut
  {NoStop}%
\bibitem [{\citenamefont {Brehmer}\ \emph {et~al.}(2015)\citenamefont
  {Brehmer}, \citenamefont {Hewett}, \citenamefont {Kopp}, \citenamefont
  {Rizzo},\ and\ \citenamefont {Tattersall}}]{Brehmer:2015cia}%
  \BibitemOpen
  \bibfield  {author} {\bibinfo {author} {\bibfnamefont {J.}~\bibnamefont
  {Brehmer}}, \bibinfo {author} {\bibfnamefont {J.}~\bibnamefont {Hewett}},
  \bibinfo {author} {\bibfnamefont {J.}~\bibnamefont {Kopp}}, \bibinfo {author}
  {\bibfnamefont {T.}~\bibnamefont {Rizzo}}, \ and\ \bibinfo {author}
  {\bibfnamefont {J.}~\bibnamefont {Tattersall}},\ }\href {\doibase
  10.1007/JHEP10(2015)182} {\bibfield  {journal} {\bibinfo  {journal} {JHEP}\
  }\textbf {\bibinfo {volume} {10}},\ \bibinfo {pages} {182} (\bibinfo {year}
  {2015})},\ \Eprint {http://arxiv.org/abs/1507.00013} {arXiv:1507.00013
  [hep-ph]} \BibitemShut {NoStop}%
\bibitem [{\citenamefont {Vasquez}(2015)}]{Vasquez:2015una}%
  \BibitemOpen
  \bibfield  {author} {\bibinfo {author} {\bibfnamefont {J.~C.}\ \bibnamefont
  {Vasquez}},\ }\href {\doibase 10.1007/JHEP09(2015)131} {\bibfield  {journal}
  {\bibinfo  {journal} {JHEP}\ }\textbf {\bibinfo {volume} {09}},\ \bibinfo
  {pages} {131} (\bibinfo {year} {2015})},\ \Eprint
  {http://arxiv.org/abs/1504.05220} {arXiv:1504.05220 [hep-ph]} \BibitemShut
  {NoStop}%
\bibitem [{\citenamefont {Coloma}\ \emph {et~al.}(2015)\citenamefont {Coloma},
  \citenamefont {Dobrescu},\ and\ \citenamefont
  {Lopez-Pavon}}]{Coloma:2015una}%
  \BibitemOpen
  \bibfield  {author} {\bibinfo {author} {\bibfnamefont {P.}~\bibnamefont
  {Coloma}}, \bibinfo {author} {\bibfnamefont {B.~A.}\ \bibnamefont
  {Dobrescu}}, \ and\ \bibinfo {author} {\bibfnamefont {J.}~\bibnamefont
  {Lopez-Pavon}},\ }\href {\doibase 10.1103/PhysRevD.92.115023} {\bibfield
  {journal} {\bibinfo  {journal} {Phys. Rev.}\ }\textbf {\bibinfo {volume}
  {D92}},\ \bibinfo {pages} {115023} (\bibinfo {year} {2015})},\ \Eprint
  {http://arxiv.org/abs/1508.04129} {arXiv:1508.04129 [hep-ph]} \BibitemShut
  {NoStop}%
\bibitem [{\citenamefont {Das}\ \emph {et~al.}(2014)\citenamefont {Das},
  \citenamefont {Bhupal~Dev},\ and\ \citenamefont {Okada}}]{Das:2014jxa}%
  \BibitemOpen
  \bibfield  {author} {\bibinfo {author} {\bibfnamefont {A.}~\bibnamefont
  {Das}}, \bibinfo {author} {\bibfnamefont {P.~S.}\ \bibnamefont {Bhupal~Dev}},
  \ and\ \bibinfo {author} {\bibfnamefont {N.}~\bibnamefont {Okada}},\ }\href
  {\doibase 10.1016/j.physletb.2014.06.058} {\bibfield  {journal} {\bibinfo
  {journal} {Phys. Lett.}\ }\textbf {\bibinfo {volume} {B735}},\ \bibinfo
  {pages} {364} (\bibinfo {year} {2014})},\ \Eprint
  {http://arxiv.org/abs/1405.0177} {arXiv:1405.0177 [hep-ph]} \BibitemShut
  {NoStop}%
\bibitem [{\citenamefont {Berger}\ \emph {et~al.}(2015)\citenamefont {Berger},
  \citenamefont {Dror},\ and\ \citenamefont {Ng}}]{Berger:2015qra}%
  \BibitemOpen
  \bibfield  {author} {\bibinfo {author} {\bibfnamefont {J.}~\bibnamefont
  {Berger}}, \bibinfo {author} {\bibfnamefont {J.~A.}\ \bibnamefont {Dror}}, \
  and\ \bibinfo {author} {\bibfnamefont {W.~H.}\ \bibnamefont {Ng}},\ }\href
  {\doibase 10.1007/JHEP09(2015)156} {\bibfield  {journal} {\bibinfo  {journal}
  {JHEP}\ }\textbf {\bibinfo {volume} {09}},\ \bibinfo {pages} {156} (\bibinfo
  {year} {2015})},\ \Eprint {http://arxiv.org/abs/1506.08213} {arXiv:1506.08213
  [hep-ph]} \BibitemShut {NoStop}%
\bibitem [{\citenamefont {Dobrescu}\ and\ \citenamefont
  {Liu}(2015{\natexlab{b}})}]{Dobrescu:2015yba}%
  \BibitemOpen
  \bibfield  {author} {\bibinfo {author} {\bibfnamefont {B.~A.}\ \bibnamefont
  {Dobrescu}}\ and\ \bibinfo {author} {\bibfnamefont {Z.}~\bibnamefont {Liu}},\
  }\href {\doibase 10.1007/JHEP10(2015)118} {\bibfield  {journal} {\bibinfo
  {journal} {JHEP}\ }\textbf {\bibinfo {volume} {10}},\ \bibinfo {pages} {118}
  (\bibinfo {year} {2015}{\natexlab{b}})},\ \Eprint
  {http://arxiv.org/abs/1507.01923} {arXiv:1507.01923 [hep-ph]} \BibitemShut
  {NoStop}%
\bibitem [{\citenamefont {Hisano}\ \emph {et~al.}(2015)\citenamefont {Hisano},
  \citenamefont {Nagata},\ and\ \citenamefont {Omura}}]{Hisano:2015gna}%
  \BibitemOpen
  \bibfield  {author} {\bibinfo {author} {\bibfnamefont {J.}~\bibnamefont
  {Hisano}}, \bibinfo {author} {\bibfnamefont {N.}~\bibnamefont {Nagata}}, \
  and\ \bibinfo {author} {\bibfnamefont {Y.}~\bibnamefont {Omura}},\ }\href
  {\doibase 10.1103/PhysRevD.92.055001} {\bibfield  {journal} {\bibinfo
  {journal} {Phys. Rev.}\ }\textbf {\bibinfo {volume} {D92}},\ \bibinfo {pages}
  {055001} (\bibinfo {year} {2015})},\ \Eprint
  {http://arxiv.org/abs/1506.03931} {arXiv:1506.03931 [hep-ph]} \BibitemShut
  {NoStop}%
\bibitem [{\citenamefont {Krauss}\ and\ \citenamefont
  {Porod}(2015)}]{Krauss:2015nba}%
  \BibitemOpen
  \bibfield  {author} {\bibinfo {author} {\bibfnamefont {M.~E.}\ \bibnamefont
  {Krauss}}\ and\ \bibinfo {author} {\bibfnamefont {W.}~\bibnamefont {Porod}},\
  }\href {\doibase 10.1103/PhysRevD.92.055019} {\bibfield  {journal} {\bibinfo
  {journal} {Phys. Rev.}\ }\textbf {\bibinfo {volume} {D92}},\ \bibinfo {pages}
  {055019} (\bibinfo {year} {2015})},\ \Eprint
  {http://arxiv.org/abs/1507.04349} {arXiv:1507.04349 [hep-ph]} \BibitemShut
  {NoStop}%
\bibitem [{\citenamefont {Dhuria}\ \emph {et~al.}(2016)\citenamefont {Dhuria},
  \citenamefont {Hati},\ and\ \citenamefont {Sarkar}}]{Dhuria:2015swa}%
  \BibitemOpen
  \bibfield  {author} {\bibinfo {author} {\bibfnamefont {M.}~\bibnamefont
  {Dhuria}}, \bibinfo {author} {\bibfnamefont {C.}~\bibnamefont {Hati}}, \ and\
  \bibinfo {author} {\bibfnamefont {U.}~\bibnamefont {Sarkar}},\ }\href
  {\doibase 10.1103/PhysRevD.93.015001} {\bibfield  {journal} {\bibinfo
  {journal} {Phys. Rev.}\ }\textbf {\bibinfo {volume} {D93}},\ \bibinfo {pages}
  {015001} (\bibinfo {year} {2016})},\ \Eprint
  {http://arxiv.org/abs/1507.08297} {arXiv:1507.08297 [hep-ph]} \BibitemShut
  {NoStop}%
\bibitem [{\citenamefont {Bandyopadhyay}\ \emph {et~al.}(2016)\citenamefont
  {Bandyopadhyay}, \citenamefont {Brahmachari},\ and\ \citenamefont
  {Raychaudhuri}}]{Bandyopadhyay:2015fka}%
  \BibitemOpen
  \bibfield  {author} {\bibinfo {author} {\bibfnamefont {T.}~\bibnamefont
  {Bandyopadhyay}}, \bibinfo {author} {\bibfnamefont {B.}~\bibnamefont
  {Brahmachari}}, \ and\ \bibinfo {author} {\bibfnamefont {A.}~\bibnamefont
  {Raychaudhuri}},\ }\href {\doibase 10.1007/JHEP02(2016)023} {\bibfield
  {journal} {\bibinfo  {journal} {JHEP}\ }\textbf {\bibinfo {volume} {02}},\
  \bibinfo {pages} {023} (\bibinfo {year} {2016})},\ \Eprint
  {http://arxiv.org/abs/1509.03232} {arXiv:1509.03232 [hep-ph]} \BibitemShut
  {NoStop}%
\bibitem [{\citenamefont {Dev}\ \emph {et~al.}(2016)\citenamefont {Dev},
  \citenamefont {Kim},\ and\ \citenamefont {Mohapatra}}]{Dev:2015kca}%
  \BibitemOpen
  \bibfield  {author} {\bibinfo {author} {\bibfnamefont {P.~S.~B.}\
  \bibnamefont {Dev}}, \bibinfo {author} {\bibfnamefont {D.}~\bibnamefont
  {Kim}}, \ and\ \bibinfo {author} {\bibfnamefont {R.~N.}\ \bibnamefont
  {Mohapatra}},\ }\href {\doibase 10.1007/JHEP01(2016)118} {\bibfield
  {journal} {\bibinfo  {journal} {JHEP}\ }\textbf {\bibinfo {volume} {01}},\
  \bibinfo {pages} {118} (\bibinfo {year} {2016})},\ \Eprint
  {http://arxiv.org/abs/1510.04328} {arXiv:1510.04328 [hep-ph]} \BibitemShut
  {NoStop}%
\bibitem [{\citenamefont {Deppisch}(2015)}]{Deppisch:2015bbh}%
  \BibitemOpen
  \bibfield  {author} {\bibinfo {author} {\bibfnamefont {F.~F.}\ \bibnamefont
  {Deppisch}},\ }\href {\doibase 10.5506/APhysPolB.46.2301} {\bibfield
  {journal} {\bibinfo  {journal} {Acta Phys. Polon.}\ }\textbf {\bibinfo
  {volume} {B46}},\ \bibinfo {pages} {2301} (\bibinfo {year}
  {2015})}\BibitemShut {NoStop}%
\bibitem [{\citenamefont {Deppisch}\ \emph
  {et~al.}(2015{\natexlab{b}})\citenamefont {Deppisch}, \citenamefont
  {Bhupal~Dev},\ and\ \citenamefont {Pilaftsis}}]{Deppisch:2015qwa}%
  \BibitemOpen
  \bibfield  {author} {\bibinfo {author} {\bibfnamefont {F.~F.}\ \bibnamefont
  {Deppisch}}, \bibinfo {author} {\bibfnamefont {P.~S.}\ \bibnamefont
  {Bhupal~Dev}}, \ and\ \bibinfo {author} {\bibfnamefont {A.}~\bibnamefont
  {Pilaftsis}},\ }\href {\doibase 10.1088/1367-2630/17/7/075019} {\bibfield
  {journal} {\bibinfo  {journal} {New J. Phys.}\ }\textbf {\bibinfo {volume}
  {17}},\ \bibinfo {pages} {075019} (\bibinfo {year} {2015}{\natexlab{b}})},\
  \Eprint {http://arxiv.org/abs/1502.06541} {arXiv:1502.06541 [hep-ph]}
  \BibitemShut {NoStop}%
\bibitem [{\citenamefont {Banerjee}\ \emph
  {et~al.}(2015{\natexlab{a}})\citenamefont {Banerjee}, \citenamefont {Mitra},\
  and\ \citenamefont {Spannowsky}}]{Banerjee:2015hoa}%
  \BibitemOpen
  \bibfield  {author} {\bibinfo {author} {\bibfnamefont {S.}~\bibnamefont
  {Banerjee}}, \bibinfo {author} {\bibfnamefont {M.}~\bibnamefont {Mitra}}, \
  and\ \bibinfo {author} {\bibfnamefont {M.}~\bibnamefont {Spannowsky}},\
  }\href {\doibase 10.1103/PhysRevD.92.055013} {\bibfield  {journal} {\bibinfo
  {journal} {Phys. Rev.}\ }\textbf {\bibinfo {volume} {D92}},\ \bibinfo {pages}
  {055013} (\bibinfo {year} {2015}{\natexlab{a}})},\ \Eprint
  {http://arxiv.org/abs/1506.06415} {arXiv:1506.06415 [hep-ph]} \BibitemShut
  {NoStop}%
\bibitem [{\citenamefont {Dhuria}\ \emph {et~al.}(2015)\citenamefont {Dhuria},
  \citenamefont {Hati}, \citenamefont {Rangarajan},\ and\ \citenamefont
  {Sarkar}}]{Dhuria:2015cfa}%
  \BibitemOpen
  \bibfield  {author} {\bibinfo {author} {\bibfnamefont {M.}~\bibnamefont
  {Dhuria}}, \bibinfo {author} {\bibfnamefont {C.}~\bibnamefont {Hati}},
  \bibinfo {author} {\bibfnamefont {R.}~\bibnamefont {Rangarajan}}, \ and\
  \bibinfo {author} {\bibfnamefont {U.}~\bibnamefont {Sarkar}},\ }\href
  {\doibase 10.1103/PhysRevD.92.031701} {\bibfield  {journal} {\bibinfo
  {journal} {Phys. Rev.}\ }\textbf {\bibinfo {volume} {D92}},\ \bibinfo {pages}
  {031701} (\bibinfo {year} {2015})},\ \Eprint
  {http://arxiv.org/abs/1503.07198} {arXiv:1503.07198 [hep-ph]} \BibitemShut
  {NoStop}%
\bibitem [{\citenamefont {Jeli\'nski}\ and\ \citenamefont
  {Kordiaczy\'nska}(2015)}]{Jelinski:2015ifw}%
  \BibitemOpen
  \bibfield  {author} {\bibinfo {author} {\bibfnamefont {T.}~\bibnamefont
  {Jeli\'nski}}\ and\ \bibinfo {author} {\bibfnamefont {M.}~\bibnamefont
  {Kordiaczy\'nska}},\ }\href {\doibase 10.5506/APhysPolB.46.2193} {\bibfield
  {journal} {\bibinfo  {journal} {Acta Phys. Polon.}\ }\textbf {\bibinfo
  {volume} {B46}},\ \bibinfo {pages} {2193} (\bibinfo {year}
  {2015})}\BibitemShut {NoStop}%
\bibitem [{\citenamefont {Ko}\ and\ \citenamefont {Nomura}(2016)}]{Ko:2015uma}%
  \BibitemOpen
  \bibfield  {author} {\bibinfo {author} {\bibfnamefont {P.}~\bibnamefont
  {Ko}}\ and\ \bibinfo {author} {\bibfnamefont {T.}~\bibnamefont {Nomura}},\
  }\href {\doibase 10.1016/j.physletb.2015.12.072} {\bibfield  {journal}
  {\bibinfo  {journal} {Phys. Lett.}\ }\textbf {\bibinfo {volume} {B753}},\
  \bibinfo {pages} {612} (\bibinfo {year} {2016})},\ \Eprint
  {http://arxiv.org/abs/1510.07872} {arXiv:1510.07872 [hep-ph]} \BibitemShut
  {NoStop}%
\bibitem [{\citenamefont {Leonardi}\ \emph {et~al.}(2015)\citenamefont
  {Leonardi}, \citenamefont {Alunni}, \citenamefont {Romeo}, \citenamefont
  {Fano},\ and\ \citenamefont {Panella}}]{Leonardi:2015qna}%
  \BibitemOpen
  \bibfield  {author} {\bibinfo {author} {\bibfnamefont {R.}~\bibnamefont
  {Leonardi}}, \bibinfo {author} {\bibfnamefont {L.}~\bibnamefont {Alunni}},
  \bibinfo {author} {\bibfnamefont {F.}~\bibnamefont {Romeo}}, \bibinfo
  {author} {\bibfnamefont {L.}~\bibnamefont {Fano}}, \ and\ \bibinfo {author}
  {\bibfnamefont {O.}~\bibnamefont {Panella}},\ }\href@noop {} {\  (\bibinfo
  {year} {2015})},\ \Eprint {http://arxiv.org/abs/1510.07988} {arXiv:1510.07988
  [hep-ph]} \BibitemShut {NoStop}%
\bibitem [{\citenamefont {Collins}\ and\ \citenamefont
  {Ng}(2016)}]{Collins:2015wua}%
  \BibitemOpen
  \bibfield  {author} {\bibinfo {author} {\bibfnamefont {J.~H.}\ \bibnamefont
  {Collins}}\ and\ \bibinfo {author} {\bibfnamefont {W.~H.}\ \bibnamefont
  {Ng}},\ }\href {\doibase 10.1007/JHEP01(2016)159} {\bibfield  {journal}
  {\bibinfo  {journal} {JHEP}\ }\textbf {\bibinfo {volume} {01}},\ \bibinfo
  {pages} {159} (\bibinfo {year} {2016})},\ \Eprint
  {http://arxiv.org/abs/1510.08083} {arXiv:1510.08083 [hep-ph]} \BibitemShut
  {NoStop}%
\bibitem [{\citenamefont {Dobrescu}\ and\ \citenamefont
  {Fox}(2015)}]{Dobrescu:2015jvn}%
  \BibitemOpen
  \bibfield  {author} {\bibinfo {author} {\bibfnamefont {B.~A.}\ \bibnamefont
  {Dobrescu}}\ and\ \bibinfo {author} {\bibfnamefont {P.~J.}\ \bibnamefont
  {Fox}},\ }\href@noop {} {\  (\bibinfo {year} {2015})},\ \Eprint
  {http://arxiv.org/abs/1511.02148} {arXiv:1511.02148 [hep-ph]} \BibitemShut
  {NoStop}%
\bibitem [{\citenamefont {Hati}\ \emph {et~al.}(2016)\citenamefont {Hati},
  \citenamefont {Kumar},\ and\ \citenamefont {Mahajan}}]{Hati:2015awg}%
  \BibitemOpen
  \bibfield  {author} {\bibinfo {author} {\bibfnamefont {C.}~\bibnamefont
  {Hati}}, \bibinfo {author} {\bibfnamefont {G.}~\bibnamefont {Kumar}}, \ and\
  \bibinfo {author} {\bibfnamefont {N.}~\bibnamefont {Mahajan}},\ }\href
  {\doibase 10.1007/JHEP01(2016)117} {\bibfield  {journal} {\bibinfo  {journal}
  {JHEP}\ }\textbf {\bibinfo {volume} {01}},\ \bibinfo {pages} {117} (\bibinfo
  {year} {2016})},\ \Eprint {http://arxiv.org/abs/1511.03290} {arXiv:1511.03290
  [hep-ph]} \BibitemShut {NoStop}%
\bibitem [{\citenamefont {Aydemir}(2016)}]{Aydemir:2015oob}%
  \BibitemOpen
  \bibfield  {author} {\bibinfo {author} {\bibfnamefont {U.}~\bibnamefont
  {Aydemir}},\ }\href {\doibase 10.1142/S0217751X16500342} {\bibfield
  {journal} {\bibinfo  {journal} {Int. J. Mod. Phys.}\ }\textbf {\bibinfo
  {volume} {A31}},\ \bibinfo {pages} {1650034} (\bibinfo {year} {2016})},\
  \Eprint {http://arxiv.org/abs/1512.00568} {arXiv:1512.00568 [hep-ph]}
  \BibitemShut {NoStop}%
\bibitem [{\citenamefont {Garcia-Cely}\ and\ \citenamefont
  {Heeck}(2015)}]{Garcia-Cely:2015quu}%
  \BibitemOpen
  \bibfield  {author} {\bibinfo {author} {\bibfnamefont {C.}~\bibnamefont
  {Garcia-Cely}}\ and\ \bibinfo {author} {\bibfnamefont {J.}~\bibnamefont
  {Heeck}},\ }\href {\doibase 10.1088/1475-7516/2016/03/021} {\  (\bibinfo
  {year} {2015}),\ 10.1088/1475-7516/2016/03/021},\ \Eprint
  {http://arxiv.org/abs/1512.03332} {arXiv:1512.03332 [hep-ph]} \BibitemShut
  {NoStop}%
\bibitem [{\citenamefont {Blumenhagen}(2016)}]{Blumenhagen:2015fqn}%
  \BibitemOpen
  \bibfield  {author} {\bibinfo {author} {\bibfnamefont {R.}~\bibnamefont
  {Blumenhagen}},\ }\href {\doibase 10.1007/JHEP01(2016)039} {\bibfield
  {journal} {\bibinfo  {journal} {JHEP}\ }\textbf {\bibinfo {volume} {01}},\
  \bibinfo {pages} {039} (\bibinfo {year} {2016})},\ \Eprint
  {http://arxiv.org/abs/1512.03154} {arXiv:1512.03154 [hep-ph]} \BibitemShut
  {NoStop}%
\bibitem [{\citenamefont {Dasgupta}\ \emph {et~al.}(2015)\citenamefont
  {Dasgupta}, \citenamefont {Mitra},\ and\ \citenamefont
  {Borah}}]{Dasgupta:2015pbr}%
  \BibitemOpen
  \bibfield  {author} {\bibinfo {author} {\bibfnamefont {A.}~\bibnamefont
  {Dasgupta}}, \bibinfo {author} {\bibfnamefont {M.}~\bibnamefont {Mitra}}, \
  and\ \bibinfo {author} {\bibfnamefont {D.}~\bibnamefont {Borah}},\
  }\href@noop {} {\  (\bibinfo {year} {2015})},\ \Eprint
  {http://arxiv.org/abs/1512.09202} {arXiv:1512.09202 [hep-ph]} \BibitemShut
  {NoStop}%
\bibitem [{\citenamefont {Shu}\ and\ \citenamefont
  {Yepes}(2015)}]{Shu:2015cxm}%
  \BibitemOpen
  \bibfield  {author} {\bibinfo {author} {\bibfnamefont {J.}~\bibnamefont
  {Shu}}\ and\ \bibinfo {author} {\bibfnamefont {J.}~\bibnamefont {Yepes}},\
  }\href@noop {} {\  (\bibinfo {year} {2015})},\ \Eprint
  {http://arxiv.org/abs/1512.09310} {arXiv:1512.09310 [hep-ph]} \BibitemShut
  {NoStop}%
\bibitem [{\citenamefont {Hati}(2016)}]{Hati:2016thk}%
  \BibitemOpen
  \bibfield  {author} {\bibinfo {author} {\bibfnamefont {C.}~\bibnamefont
  {Hati}},\ }\href {\doibase 10.1103/PhysRevD.93.075002} {\bibfield  {journal}
  {\bibinfo  {journal} {Phys. Rev.}\ }\textbf {\bibinfo {volume} {D93}},\
  \bibinfo {pages} {075002} (\bibinfo {year} {2016})},\ \Eprint
  {http://arxiv.org/abs/1601.02457} {arXiv:1601.02457 [hep-ph]} \BibitemShut
  {NoStop}%
\bibitem [{\citenamefont {Das}\ \emph {et~al.}(2016{\natexlab{a}})\citenamefont
  {Das}, \citenamefont {Nagata},\ and\ \citenamefont {Okada}}]{Das:2016akd}%
  \BibitemOpen
  \bibfield  {author} {\bibinfo {author} {\bibfnamefont {A.}~\bibnamefont
  {Das}}, \bibinfo {author} {\bibfnamefont {N.}~\bibnamefont {Nagata}}, \ and\
  \bibinfo {author} {\bibfnamefont {N.}~\bibnamefont {Okada}},\ }\href
  {\doibase 10.1007/JHEP03(2016)049} {\bibfield  {journal} {\bibinfo  {journal}
  {JHEP}\ }\textbf {\bibinfo {volume} {03}},\ \bibinfo {pages} {049} (\bibinfo
  {year} {2016}{\natexlab{a}})},\ \Eprint {http://arxiv.org/abs/1601.05079}
  {arXiv:1601.05079 [hep-ph]} \BibitemShut {NoStop}%
\bibitem [{\citenamefont {Shu}\ and\ \citenamefont
  {Yepes}(2016)}]{Shu:2016exh}%
  \BibitemOpen
  \bibfield  {author} {\bibinfo {author} {\bibfnamefont {J.}~\bibnamefont
  {Shu}}\ and\ \bibinfo {author} {\bibfnamefont {J.}~\bibnamefont {Yepes}},\
  }\href@noop {} {\  (\bibinfo {year} {2016})},\ \Eprint
  {http://arxiv.org/abs/1601.06891} {arXiv:1601.06891 [hep-ph]} \BibitemShut
  {NoStop}%
\bibitem [{\citenamefont {Datta}\ \emph {et~al.}(1994)\citenamefont {Datta},
  \citenamefont {Guchait},\ and\ \citenamefont {Pilaftsis}}]{Datta:1993nm}%
  \BibitemOpen
  \bibfield  {author} {\bibinfo {author} {\bibfnamefont {A.}~\bibnamefont
  {Datta}}, \bibinfo {author} {\bibfnamefont {M.}~\bibnamefont {Guchait}}, \
  and\ \bibinfo {author} {\bibfnamefont {A.}~\bibnamefont {Pilaftsis}},\ }\href
  {\doibase 10.1103/PhysRevD.50.3195} {\bibfield  {journal} {\bibinfo
  {journal} {Phys. Rev.}\ }\textbf {\bibinfo {volume} {D50}},\ \bibinfo {pages}
  {3195} (\bibinfo {year} {1994})},\ \Eprint
  {http://arxiv.org/abs/hep-ph/9311257} {arXiv:hep-ph/9311257 [hep-ph]}
  \BibitemShut {NoStop}%
\bibitem [{\citenamefont {Ferrari}\ \emph {et~al.}(2000)\citenamefont
  {Ferrari}, \citenamefont {Collot}, \citenamefont {Andrieux}, \citenamefont
  {Belhorma}, \citenamefont {de~Saintignon} \emph {et~al.}}]{Ferrari:2000sp}%
  \BibitemOpen
  \bibfield  {author} {\bibinfo {author} {\bibfnamefont {A.}~\bibnamefont
  {Ferrari}}, \bibinfo {author} {\bibfnamefont {J.}~\bibnamefont {Collot}},
  \bibinfo {author} {\bibfnamefont {M.-L.}\ \bibnamefont {Andrieux}}, \bibinfo
  {author} {\bibfnamefont {B.}~\bibnamefont {Belhorma}}, \bibinfo {author}
  {\bibfnamefont {P.}~\bibnamefont {de~Saintignon}},  \emph {et~al.},\ }\href
  {\doibase 10.1103/PhysRevD.62.013001} {\bibfield  {journal} {\bibinfo
  {journal} {Phys.Rev.}\ }\textbf {\bibinfo {volume} {D62}},\ \bibinfo {pages}
  {013001} (\bibinfo {year} {2000})}\BibitemShut {NoStop}%
\bibitem [{\citenamefont {Kersten}\ and\ \citenamefont
  {Smirnov}(2007)}]{Kersten:2007vk}%
  \BibitemOpen
  \bibfield  {author} {\bibinfo {author} {\bibfnamefont {J.}~\bibnamefont
  {Kersten}}\ and\ \bibinfo {author} {\bibfnamefont {A.~{\relax Yu}.}\
  \bibnamefont {Smirnov}},\ }\href {\doibase 10.1103/PhysRevD.76.073005}
  {\bibfield  {journal} {\bibinfo  {journal} {Phys. Rev.}\ }\textbf {\bibinfo
  {volume} {D76}},\ \bibinfo {pages} {073005} (\bibinfo {year} {2007})},\
  \Eprint {http://arxiv.org/abs/0705.3221} {arXiv:0705.3221 [hep-ph]}
  \BibitemShut {NoStop}%
\bibitem [{\citenamefont {del Aguila}\ \emph
  {et~al.}(2007{\natexlab{a}})\citenamefont {del Aguila}, \citenamefont
  {Aguilar-Saavedra},\ and\ \citenamefont {Pittau}}]{delAguila:2007qnc}%
  \BibitemOpen
  \bibfield  {author} {\bibinfo {author} {\bibfnamefont {F.}~\bibnamefont {del
  Aguila}}, \bibinfo {author} {\bibfnamefont {J.~A.}\ \bibnamefont
  {Aguilar-Saavedra}}, \ and\ \bibinfo {author} {\bibfnamefont
  {R.}~\bibnamefont {Pittau}},\ }\href {\doibase 10.1088/1126-6708/2007/10/047}
  {\bibfield  {journal} {\bibinfo  {journal} {JHEP}\ }\textbf {\bibinfo
  {volume} {10}},\ \bibinfo {pages} {047} (\bibinfo {year}
  {2007}{\natexlab{a}})},\ \Eprint {http://arxiv.org/abs/hep-ph/0703261}
  {arXiv:hep-ph/0703261 [hep-ph]} \BibitemShut {NoStop}%
\bibitem [{\citenamefont {del Aguila}\ and\ \citenamefont
  {Aguilar-Saavedra}(2009)}]{delAguila:2008hw}%
  \BibitemOpen
  \bibfield  {author} {\bibinfo {author} {\bibfnamefont {F.}~\bibnamefont {del
  Aguila}}\ and\ \bibinfo {author} {\bibfnamefont {J.~A.}\ \bibnamefont
  {Aguilar-Saavedra}},\ }\href {\doibase 10.1016/j.physletb.2009.01.010}
  {\bibfield  {journal} {\bibinfo  {journal} {Phys. Lett.}\ }\textbf {\bibinfo
  {volume} {B672}},\ \bibinfo {pages} {158} (\bibinfo {year} {2009})},\ \Eprint
  {http://arxiv.org/abs/0809.2096} {arXiv:0809.2096 [hep-ph]} \BibitemShut
  {NoStop}%
\bibitem [{\citenamefont {Maiezza}\ \emph {et~al.}(2010)\citenamefont
  {Maiezza}, \citenamefont {Nemevsek}, \citenamefont {Nesti},\ and\
  \citenamefont {Senjanovic}}]{Maiezza:2010ic}%
  \BibitemOpen
  \bibfield  {author} {\bibinfo {author} {\bibfnamefont {A.}~\bibnamefont
  {Maiezza}}, \bibinfo {author} {\bibfnamefont {M.}~\bibnamefont {Nemevsek}},
  \bibinfo {author} {\bibfnamefont {F.}~\bibnamefont {Nesti}}, \ and\ \bibinfo
  {author} {\bibfnamefont {G.}~\bibnamefont {Senjanovic}},\ }\href {\doibase
  10.1103/PhysRevD.82.055022} {\bibfield  {journal} {\bibinfo  {journal}
  {Phys.Rev.}\ }\textbf {\bibinfo {volume} {D82}},\ \bibinfo {pages} {055022}
  (\bibinfo {year} {2010})},\ \Eprint {http://arxiv.org/abs/1005.5160}
  {arXiv:1005.5160 [hep-ph]} \BibitemShut {NoStop}%
\bibitem [{\citenamefont {Nemevsek}\ \emph {et~al.}(2011)\citenamefont
  {Nemevsek}, \citenamefont {Nesti}, \citenamefont {Senjanovic},\ and\
  \citenamefont {Zhang}}]{Nemevsek:2011hz}%
  \BibitemOpen
  \bibfield  {author} {\bibinfo {author} {\bibfnamefont {M.}~\bibnamefont
  {Nemevsek}}, \bibinfo {author} {\bibfnamefont {F.}~\bibnamefont {Nesti}},
  \bibinfo {author} {\bibfnamefont {G.}~\bibnamefont {Senjanovic}}, \ and\
  \bibinfo {author} {\bibfnamefont {Y.}~\bibnamefont {Zhang}},\ }\href
  {\doibase 10.1103/PhysRevD.83.115014} {\bibfield  {journal} {\bibinfo
  {journal} {Phys.Rev.}\ }\textbf {\bibinfo {volume} {D83}},\ \bibinfo {pages}
  {115014} (\bibinfo {year} {2011})},\ \Eprint {http://arxiv.org/abs/1103.1627}
  {arXiv:1103.1627 [hep-ph]} \BibitemShut {NoStop}%
\bibitem [{\citenamefont {Chen}\ and\ \citenamefont {Dev}(2012)}]{Chen:2011hc}%
  \BibitemOpen
  \bibfield  {author} {\bibinfo {author} {\bibfnamefont {C.-Y.}\ \bibnamefont
  {Chen}}\ and\ \bibinfo {author} {\bibfnamefont {P.~S.~B.}\ \bibnamefont
  {Dev}},\ }\href {\doibase 10.1103/PhysRevD.85.093018} {\bibfield  {journal}
  {\bibinfo  {journal} {Phys. Rev.}\ }\textbf {\bibinfo {volume} {D85}},\
  \bibinfo {pages} {093018} (\bibinfo {year} {2012})},\ \Eprint
  {http://arxiv.org/abs/1112.6419} {arXiv:1112.6419 [hep-ph]} \BibitemShut
  {NoStop}%
\bibitem [{\citenamefont {Das}\ \emph {et~al.}(2012)\citenamefont {Das},
  \citenamefont {Deppisch}, \citenamefont {Kittel},\ and\ \citenamefont
  {Valle}}]{Das:2012ii}%
  \BibitemOpen
  \bibfield  {author} {\bibinfo {author} {\bibfnamefont {S.}~\bibnamefont
  {Das}}, \bibinfo {author} {\bibfnamefont {F.}~\bibnamefont {Deppisch}},
  \bibinfo {author} {\bibfnamefont {O.}~\bibnamefont {Kittel}}, \ and\ \bibinfo
  {author} {\bibfnamefont {J.}~\bibnamefont {Valle}},\ }\href {\doibase
  10.1103/PhysRevD.86.055006} {\bibfield  {journal} {\bibinfo  {journal}
  {Phys.Rev.}\ }\textbf {\bibinfo {volume} {D86}},\ \bibinfo {pages} {055006}
  (\bibinfo {year} {2012})},\ \Eprint {http://arxiv.org/abs/1206.0256}
  {arXiv:1206.0256 [hep-ph]} \BibitemShut {NoStop}%
\bibitem [{\citenamefont {Chakrabortty}\ \emph
  {et~al.}(2012{\natexlab{a}})\citenamefont {Chakrabortty}, \citenamefont
  {Gluza}, \citenamefont {Sevillano},\ and\ \citenamefont
  {Szafron}}]{Chakrabortty:2012pp}%
  \BibitemOpen
  \bibfield  {author} {\bibinfo {author} {\bibfnamefont {J.}~\bibnamefont
  {Chakrabortty}}, \bibinfo {author} {\bibfnamefont {J.}~\bibnamefont {Gluza}},
  \bibinfo {author} {\bibfnamefont {R.}~\bibnamefont {Sevillano}}, \ and\
  \bibinfo {author} {\bibfnamefont {R.}~\bibnamefont {Szafron}},\ }\href
  {\doibase 10.1007/JHEP07(2012)038} {\bibfield  {journal} {\bibinfo  {journal}
  {JHEP}\ }\textbf {\bibinfo {volume} {1207}},\ \bibinfo {pages} {038}
  (\bibinfo {year} {2012}{\natexlab{a}})},\ \Eprint
  {http://arxiv.org/abs/1204.0736} {arXiv:1204.0736 [hep-ph]} \BibitemShut
  {NoStop}%
\bibitem [{\citenamefont {Han}\ \emph {et~al.}(2013)\citenamefont {Han},
  \citenamefont {Lewis}, \citenamefont {Ruiz},\ and\ \citenamefont
  {Si}}]{Han:2012vk}%
  \BibitemOpen
  \bibfield  {author} {\bibinfo {author} {\bibfnamefont {T.}~\bibnamefont
  {Han}}, \bibinfo {author} {\bibfnamefont {I.}~\bibnamefont {Lewis}}, \bibinfo
  {author} {\bibfnamefont {R.}~\bibnamefont {Ruiz}}, \ and\ \bibinfo {author}
  {\bibfnamefont {Z.-g.}\ \bibnamefont {Si}},\ }\href {\doibase
  10.1103/PhysRevD.87.035011, 10.1103/PhysRevD.87.039906} {\bibfield  {journal}
  {\bibinfo  {journal} {Phys. Rev.}\ }\textbf {\bibinfo {volume} {D87}},\
  \bibinfo {pages} {035011} (\bibinfo {year} {2013})},\ \bibinfo {note}
  {[Erratum: Phys. Rev.D87,no.3,039906(2013)]},\ \Eprint
  {http://arxiv.org/abs/1211.6447} {arXiv:1211.6447 [hep-ph]} \BibitemShut
  {NoStop}%
\bibitem [{\citenamefont {Aguilar-Saavedra}\ and\ \citenamefont
  {Joaquim}(2012)}]{AguilarSaavedra:2012gf}%
  \BibitemOpen
  \bibfield  {author} {\bibinfo {author} {\bibfnamefont {J.~A.}\ \bibnamefont
  {Aguilar-Saavedra}}\ and\ \bibinfo {author} {\bibfnamefont {F.~R.}\
  \bibnamefont {Joaquim}},\ }\href {\doibase 10.1103/PhysRevD.86.073005}
  {\bibfield  {journal} {\bibinfo  {journal} {Phys. Rev.}\ }\textbf {\bibinfo
  {volume} {D86}},\ \bibinfo {pages} {073005} (\bibinfo {year} {2012})},\
  \Eprint {http://arxiv.org/abs/1207.4193} {arXiv:1207.4193 [hep-ph]}
  \BibitemShut {NoStop}%
\bibitem [{\citenamefont {Dev}\ and\ \citenamefont
  {Mohapatra}(2013)}]{Dev:2013vba}%
  \BibitemOpen
  \bibfield  {author} {\bibinfo {author} {\bibfnamefont {P.~S.~B.}\
  \bibnamefont {Dev}}\ and\ \bibinfo {author} {\bibfnamefont {R.~N.}\
  \bibnamefont {Mohapatra}}\ }(\bibinfo {year} {2013})\ \Eprint
  {http://arxiv.org/abs/1308.2151} {arXiv:1308.2151 [hep-ph]} \BibitemShut
  {NoStop}%
\bibitem [{\citenamefont {Kayser}(1982)}]{Kayser:1982br}%
  \BibitemOpen
  \bibfield  {author} {\bibinfo {author} {\bibfnamefont {B.}~\bibnamefont
  {Kayser}},\ }\href {\doibase 10.1103/PhysRevD.26.1662} {\bibfield  {journal}
  {\bibinfo  {journal} {Phys. Rev.}\ }\textbf {\bibinfo {volume} {D26}},\
  \bibinfo {pages} {1662} (\bibinfo {year} {1982})}\BibitemShut {NoStop}%
\bibitem [{\citenamefont {Dicus}\ \emph {et~al.}(1991)\citenamefont {Dicus},
  \citenamefont {Karatas},\ and\ \citenamefont {Roy}}]{Dicus:1991fk}%
  \BibitemOpen
  \bibfield  {author} {\bibinfo {author} {\bibfnamefont {D.~A.}\ \bibnamefont
  {Dicus}}, \bibinfo {author} {\bibfnamefont {D.~D.}\ \bibnamefont {Karatas}},
  \ and\ \bibinfo {author} {\bibfnamefont {P.}~\bibnamefont {Roy}},\ }\href
  {\doibase 10.1103/PhysRevD.44.2033} {\bibfield  {journal} {\bibinfo
  {journal} {Phys. Rev.}\ }\textbf {\bibinfo {volume} {D44}},\ \bibinfo {pages}
  {2033} (\bibinfo {year} {1991})}\BibitemShut {NoStop}%
\bibitem [{\citenamefont {Datta}\ and\ \citenamefont
  {Pilaftsis}(1992)}]{Datta:1991mf}%
  \BibitemOpen
  \bibfield  {author} {\bibinfo {author} {\bibfnamefont {A.}~\bibnamefont
  {Datta}}\ and\ \bibinfo {author} {\bibfnamefont {A.}~\bibnamefont
  {Pilaftsis}},\ }\href {\doibase 10.1016/0370-2693(92)90727-L} {\bibfield
  {journal} {\bibinfo  {journal} {Phys. Lett.}\ }\textbf {\bibinfo {volume}
  {B278}},\ \bibinfo {pages} {162} (\bibinfo {year} {1992})}\BibitemShut
  {NoStop}%
\bibitem [{\citenamefont {Datta}\ \emph {et~al.}(1993)\citenamefont {Datta},
  \citenamefont {Guchait},\ and\ \citenamefont {Roy}}]{Datta:1992qw}%
  \BibitemOpen
  \bibfield  {author} {\bibinfo {author} {\bibfnamefont {A.}~\bibnamefont
  {Datta}}, \bibinfo {author} {\bibfnamefont {M.}~\bibnamefont {Guchait}}, \
  and\ \bibinfo {author} {\bibfnamefont {D.~P.}\ \bibnamefont {Roy}},\ }\href
  {\doibase 10.1103/PhysRevD.47.961} {\bibfield  {journal} {\bibinfo  {journal}
  {Phys. Rev.}\ }\textbf {\bibinfo {volume} {D47}},\ \bibinfo {pages} {961}
  (\bibinfo {year} {1993})},\ \Eprint {http://arxiv.org/abs/hep-ph/9208228}
  {arXiv:hep-ph/9208228 [hep-ph]} \BibitemShut {NoStop}%
\bibitem [{\citenamefont {Gluza}\ and\ \citenamefont
  {Zra{\l}ek}(1993)}]{Gluza:1993gf}%
  \BibitemOpen
  \bibfield  {author} {\bibinfo {author} {\bibfnamefont {J.}~\bibnamefont
  {Gluza}}\ and\ \bibinfo {author} {\bibfnamefont {M.}~\bibnamefont
  {Zra{\l}ek}},\ }\href {\doibase 10.1103/PhysRevD.48.5093} {\bibfield
  {journal} {\bibinfo  {journal} {Phys.Rev.}\ }\textbf {\bibinfo {volume}
  {D48}},\ \bibinfo {pages} {5093} (\bibinfo {year} {1993})}\BibitemShut
  {NoStop}%
\bibitem [{\citenamefont {Gluza}\ and\ \citenamefont
  {Zra{\l}ek}(1996)}]{Gluza:1995js}%
  \BibitemOpen
  \bibfield  {author} {\bibinfo {author} {\bibfnamefont {J.}~\bibnamefont
  {Gluza}}\ and\ \bibinfo {author} {\bibfnamefont {M.}~\bibnamefont
  {Zra{\l}ek}},\ }\href {\doibase 10.1016/0370-2693(96)00074-3} {\bibfield
  {journal} {\bibinfo  {journal} {Phys. Lett.}\ }\textbf {\bibinfo {volume}
  {B372}},\ \bibinfo {pages} {259} (\bibinfo {year} {1996})},\ \Eprint
  {http://arxiv.org/abs/hep-ph/9510407} {arXiv:hep-ph/9510407 [hep-ph]}
  \BibitemShut {NoStop}%
\bibitem [{\citenamefont {Gluza}(1997)}]{Gluza:1997kg}%
  \BibitemOpen
  \bibfield  {author} {\bibinfo {author} {\bibfnamefont {J.}~\bibnamefont
  {Gluza}},\ }\href {\doibase 10.1016/S0370-2693(97)00381-X} {\bibfield
  {journal} {\bibinfo  {journal} {Phys. Lett.}\ }\textbf {\bibinfo {volume}
  {B403}},\ \bibinfo {pages} {304} (\bibinfo {year} {1997})},\ \Eprint
  {http://arxiv.org/abs/hep-ph/9704202} {arXiv:hep-ph/9704202 [hep-ph]}
  \BibitemShut {NoStop}%
\bibitem [{\citenamefont {Gluza}\ \emph {et~al.}(1997)\citenamefont {Gluza},
  \citenamefont {Maalampi}, \citenamefont {Raidal},\ and\ \citenamefont
  {Zra{\l}ek}}]{Gluza:1997ts}%
  \BibitemOpen
  \bibfield  {author} {\bibinfo {author} {\bibfnamefont {J.}~\bibnamefont
  {Gluza}}, \bibinfo {author} {\bibfnamefont {J.}~\bibnamefont {Maalampi}},
  \bibinfo {author} {\bibfnamefont {M.}~\bibnamefont {Raidal}}, \ and\ \bibinfo
  {author} {\bibfnamefont {M.}~\bibnamefont {Zra{\l}ek}},\ }\href {\doibase
  10.1016/S0370-2693(97)00713-2} {\bibfield  {journal} {\bibinfo  {journal}
  {Phys.Lett.}\ }\textbf {\bibinfo {volume} {B407}},\ \bibinfo {pages} {45}
  (\bibinfo {year} {1997})},\ \Eprint {http://arxiv.org/abs/hep-ph/9703215}
  {arXiv:hep-ph/9703215 [hep-ph]} \BibitemShut {NoStop}%
\bibitem [{\citenamefont {del Aguila}\ \emph {et~al.}(2005)\citenamefont {del
  Aguila}, \citenamefont {Aguilar-Saavedra}, \citenamefont {Martinez de~la
  Ossa},\ and\ \citenamefont {Meloni}}]{delAguila:2005ssc}%
  \BibitemOpen
  \bibfield  {author} {\bibinfo {author} {\bibfnamefont {F.}~\bibnamefont {del
  Aguila}}, \bibinfo {author} {\bibfnamefont {J.~A.}\ \bibnamefont
  {Aguilar-Saavedra}}, \bibinfo {author} {\bibfnamefont {A.}~\bibnamefont
  {Martinez de~la Ossa}}, \ and\ \bibinfo {author} {\bibfnamefont
  {D.}~\bibnamefont {Meloni}},\ }\href {\doibase
  10.1016/j.j.physletb.2005.03.054, 10.1016/j.physletb.2005.03.054} {\bibfield
  {journal} {\bibinfo  {journal} {Phys. Lett.}\ }\textbf {\bibinfo {volume}
  {B613}},\ \bibinfo {pages} {170} (\bibinfo {year} {2005})},\ \Eprint
  {http://arxiv.org/abs/hep-ph/0502189} {arXiv:hep-ph/0502189 [hep-ph]}
  \BibitemShut {NoStop}%
\bibitem [{\citenamefont {Han}\ and\ \citenamefont {Zhang}(2006)}]{Han:2006ip}%
  \BibitemOpen
  \bibfield  {author} {\bibinfo {author} {\bibfnamefont {T.}~\bibnamefont
  {Han}}\ and\ \bibinfo {author} {\bibfnamefont {B.}~\bibnamefont {Zhang}},\
  }\href {\doibase 10.1103/PhysRevLett.97.171804} {\bibfield  {journal}
  {\bibinfo  {journal} {Phys. Rev. Lett.}\ }\textbf {\bibinfo {volume} {97}},\
  \bibinfo {pages} {171804} (\bibinfo {year} {2006})},\ \Eprint
  {http://arxiv.org/abs/hep-ph/0604064} {arXiv:hep-ph/0604064 [hep-ph]}
  \BibitemShut {NoStop}%
\bibitem [{\citenamefont {Atre}\ \emph {et~al.}(2009)\citenamefont {Atre},
  \citenamefont {Han}, \citenamefont {Pascoli},\ and\ \citenamefont
  {Zhang}}]{Atre:2009rg}%
  \BibitemOpen
  \bibfield  {author} {\bibinfo {author} {\bibfnamefont {A.}~\bibnamefont
  {Atre}}, \bibinfo {author} {\bibfnamefont {T.}~\bibnamefont {Han}}, \bibinfo
  {author} {\bibfnamefont {S.}~\bibnamefont {Pascoli}}, \ and\ \bibinfo
  {author} {\bibfnamefont {B.}~\bibnamefont {Zhang}},\ }\href {\doibase
  10.1088/1126-6708/2009/05/030} {\bibfield  {journal} {\bibinfo  {journal}
  {JHEP}\ }\textbf {\bibinfo {volume} {05}},\ \bibinfo {pages} {030} (\bibinfo
  {year} {2009})},\ \Eprint {http://arxiv.org/abs/0901.3589} {arXiv:0901.3589
  [hep-ph]} \BibitemShut {NoStop}%
\bibitem [{\citenamefont {Ibarra}\ \emph {et~al.}(2010)\citenamefont {Ibarra},
  \citenamefont {Molinaro},\ and\ \citenamefont {Petcov}}]{Ibarra:2010xw}%
  \BibitemOpen
  \bibfield  {author} {\bibinfo {author} {\bibfnamefont {A.}~\bibnamefont
  {Ibarra}}, \bibinfo {author} {\bibfnamefont {E.}~\bibnamefont {Molinaro}}, \
  and\ \bibinfo {author} {\bibfnamefont {S.~T.}\ \bibnamefont {Petcov}},\
  }\href {\doibase 10.1007/JHEP09(2010)108} {\bibfield  {journal} {\bibinfo
  {journal} {JHEP}\ }\textbf {\bibinfo {volume} {09}},\ \bibinfo {pages} {108}
  (\bibinfo {year} {2010})},\ \Eprint {http://arxiv.org/abs/1007.2378}
  {arXiv:1007.2378 [hep-ph]} \BibitemShut {NoStop}%
\bibitem [{\citenamefont {Adhikari}\ and\ \citenamefont
  {Raychaudhuri}(2011)}]{Adhikari:2010yt}%
  \BibitemOpen
  \bibfield  {author} {\bibinfo {author} {\bibfnamefont {R.}~\bibnamefont
  {Adhikari}}\ and\ \bibinfo {author} {\bibfnamefont {A.}~\bibnamefont
  {Raychaudhuri}},\ }\href {\doibase 10.1103/PhysRevD.84.033002} {\bibfield
  {journal} {\bibinfo  {journal} {Phys. Rev.}\ }\textbf {\bibinfo {volume}
  {D84}},\ \bibinfo {pages} {033002} (\bibinfo {year} {2011})},\ \Eprint
  {http://arxiv.org/abs/1004.5111} {arXiv:1004.5111 [hep-ph]} \BibitemShut
  {NoStop}%
\bibitem [{\citenamefont {Cely}\ \emph {et~al.}(2013)\citenamefont {Cely},
  \citenamefont {Ibarra}, \citenamefont {Molinaro},\ and\ \citenamefont
  {Petcov}}]{Cely:2012bz}%
  \BibitemOpen
  \bibfield  {author} {\bibinfo {author} {\bibfnamefont {C.~G.}\ \bibnamefont
  {Cely}}, \bibinfo {author} {\bibfnamefont {A.}~\bibnamefont {Ibarra}},
  \bibinfo {author} {\bibfnamefont {E.}~\bibnamefont {Molinaro}}, \ and\
  \bibinfo {author} {\bibfnamefont {S.~T.}\ \bibnamefont {Petcov}},\ }\href
  {\doibase 10.1016/j.physletb.2012.11.026} {\bibfield  {journal} {\bibinfo
  {journal} {Phys. Lett.}\ }\textbf {\bibinfo {volume} {B718}},\ \bibinfo
  {pages} {957} (\bibinfo {year} {2013})},\ \Eprint
  {http://arxiv.org/abs/1208.3654} {arXiv:1208.3654 [hep-ph]} \BibitemShut
  {NoStop}%
\bibitem [{\citenamefont {Bhupal~Dev}\ \emph {et~al.}(2012)\citenamefont
  {Bhupal~Dev}, \citenamefont {Franceschini},\ and\ \citenamefont
  {Mohapatra}}]{BhupalDev:2012zg}%
  \BibitemOpen
  \bibfield  {author} {\bibinfo {author} {\bibfnamefont {P.~S.}\ \bibnamefont
  {Bhupal~Dev}}, \bibinfo {author} {\bibfnamefont {R.}~\bibnamefont
  {Franceschini}}, \ and\ \bibinfo {author} {\bibfnamefont {R.~N.}\
  \bibnamefont {Mohapatra}},\ }\href {\doibase 10.1103/PhysRevD.86.093010}
  {\bibfield  {journal} {\bibinfo  {journal} {Phys. Rev.}\ }\textbf {\bibinfo
  {volume} {D86}},\ \bibinfo {pages} {093010} (\bibinfo {year} {2012})},\
  \Eprint {http://arxiv.org/abs/1207.2756} {arXiv:1207.2756 [hep-ph]}
  \BibitemShut {NoStop}%
\bibitem [{\citenamefont {Das}\ and\ \citenamefont {Okada}(2013)}]{Das:2012ze}%
  \BibitemOpen
  \bibfield  {author} {\bibinfo {author} {\bibfnamefont {A.}~\bibnamefont
  {Das}}\ and\ \bibinfo {author} {\bibfnamefont {N.}~\bibnamefont {Okada}},\
  }\href {\doibase 10.1103/PhysRevD.88.113001} {\bibfield  {journal} {\bibinfo
  {journal} {Phys. Rev.}\ }\textbf {\bibinfo {volume} {D88}},\ \bibinfo {pages}
  {113001} (\bibinfo {year} {2013})},\ \Eprint {http://arxiv.org/abs/1207.3734}
  {arXiv:1207.3734 [hep-ph]} \BibitemShut {NoStop}%
\bibitem [{\citenamefont {Dev}\ \emph {et~al.}(2014)\citenamefont {Dev},
  \citenamefont {Pilaftsis},\ and\ \citenamefont {Yang}}]{Dev:2013wba}%
  \BibitemOpen
  \bibfield  {author} {\bibinfo {author} {\bibfnamefont {P.~S.~B.}\
  \bibnamefont {Dev}}, \bibinfo {author} {\bibfnamefont {A.}~\bibnamefont
  {Pilaftsis}}, \ and\ \bibinfo {author} {\bibfnamefont {U.-k.}\ \bibnamefont
  {Yang}},\ }\href {\doibase 10.1103/PhysRevLett.112.081801} {\bibfield
  {journal} {\bibinfo  {journal} {Phys. Rev. Lett.}\ }\textbf {\bibinfo
  {volume} {112}},\ \bibinfo {pages} {081801} (\bibinfo {year} {2014})},\
  \Eprint {http://arxiv.org/abs/1308.2209} {arXiv:1308.2209 [hep-ph]}
  \BibitemShut {NoStop}%
\bibitem [{\citenamefont {Helo}\ \emph {et~al.}(2014)\citenamefont {Helo},
  \citenamefont {Hirsch},\ and\ \citenamefont {Kovalenko}}]{Helo:2013esa}%
  \BibitemOpen
  \bibfield  {author} {\bibinfo {author} {\bibfnamefont {J.~C.}\ \bibnamefont
  {Helo}}, \bibinfo {author} {\bibfnamefont {M.}~\bibnamefont {Hirsch}}, \ and\
  \bibinfo {author} {\bibfnamefont {S.}~\bibnamefont {Kovalenko}},\ }\href
  {\doibase 10.1103/PhysRevD.89.073005} {\bibfield  {journal} {\bibinfo
  {journal} {Phys. Rev.}\ }\textbf {\bibinfo {volume} {D89}},\ \bibinfo {pages}
  {073005} (\bibinfo {year} {2014})},\ \Eprint {http://arxiv.org/abs/1312.2900}
  {arXiv:1312.2900 [hep-ph]} \BibitemShut {NoStop}%
\bibitem [{\citenamefont {Alva}\ \emph {et~al.}(2015)\citenamefont {Alva},
  \citenamefont {Han},\ and\ \citenamefont {Ruiz}}]{Alva:2014gxa}%
  \BibitemOpen
  \bibfield  {author} {\bibinfo {author} {\bibfnamefont {D.}~\bibnamefont
  {Alva}}, \bibinfo {author} {\bibfnamefont {T.}~\bibnamefont {Han}}, \ and\
  \bibinfo {author} {\bibfnamefont {R.}~\bibnamefont {Ruiz}},\ }\href {\doibase
  10.1007/JHEP02(2015)072} {\bibfield  {journal} {\bibinfo  {journal} {JHEP}\
  }\textbf {\bibinfo {volume} {02}},\ \bibinfo {pages} {072} (\bibinfo {year}
  {2015})},\ \Eprint {http://arxiv.org/abs/1411.7305} {arXiv:1411.7305
  [hep-ph]} \BibitemShut {NoStop}%
\bibitem [{\citenamefont {Antusch}\ and\ \citenamefont
  {Fischer}(2015)}]{Antusch:2015mia}%
  \BibitemOpen
  \bibfield  {author} {\bibinfo {author} {\bibfnamefont {S.}~\bibnamefont
  {Antusch}}\ and\ \bibinfo {author} {\bibfnamefont {O.}~\bibnamefont
  {Fischer}},\ }\href {\doibase 10.1007/JHEP05(2015)053} {\bibfield  {journal}
  {\bibinfo  {journal} {JHEP}\ }\textbf {\bibinfo {volume} {05}},\ \bibinfo
  {pages} {053} (\bibinfo {year} {2015})},\ \Eprint
  {http://arxiv.org/abs/1502.05915} {arXiv:1502.05915 [hep-ph]} \BibitemShut
  {NoStop}%
\bibitem [{\citenamefont {Banerjee}\ \emph
  {et~al.}(2015{\natexlab{b}})\citenamefont {Banerjee}, \citenamefont {Dev},
  \citenamefont {Ibarra}, \citenamefont {Mandal},\ and\ \citenamefont
  {Mitra}}]{Banerjee:2015gca}%
  \BibitemOpen
  \bibfield  {author} {\bibinfo {author} {\bibfnamefont {S.}~\bibnamefont
  {Banerjee}}, \bibinfo {author} {\bibfnamefont {P.~S.~B.}\ \bibnamefont
  {Dev}}, \bibinfo {author} {\bibfnamefont {A.}~\bibnamefont {Ibarra}},
  \bibinfo {author} {\bibfnamefont {T.}~\bibnamefont {Mandal}}, \ and\ \bibinfo
  {author} {\bibfnamefont {M.}~\bibnamefont {Mitra}},\ }\href {\doibase
  10.1103/PhysRevD.92.075002} {\bibfield  {journal} {\bibinfo  {journal} {Phys.
  Rev.}\ }\textbf {\bibinfo {volume} {D92}},\ \bibinfo {pages} {075002}
  (\bibinfo {year} {2015}{\natexlab{b}})},\ \Eprint
  {http://arxiv.org/abs/1503.05491} {arXiv:1503.05491 [hep-ph]} \BibitemShut
  {NoStop}%
\bibitem [{\citenamefont {Arganda}\ \emph {et~al.}(2016)\citenamefont
  {Arganda}, \citenamefont {Herrero}, \citenamefont {Marcano},\ and\
  \citenamefont {Weiland}}]{Arganda:2015ija}%
  \BibitemOpen
  \bibfield  {author} {\bibinfo {author} {\bibfnamefont {E.}~\bibnamefont
  {Arganda}}, \bibinfo {author} {\bibfnamefont {M.~J.}\ \bibnamefont
  {Herrero}}, \bibinfo {author} {\bibfnamefont {X.}~\bibnamefont {Marcano}}, \
  and\ \bibinfo {author} {\bibfnamefont {C.}~\bibnamefont {Weiland}},\ }\href
  {\doibase 10.1016/j.physletb.2015.11.013} {\bibfield  {journal} {\bibinfo
  {journal} {Phys. Lett.}\ }\textbf {\bibinfo {volume} {B752}},\ \bibinfo
  {pages} {46} (\bibinfo {year} {2016})},\ \Eprint
  {http://arxiv.org/abs/1508.05074} {arXiv:1508.05074 [hep-ph]} \BibitemShut
  {NoStop}%
\bibitem [{\citenamefont {Das}\ \emph {et~al.}(2016{\natexlab{b}})\citenamefont
  {Das}, \citenamefont {Konar},\ and\ \citenamefont {Majhi}}]{Das:2016hof}%
  \BibitemOpen
  \bibfield  {author} {\bibinfo {author} {\bibfnamefont {A.}~\bibnamefont
  {Das}}, \bibinfo {author} {\bibfnamefont {P.}~\bibnamefont {Konar}}, \ and\
  \bibinfo {author} {\bibfnamefont {S.}~\bibnamefont {Majhi}},\ }\href@noop {}
  {\  (\bibinfo {year} {2016}{\natexlab{b}})},\ \Eprint
  {http://arxiv.org/abs/1604.00608} {arXiv:1604.00608 [hep-ph]} \BibitemShut
  {NoStop}%
\bibitem [{\citenamefont {Aad}\ \emph {et~al.}(2015)\citenamefont {Aad} \emph
  {et~al.}}]{Aad:2015xaa}%
  \BibitemOpen
  \bibfield  {author} {\bibinfo {author} {\bibfnamefont {G.}~\bibnamefont
  {Aad}} \emph {et~al.} (\bibinfo {collaboration} {ATLAS}),\ }\href {\doibase
  10.1007/JHEP07(2015)162} {\bibfield  {journal} {\bibinfo  {journal} {JHEP}\
  }\textbf {\bibinfo {volume} {07}},\ \bibinfo {pages} {162} (\bibinfo {year}
  {2015})},\ \Eprint {http://arxiv.org/abs/1506.06020} {arXiv:1506.06020
  [hep-ex]} \BibitemShut {NoStop}%
\bibitem [{\citenamefont {Bambhaniya}\ \emph {et~al.}(2014)\citenamefont
  {Bambhaniya}, \citenamefont {Chakrabortty}, \citenamefont {Gluza},
  \citenamefont {Kordiaczy\'nska},\ and\ \citenamefont
  {Szafron}}]{Bambhaniya:2013wza}%
  \BibitemOpen
  \bibfield  {author} {\bibinfo {author} {\bibfnamefont {G.}~\bibnamefont
  {Bambhaniya}}, \bibinfo {author} {\bibfnamefont {J.}~\bibnamefont
  {Chakrabortty}}, \bibinfo {author} {\bibfnamefont {J.}~\bibnamefont {Gluza}},
  \bibinfo {author} {\bibfnamefont {M.}~\bibnamefont {Kordiaczy\'nska}}, \ and\
  \bibinfo {author} {\bibfnamefont {R.}~\bibnamefont {Szafron}},\ }\href
  {\doibase 10.1007/JHEP05(2014)033} {\bibfield  {journal} {\bibinfo  {journal}
  {JHEP}\ }\textbf {\bibinfo {volume} {1405}},\ \bibinfo {pages} {033}
  (\bibinfo {year} {2014})},\ \Eprint {http://arxiv.org/abs/1311.4144}
  {arXiv:1311.4144 [hep-ph]} \BibitemShut {NoStop}%
\bibitem [{\citenamefont {Belanger}\ \emph {et~al.}(1996)\citenamefont
  {Belanger}, \citenamefont {Boudjema}, \citenamefont {London},\ and\
  \citenamefont {Nadeau}}]{Belanger:1995nh}%
  \BibitemOpen
  \bibfield  {author} {\bibinfo {author} {\bibfnamefont {G.}~\bibnamefont
  {Belanger}}, \bibinfo {author} {\bibfnamefont {F.}~\bibnamefont {Boudjema}},
  \bibinfo {author} {\bibfnamefont {D.}~\bibnamefont {London}}, \ and\ \bibinfo
  {author} {\bibfnamefont {H.}~\bibnamefont {Nadeau}},\ }\href {\doibase
  10.1103/PhysRevD.53.6292} {\bibfield  {journal} {\bibinfo  {journal} {Phys.
  Rev.}\ }\textbf {\bibinfo {volume} {D53}},\ \bibinfo {pages} {6292} (\bibinfo
  {year} {1996})},\ \Eprint {http://arxiv.org/abs/hep-ph/9508317}
  {arXiv:hep-ph/9508317 [hep-ph]} \BibitemShut {NoStop}%
\bibitem [{\citenamefont {Gluza}\ and\ \citenamefont
  {Zra{\l}ek}(1995{\natexlab{a}})}]{Gluza:1995ky}%
  \BibitemOpen
  \bibfield  {author} {\bibinfo {author} {\bibfnamefont {J.}~\bibnamefont
  {Gluza}}\ and\ \bibinfo {author} {\bibfnamefont {M.}~\bibnamefont
  {Zra{\l}ek}},\ }\href {\doibase 10.1103/PhysRevD.52.6238} {\bibfield
  {journal} {\bibinfo  {journal} {Phys.Rev.}\ }\textbf {\bibinfo {volume}
  {D52}},\ \bibinfo {pages} {6238} (\bibinfo {year} {1995}{\natexlab{a}})},\
  \Eprint {http://arxiv.org/abs/hep-ph/9502284} {arXiv:hep-ph/9502284 [hep-ph]}
  \BibitemShut {NoStop}%
\bibitem [{\citenamefont {Gluza}\ and\ \citenamefont
  {Zra{\l}ek}(1995{\natexlab{b}})}]{Gluza:1995ix}%
  \BibitemOpen
  \bibfield  {author} {\bibinfo {author} {\bibfnamefont {J.}~\bibnamefont
  {Gluza}}\ and\ \bibinfo {author} {\bibfnamefont {M.}~\bibnamefont
  {Zra{\l}ek}},\ }\href {\doibase 10.1016/0370-2693(95)01158-M} {\bibfield
  {journal} {\bibinfo  {journal} {Phys.Lett.}\ }\textbf {\bibinfo {volume}
  {B362}},\ \bibinfo {pages} {148} (\bibinfo {year} {1995}{\natexlab{b}})},\
  \Eprint {http://arxiv.org/abs/hep-ph/9507269} {arXiv:hep-ph/9507269 [hep-ph]}
  \BibitemShut {NoStop}%
\bibitem [{\citenamefont {Asaka}\ and\ \citenamefont
  {Tsuyuki}(2015)}]{Asaka:2015oia}%
  \BibitemOpen
  \bibfield  {author} {\bibinfo {author} {\bibfnamefont {T.}~\bibnamefont
  {Asaka}}\ and\ \bibinfo {author} {\bibfnamefont {T.}~\bibnamefont
  {Tsuyuki}},\ }\href {\doibase 10.1103/PhysRevD.92.094012} {\bibfield
  {journal} {\bibinfo  {journal} {Phys. Rev.}\ }\textbf {\bibinfo {volume}
  {D92}},\ \bibinfo {pages} {094012} (\bibinfo {year} {2015})},\ \Eprint
  {http://arxiv.org/abs/1508.04937} {arXiv:1508.04937 [hep-ph]} \BibitemShut
  {NoStop}%
\bibitem [{\citenamefont {Lee}\ and\ \citenamefont
  {Shrock}(1977)}]{Lee:1977tib}%
  \BibitemOpen
  \bibfield  {author} {\bibinfo {author} {\bibfnamefont {B.~W.}\ \bibnamefont
  {Lee}}\ and\ \bibinfo {author} {\bibfnamefont {R.~E.}\ \bibnamefont
  {Shrock}},\ }\href {\doibase 10.1103/PhysRevD.16.1444} {\bibfield  {journal}
  {\bibinfo  {journal} {Phys. Rev.}\ }\textbf {\bibinfo {volume} {D16}},\
  \bibinfo {pages} {1444} (\bibinfo {year} {1977})}\BibitemShut {NoStop}%
\bibitem [{\citenamefont {Olive}\ \emph {et~al.}(2014)\citenamefont {Olive}
  \emph {et~al.}}]{Agashe:2014kda}%
  \BibitemOpen
  \bibfield  {author} {\bibinfo {author} {\bibfnamefont {K.~A.}\ \bibnamefont
  {Olive}} \emph {et~al.} (\bibinfo {collaboration} {Particle Data Group}),\
  }\href {\doibase 10.1088/1674-1137/38/9/090001} {\bibfield  {journal}
  {\bibinfo  {journal} {Chin. Phys.}\ }\textbf {\bibinfo {volume} {C38}},\
  \bibinfo {pages} {090001} (\bibinfo {year} {2014})}\BibitemShut {NoStop}%
\bibitem [{\citenamefont {Baldini}\ \emph {et~al.}(2013)\citenamefont {Baldini}
  \emph {et~al.}}]{Baldini:2013ke}%
  \BibitemOpen
  \bibfield  {author} {\bibinfo {author} {\bibfnamefont {A.~M.}\ \bibnamefont
  {Baldini}} \emph {et~al.},\ }\href@noop {} {\  (\bibinfo {year} {2013})},\
  \Eprint {http://arxiv.org/abs/1301.7225} {arXiv:1301.7225 [physics.ins-det]}
  \BibitemShut {NoStop}%
\bibitem [{\citenamefont {Aushev}\ \emph {et~al.}(2010)\citenamefont {Aushev}
  \emph {et~al.}}]{Aushev:2010bq}%
  \BibitemOpen
  \bibfield  {author} {\bibinfo {author} {\bibfnamefont {T.}~\bibnamefont
  {Aushev}} \emph {et~al.},\ }\href@noop {} {\  (\bibinfo {year} {2010})},\
  \Eprint {http://arxiv.org/abs/1002.5012} {arXiv:1002.5012 [hep-ex]}
  \BibitemShut {NoStop}%
\bibitem [{\citenamefont {Bartoszek}\ \emph {et~al.}(2014)\citenamefont
  {Bartoszek} \emph {et~al.}}]{Bartoszek:2014mya}%
  \BibitemOpen
  \bibfield  {author} {\bibinfo {author} {\bibfnamefont {L.}~\bibnamefont
  {Bartoszek}} \emph {et~al.} (\bibinfo {collaboration} {Mu2e}),\ }\href@noop
  {} {\  (\bibinfo {year} {2014})},\ \Eprint {http://arxiv.org/abs/1501.05241}
  {arXiv:1501.05241 [physics.ins-det]} \BibitemShut {NoStop}%
\bibitem [{\citenamefont {Bu}\ \emph {et~al.}(2008)\citenamefont {Bu},
  \citenamefont {Liao},\ and\ \citenamefont {Liu}}]{Bu:2008fx}%
  \BibitemOpen
  \bibfield  {author} {\bibinfo {author} {\bibfnamefont {J.-P.}\ \bibnamefont
  {Bu}}, \bibinfo {author} {\bibfnamefont {Y.}~\bibnamefont {Liao}}, \ and\
  \bibinfo {author} {\bibfnamefont {J.-Y.}\ \bibnamefont {Liu}},\ }\href
  {\doibase 10.1016/j.physletb.2008.05.059} {\bibfield  {journal} {\bibinfo
  {journal} {Phys. Lett.}\ }\textbf {\bibinfo {volume} {B665}},\ \bibinfo
  {pages} {39} (\bibinfo {year} {2008})},\ \Eprint
  {http://arxiv.org/abs/0802.3241} {arXiv:0802.3241 [hep-ph]} \BibitemShut
  {NoStop}%
\bibitem [{\citenamefont {Langacker}\ and\ \citenamefont
  {London}(1988)}]{Langacker:1988up}%
  \BibitemOpen
  \bibfield  {author} {\bibinfo {author} {\bibfnamefont {P.}~\bibnamefont
  {Langacker}}\ and\ \bibinfo {author} {\bibfnamefont {D.}~\bibnamefont
  {London}},\ }\href {\doibase 10.1103/PhysRevD.38.907} {\bibfield  {journal}
  {\bibinfo  {journal} {Phys. Rev.}\ }\textbf {\bibinfo {volume} {D38}},\
  \bibinfo {pages} {907} (\bibinfo {year} {1988})}\BibitemShut {NoStop}%
\bibitem [{\citenamefont {del Aguila}\ \emph {et~al.}(1997)\citenamefont {del
  Aguila}, \citenamefont {Aguilar-Saavedra},\ and\ \citenamefont
  {Zra{\l}ek}}]{delAguila:1996ex}%
  \BibitemOpen
  \bibfield  {author} {\bibinfo {author} {\bibfnamefont {F.}~\bibnamefont {del
  Aguila}}, \bibinfo {author} {\bibfnamefont {J.~A.}\ \bibnamefont
  {Aguilar-Saavedra}}, \ and\ \bibinfo {author} {\bibfnamefont
  {M.}~\bibnamefont {Zra{\l}ek}},\ }\href {\doibase
  10.1016/S0010-4655(96)00159-2} {\bibfield  {journal} {\bibinfo  {journal}
  {Comput. Phys. Commun.}\ }\textbf {\bibinfo {volume} {100}},\ \bibinfo
  {pages} {231} (\bibinfo {year} {1997})},\ \Eprint
  {http://arxiv.org/abs/hep-ph/9607311} {arXiv:hep-ph/9607311 [hep-ph]}
  \BibitemShut {NoStop}%
\bibitem [{\citenamefont {Wolfenstein}(1981)}]{Wolfenstein:1981rk}%
  \BibitemOpen
  \bibfield  {author} {\bibinfo {author} {\bibfnamefont {L.}~\bibnamefont
  {Wolfenstein}},\ }\href {\doibase 10.1016/0370-2693(81)91151-5} {\bibfield
  {journal} {\bibinfo  {journal} {Phys. Lett.}\ }\textbf {\bibinfo {volume}
  {B107}},\ \bibinfo {pages} {77} (\bibinfo {year} {1981})}\BibitemShut
  {NoStop}%
\bibitem [{\citenamefont {Wilczek}(2009)}]{wilczek}%
  \BibitemOpen
  \bibfield  {author} {\bibinfo {author} {\bibfnamefont {F.}~\bibnamefont
  {Wilczek}},\ }\href@noop {} {\bibfield  {journal} {\bibinfo  {journal}
  {Nature Physics}\ }\textbf {\bibinfo {volume} {5}},\ \bibinfo {pages} {614}
  (\bibinfo {year} {2009})}\BibitemShut {NoStop}%
\bibitem [{\citenamefont {Elliott}\ and\ \citenamefont
  {Franz}(2015)}]{Elliott:2014iha}%
  \BibitemOpen
  \bibfield  {author} {\bibinfo {author} {\bibfnamefont {S.~R.}\ \bibnamefont
  {Elliott}}\ and\ \bibinfo {author} {\bibfnamefont {M.}~\bibnamefont
  {Franz}},\ }\href {\doibase 10.1103/RevModPhys.87.137} {\bibfield  {journal}
  {\bibinfo  {journal} {Rev. Mod. Phys.}\ }\textbf {\bibinfo {volume} {87}},\
  \bibinfo {pages} {137} (\bibinfo {year} {2015})},\ \Eprint
  {http://arxiv.org/abs/1403.4976} {arXiv:1403.4976 [cond-mat.supr-con]}
  \BibitemShut {NoStop}%
\bibitem [{\citenamefont {{Kitaev}}(2001)}]{2001PhyU...44..131K}%
  \BibitemOpen
  \bibfield  {author} {\bibinfo {author} {\bibfnamefont {A.~Y.}\ \bibnamefont
  {{Kitaev}}},\ }\href {\doibase 10.1070/1063-7869/44/10S/S29} {\bibfield
  {journal} {\bibinfo  {journal} {Physics Uspekhi}\ }\textbf {\bibinfo {volume}
  {44}},\ \bibinfo {pages} {131} (\bibinfo {year} {2001})},\ \Eprint
  {http://arxiv.org/abs/cond-mat/0010440} {cond-mat/0010440} \BibitemShut
  {NoStop}%
\bibitem [{\citenamefont {Bertone}\ \emph {et~al.}(2005)\citenamefont
  {Bertone}, \citenamefont {Hooper},\ and\ \citenamefont
  {Silk}}]{Bertone:2004pz}%
  \BibitemOpen
  \bibfield  {author} {\bibinfo {author} {\bibfnamefont {G.}~\bibnamefont
  {Bertone}}, \bibinfo {author} {\bibfnamefont {D.}~\bibnamefont {Hooper}}, \
  and\ \bibinfo {author} {\bibfnamefont {J.}~\bibnamefont {Silk}},\ }\href
  {\doibase 10.1016/j.physrep.2004.08.031} {\bibfield  {journal} {\bibinfo
  {journal} {Phys. Rept.}\ }\textbf {\bibinfo {volume} {405}},\ \bibinfo
  {pages} {279} (\bibinfo {year} {2005})},\ \Eprint
  {http://arxiv.org/abs/hep-ph/0404175} {arXiv:hep-ph/0404175 [hep-ph]}
  \BibitemShut {NoStop}%
\bibitem [{\citenamefont {Van~der Waerden}( 109)}]{waerden1929}%
  \BibitemOpen
  \bibfield  {author} {\bibinfo {author} {\bibfnamefont {B.}~\bibnamefont
  {Van~der Waerden}},\ }\href@noop {} {\emph {\bibinfo {title}
  {{Spinoranalyse}}}}\ (\bibinfo {year} {1929: 100-109})\BibitemShut {NoStop}%
\bibitem [{\citenamefont {Weyl}(1929)}]{Weyl:1929fm}%
  \BibitemOpen
  \bibfield  {author} {\bibinfo {author} {\bibfnamefont {H.}~\bibnamefont
  {Weyl}},\ }\href {\doibase 10.1007/BF01339504} {\bibfield  {journal}
  {\bibinfo  {journal} {Z. Phys.}\ }\textbf {\bibinfo {volume} {56}},\ \bibinfo
  {pages} {330} (\bibinfo {year} {1929})},\ \bibinfo {note} {[Surveys High
  Energ. Phys.5,261(1986)]}\BibitemShut {NoStop}%
\bibitem [{\citenamefont {Dirac}(1928{\natexlab{a}})}]{Dirac:1928hu}%
  \BibitemOpen
  \bibfield  {author} {\bibinfo {author} {\bibfnamefont {P.~A.~M.}\
  \bibnamefont {Dirac}},\ }\href {\doibase 10.1098/rspa.1928.0023} {\bibfield
  {journal} {\bibinfo  {journal} {Proc. Roy. Soc. Lond.}\ }\textbf {\bibinfo
  {volume} {A117}},\ \bibinfo {pages} {610} (\bibinfo {year}
  {1928}{\natexlab{a}})}\BibitemShut {NoStop}%
\bibitem [{\citenamefont {Dirac}(1928{\natexlab{b}})}]{Dirac:1928ej}%
  \BibitemOpen
  \bibfield  {author} {\bibinfo {author} {\bibfnamefont {P.~A.~M.}\
  \bibnamefont {Dirac}},\ }\href {\doibase 10.1098/rspa.1928.0056} {\bibfield
  {journal} {\bibinfo  {journal} {Proc. Roy. Soc. Lond.}\ }\textbf {\bibinfo
  {volume} {A118}},\ \bibinfo {pages} {351} (\bibinfo {year}
  {1928}{\natexlab{b}})}\BibitemShut {NoStop}%
\bibitem [{\citenamefont {Majorana}(1937)}]{Majorana:1937vz}%
  \BibitemOpen
  \bibfield  {author} {\bibinfo {author} {\bibfnamefont {E.}~\bibnamefont
  {Majorana}},\ }\href {\doibase 10.1007/BF02961314} {\bibfield  {journal}
  {\bibinfo  {journal} {Nuovo Cim.}\ }\textbf {\bibinfo {volume} {14}},\
  \bibinfo {pages} {171} (\bibinfo {year} {1937})}\BibitemShut {NoStop}%
\bibitem [{\citenamefont {Maggiore}(2005)}]{Maggiore:2005qv}%
  \BibitemOpen
  \bibfield  {author} {\bibinfo {author} {\bibfnamefont {M.}~\bibnamefont
  {Maggiore}},\ }\href@noop {} {\emph {\bibinfo {title} {{A Modern introduction
  to quantum field theory}}}}\ (\bibinfo {year} {Oxford University Press,
  2005})\BibitemShut {NoStop}%
\bibitem [{\citenamefont {Dib}\ and\ \citenamefont {Kim}(2015)}]{Dib:2015oka}%
  \BibitemOpen
  \bibfield  {author} {\bibinfo {author} {\bibfnamefont {C.~O.}\ \bibnamefont
  {Dib}}\ and\ \bibinfo {author} {\bibfnamefont {C.~S.}\ \bibnamefont {Kim}},\
  }\href {\doibase 10.1103/PhysRevD.92.093009} {\bibfield  {journal} {\bibinfo
  {journal} {Phys. Rev.}\ }\textbf {\bibinfo {volume} {D92}},\ \bibinfo {pages}
  {093009} (\bibinfo {year} {2015})},\ \Eprint
  {http://arxiv.org/abs/1509.05981} {arXiv:1509.05981 [hep-ph]} \BibitemShut
  {NoStop}%
\bibitem [{\citenamefont {Racah}(1937)}]{Racah:1937qq}%
  \BibitemOpen
  \bibfield  {author} {\bibinfo {author} {\bibfnamefont {G.}~\bibnamefont
  {Racah}},\ }\href {\doibase 10.1007/BF02961321} {\bibfield  {journal}
  {\bibinfo  {journal} {Nuovo Cim.}\ }\textbf {\bibinfo {volume} {14}},\
  \bibinfo {pages} {322} (\bibinfo {year} {1937})}\BibitemShut {NoStop}%
\bibitem [{\citenamefont {Furry}(1939)}]{Furry:1939qr}%
  \BibitemOpen
  \bibfield  {author} {\bibinfo {author} {\bibfnamefont {W.~H.}\ \bibnamefont
  {Furry}},\ }\href {\doibase 10.1103/PhysRev.56.1184} {\bibfield  {journal}
  {\bibinfo  {journal} {Phys. Rev.}\ }\textbf {\bibinfo {volume} {56}},\
  \bibinfo {pages} {1184} (\bibinfo {year} {1939})}\BibitemShut {NoStop}%
\bibitem [{\citenamefont {Zra{\l}ek}(1997)}]{Zralek1997sa}%
  \BibitemOpen
  \bibfield  {author} {\bibinfo {author} {\bibfnamefont {M.}~\bibnamefont
  {Zra{\l}ek}},\ }\href@noop {} {\bibfield  {journal} {\bibinfo  {journal}
  {Acta Phys. Polon.}\ }\textbf {\bibinfo {volume} {B28}},\ \bibinfo {pages}
  {2225} (\bibinfo {year} {1997})},\ \Eprint
  {http://arxiv.org/abs/hep-ph/9711506} {arXiv:hep-ph/9711506 [hep-ph]}
  \BibitemShut {NoStop}%
\bibitem [{\citenamefont {Agostini}\ \emph {et~al.}(2013)\citenamefont
  {Agostini} \emph {et~al.}}]{Agostini:2013mzu}%
  \BibitemOpen
  \bibfield  {author} {\bibinfo {author} {\bibfnamefont {M.}~\bibnamefont
  {Agostini}} \emph {et~al.} (\bibinfo {collaboration} {GERDA}),\ }\href
  {\doibase 10.1103/PhysRevLett.111.122503} {\bibfield  {journal} {\bibinfo
  {journal} {Phys. Rev. Lett.}\ }\textbf {\bibinfo {volume} {111}},\ \bibinfo
  {pages} {122503} (\bibinfo {year} {2013})},\ \Eprint
  {http://arxiv.org/abs/1307.4720} {arXiv:1307.4720 [nucl-ex]} \BibitemShut
  {NoStop}%
\bibitem [{\citenamefont {Gando}\ \emph {et~al.}(2013)\citenamefont {Gando}
  \emph {et~al.}}]{Gando:2012zm}%
  \BibitemOpen
  \bibfield  {author} {\bibinfo {author} {\bibfnamefont {A.}~\bibnamefont
  {Gando}} \emph {et~al.} (\bibinfo {collaboration} {KamLAND-Zen}),\ }\href
  {\doibase 10.1103/PhysRevLett.110.062502} {\bibfield  {journal} {\bibinfo
  {journal} {Phys. Rev. Lett.}\ }\textbf {\bibinfo {volume} {110}},\ \bibinfo
  {pages} {062502} (\bibinfo {year} {2013})},\ \Eprint
  {http://arxiv.org/abs/1211.3863} {arXiv:1211.3863 [hep-ex]} \BibitemShut
  {NoStop}%
\bibitem [{\citenamefont {Albert}\ \emph {et~al.}(2014)\citenamefont {Albert}
  \emph {et~al.}}]{Albert:2014awa}%
  \BibitemOpen
  \bibfield  {author} {\bibinfo {author} {\bibfnamefont {J.~B.}\ \bibnamefont
  {Albert}} \emph {et~al.} (\bibinfo {collaboration} {EXO-200}),\ }\href
  {\doibase 10.1038/nature13432} {\bibfield  {journal} {\bibinfo  {journal}
  {Nature}\ }\textbf {\bibinfo {volume} {510}},\ \bibinfo {pages} {229}
  (\bibinfo {year} {2014})},\ \Eprint {http://arxiv.org/abs/1402.6956}
  {arXiv:1402.6956 [nucl-ex]} \BibitemShut {NoStop}%
\bibitem [{\citenamefont {Alfonso}\ \emph {et~al.}(2015)\citenamefont {Alfonso}
  \emph {et~al.}}]{Alfonso:2015wka}%
  \BibitemOpen
  \bibfield  {author} {\bibinfo {author} {\bibfnamefont {K.}~\bibnamefont
  {Alfonso}} \emph {et~al.} (\bibinfo {collaboration} {CUORE}),\ }\href
  {\doibase 10.1103/PhysRevLett.115.102502} {\bibfield  {journal} {\bibinfo
  {journal} {Phys. Rev. Lett.}\ }\textbf {\bibinfo {volume} {115}},\ \bibinfo
  {pages} {102502} (\bibinfo {year} {2015})},\ \Eprint
  {http://arxiv.org/abs/1504.02454} {arXiv:1504.02454 [nucl-ex]} \BibitemShut
  {NoStop}%
\bibitem [{\citenamefont {Majorovits}(2015)}]{Majorovits:2015vka}%
  \BibitemOpen
  \bibfield  {author} {\bibinfo {author} {\bibfnamefont {B.}~\bibnamefont
  {Majorovits}} (\bibinfo {collaboration} {GERDA}),\ }\href {\doibase
  10.1063/1.4928005} {\bibfield  {journal} {\bibinfo  {journal} {AIP Conf.
  Proc.}\ }\textbf {\bibinfo {volume} {1672}},\ \bibinfo {pages} {110003}
  (\bibinfo {year} {2015})},\ \Eprint {http://arxiv.org/abs/1506.00415}
  {arXiv:1506.00415 [hep-ex]} \BibitemShut {NoStop}%
\bibitem [{\citenamefont {Albert}(2014)}]{Albert:2014afa}%
  \BibitemOpen
  \bibfield  {author} {\bibinfo {author} {\bibfnamefont {J.}~\bibnamefont
  {Albert}},\ }\href {\doibase 10.1051/epjconf/20146608001} {\bibfield
  {journal} {\bibinfo  {journal} {EPJ Web Conf.}\ }\textbf {\bibinfo {volume}
  {66}},\ \bibinfo {pages} {08001} (\bibinfo {year} {2014})}\BibitemShut
  {NoStop}%
\bibitem [{\citenamefont {Wang}\ \emph {et~al.}(2015)\citenamefont {Wang} \emph
  {et~al.}}]{Wang:2015raa}%
  \BibitemOpen
  \bibfield  {author} {\bibinfo {author} {\bibfnamefont {G.}~\bibnamefont
  {Wang}} \emph {et~al.} (\bibinfo {collaboration} {CUPID}),\ }\href@noop {} {\
   (\bibinfo {year} {2015})},\ \Eprint {http://arxiv.org/abs/1504.03599}
  {arXiv:1504.03599 [physics.ins-det]} \BibitemShut {NoStop}%
\bibitem [{\citenamefont {Czakon}\ \emph {et~al.}(2002)\citenamefont {Czakon},
  \citenamefont {Gluza}, \citenamefont {Studnik},\ and\ \citenamefont
  {Zra{\l}ek}}]{Czakon:2001uh}%
  \BibitemOpen
  \bibfield  {author} {\bibinfo {author} {\bibfnamefont {M.}~\bibnamefont
  {Czakon}}, \bibinfo {author} {\bibfnamefont {J.}~\bibnamefont {Gluza}},
  \bibinfo {author} {\bibfnamefont {J.}~\bibnamefont {Studnik}}, \ and\
  \bibinfo {author} {\bibfnamefont {M.}~\bibnamefont {Zra{\l}ek}},\ }\href
  {\doibase 10.1103/PhysRevD.65.053008} {\bibfield  {journal} {\bibinfo
  {journal} {Phys. Rev.}\ }\textbf {\bibinfo {volume} {D65}},\ \bibinfo {pages}
  {053008} (\bibinfo {year} {2002})},\ \Eprint
  {http://arxiv.org/abs/hep-ph/0110166} {arXiv:hep-ph/0110166 [hep-ph]}
  \BibitemShut {NoStop}%
\bibitem [{\citenamefont {Prezeau}\ \emph {et~al.}(2003)\citenamefont
  {Prezeau}, \citenamefont {Ramsey-Musolf},\ and\ \citenamefont
  {Vogel}}]{Prezeau:2003xn}%
  \BibitemOpen
  \bibfield  {author} {\bibinfo {author} {\bibfnamefont {G.}~\bibnamefont
  {Prezeau}}, \bibinfo {author} {\bibfnamefont {M.}~\bibnamefont
  {Ramsey-Musolf}}, \ and\ \bibinfo {author} {\bibfnamefont {P.}~\bibnamefont
  {Vogel}},\ }\href {\doibase 10.1103/PhysRevD.68.034016} {\bibfield  {journal}
  {\bibinfo  {journal} {Phys. Rev.}\ }\textbf {\bibinfo {volume} {D68}},\
  \bibinfo {pages} {034016} (\bibinfo {year} {2003})},\ \Eprint
  {http://arxiv.org/abs/hep-ph/0303205} {arXiv:hep-ph/0303205 [hep-ph]}
  \BibitemShut {NoStop}%
\bibitem [{\citenamefont {Minakata}\ \emph {et~al.}(2015)\citenamefont
  {Minakata}, \citenamefont {Nunokawa},\ and\ \citenamefont
  {Quiroga}}]{Minakata:2014jba}%
  \BibitemOpen
  \bibfield  {author} {\bibinfo {author} {\bibfnamefont {H.}~\bibnamefont
  {Minakata}}, \bibinfo {author} {\bibfnamefont {H.}~\bibnamefont {Nunokawa}},
  \ and\ \bibinfo {author} {\bibfnamefont {A.~A.}\ \bibnamefont {Quiroga}},\
  }\href {\doibase 10.1093/ptep/ptv010} {\bibfield  {journal} {\bibinfo
  {journal} {PTEP}\ }\textbf {\bibinfo {volume} {2015}},\ \bibinfo {pages}
  {033B03} (\bibinfo {year} {2015})},\ \Eprint {http://arxiv.org/abs/1402.6014}
  {arXiv:1402.6014 [hep-ph]} \BibitemShut {NoStop}%
\bibitem [{\citenamefont {Rodejohann}(2012)}]{Rodejohann:2012xd}%
  \BibitemOpen
  \bibfield  {author} {\bibinfo {author} {\bibfnamefont {W.}~\bibnamefont
  {Rodejohann}},\ }\href {\doibase 10.1088/0954-3899/39/12/124008} {\bibfield
  {journal} {\bibinfo  {journal} {J. Phys.}\ }\textbf {\bibinfo {volume}
  {G39}},\ \bibinfo {pages} {124008} (\bibinfo {year} {2012})},\ \Eprint
  {http://arxiv.org/abs/1206.2560} {arXiv:1206.2560 [hep-ph]} \BibitemShut
  {NoStop}%
\bibitem [{\citenamefont {Mohapatra}\ \emph {et~al.}(2007)\citenamefont
  {Mohapatra} \emph {et~al.}}]{Mohapatra:2005wg}%
  \BibitemOpen
  \bibfield  {author} {\bibinfo {author} {\bibfnamefont {R.~N.}\ \bibnamefont
  {Mohapatra}} \emph {et~al.},\ }\href {\doibase 10.1088/0034-4885/70/11/R02}
  {\bibfield  {journal} {\bibinfo  {journal} {Rept. Prog. Phys.}\ }\textbf
  {\bibinfo {volume} {70}},\ \bibinfo {pages} {1757} (\bibinfo {year}
  {2007})},\ \Eprint {http://arxiv.org/abs/hep-ph/0510213}
  {arXiv:hep-ph/0510213 [hep-ph]} \BibitemShut {NoStop}%
\bibitem [{\citenamefont {Chakrabortty}\ \emph
  {et~al.}(2012{\natexlab{b}})\citenamefont {Chakrabortty}, \citenamefont
  {Devi}, \citenamefont {Goswami},\ and\ \citenamefont
  {Patra}}]{Chakrabortty:2012mh}%
  \BibitemOpen
  \bibfield  {author} {\bibinfo {author} {\bibfnamefont {J.}~\bibnamefont
  {Chakrabortty}}, \bibinfo {author} {\bibfnamefont {H.~Z.}\ \bibnamefont
  {Devi}}, \bibinfo {author} {\bibfnamefont {S.}~\bibnamefont {Goswami}}, \
  and\ \bibinfo {author} {\bibfnamefont {S.}~\bibnamefont {Patra}},\ }\href
  {\doibase 10.1007/JHEP08(2012)008} {\bibfield  {journal} {\bibinfo  {journal}
  {JHEP}\ }\textbf {\bibinfo {volume} {1208}},\ \bibinfo {pages} {008}
  (\bibinfo {year} {2012}{\natexlab{b}})},\ \Eprint
  {http://arxiv.org/abs/1204.2527} {arXiv:1204.2527 [hep-ph]} \BibitemShut
  {NoStop}%
\bibitem [{\citenamefont {Mahajan}(2014{\natexlab{a}})}]{Mahajan:2013ixa}%
  \BibitemOpen
  \bibfield  {author} {\bibinfo {author} {\bibfnamefont {N.}~\bibnamefont
  {Mahajan}},\ }\href {\doibase 10.1103/PhysRevLett.112.031804} {\bibfield
  {journal} {\bibinfo  {journal} {Phys. Rev. Lett.}\ }\textbf {\bibinfo
  {volume} {112}},\ \bibinfo {pages} {031804} (\bibinfo {year}
  {2014}{\natexlab{a}})},\ \Eprint {http://arxiv.org/abs/1310.1064}
  {arXiv:1310.1064 [hep-ph]} \BibitemShut {NoStop}%
\bibitem [{\citenamefont {Mahajan}(2014{\natexlab{b}})}]{Mahajan:2014nca}%
  \BibitemOpen
  \bibfield  {author} {\bibinfo {author} {\bibfnamefont {N.}~\bibnamefont
  {Mahajan}},\ }\href {\doibase 10.1103/PhysRevD.90.035015} {\bibfield
  {journal} {\bibinfo  {journal} {Phys. Rev.}\ }\textbf {\bibinfo {volume}
  {D90}},\ \bibinfo {pages} {035015} (\bibinfo {year} {2014}{\natexlab{b}})},\
  \Eprint {http://arxiv.org/abs/1406.2606} {arXiv:1406.2606 [hep-ph]}
  \BibitemShut {NoStop}%
\bibitem [{\citenamefont {Bhupal~Dev}\ \emph {et~al.}(2015)\citenamefont
  {Bhupal~Dev}, \citenamefont {Goswami},\ and\ \citenamefont
  {Mitra}}]{Dev:2014xea}%
  \BibitemOpen
  \bibfield  {author} {\bibinfo {author} {\bibfnamefont {P.}~\bibnamefont
  {Bhupal~Dev}}, \bibinfo {author} {\bibfnamefont {S.}~\bibnamefont {Goswami}},
  \ and\ \bibinfo {author} {\bibfnamefont {M.}~\bibnamefont {Mitra}},\ }\href
  {\doibase 10.1103/PhysRevD.91.113004} {\bibfield  {journal} {\bibinfo
  {journal} {Phys. Rev.}\ }\textbf {\bibinfo {volume} {D91}},\ \bibinfo {pages}
  {113004} (\bibinfo {year} {2015})},\ \Eprint {http://arxiv.org/abs/1405.1399}
  {arXiv:1405.1399 [hep-ph]} \BibitemShut {NoStop}%
\bibitem [{\citenamefont {Ge}\ and\ \citenamefont
  {Rodejohann}(2015)}]{Ge:2015bfa}%
  \BibitemOpen
  \bibfield  {author} {\bibinfo {author} {\bibfnamefont {S.-F.}\ \bibnamefont
  {Ge}}\ and\ \bibinfo {author} {\bibfnamefont {W.}~\bibnamefont
  {Rodejohann}},\ }\href {\doibase 10.1103/PhysRevD.92.093006} {\bibfield
  {journal} {\bibinfo  {journal} {Phys. Rev.}\ }\textbf {\bibinfo {volume}
  {D92}},\ \bibinfo {pages} {093006} (\bibinfo {year} {2015})},\ \Eprint
  {http://arxiv.org/abs/1507.05514} {arXiv:1507.05514 [hep-ph]} \BibitemShut
  {NoStop}%
\bibitem [{\citenamefont {Gonzalez}\ \emph {et~al.}(2016)\citenamefont
  {Gonzalez}, \citenamefont {Hirsch},\ and\ \citenamefont
  {Kovalenko}}]{Gonzalez:2015ady}%
  \BibitemOpen
  \bibfield  {author} {\bibinfo {author} {\bibfnamefont {M.}~\bibnamefont
  {Gonzalez}}, \bibinfo {author} {\bibfnamefont {M.}~\bibnamefont {Hirsch}}, \
  and\ \bibinfo {author} {\bibfnamefont {S.~G.}\ \bibnamefont {Kovalenko}},\
  }\href {\doibase 10.1103/PhysRevD.93.013017} {\bibfield  {journal} {\bibinfo
  {journal} {Phys. Rev.}\ }\textbf {\bibinfo {volume} {D93}},\ \bibinfo {pages}
  {013017} (\bibinfo {year} {2016})},\ \Eprint
  {http://arxiv.org/abs/1511.03945} {arXiv:1511.03945 [hep-ph]} \BibitemShut
  {NoStop}%
\bibitem [{\citenamefont {Horoi}\ and\ \citenamefont
  {Neacsu}(2015)}]{Horoi:2015gdv}%
  \BibitemOpen
  \bibfield  {author} {\bibinfo {author} {\bibfnamefont {M.}~\bibnamefont
  {Horoi}}\ and\ \bibinfo {author} {\bibfnamefont {A.}~\bibnamefont {Neacsu}},\
  }\href@noop {} {\  (\bibinfo {year} {2015})},\ \Eprint
  {http://arxiv.org/abs/1511.00670} {arXiv:1511.00670 [hep-ph]} \BibitemShut
  {NoStop}%
\bibitem [{\citenamefont {Bambhaniya}\ \emph {et~al.}(2015)\citenamefont
  {Bambhaniya}, \citenamefont {Dev}, \citenamefont {Goswami},\ and\
  \citenamefont {Mitra}}]{Bambhaniya:2015ipg}%
  \BibitemOpen
  \bibfield  {author} {\bibinfo {author} {\bibfnamefont {G.}~\bibnamefont
  {Bambhaniya}}, \bibinfo {author} {\bibfnamefont {P.~S.~B.}\ \bibnamefont
  {Dev}}, \bibinfo {author} {\bibfnamefont {S.}~\bibnamefont {Goswami}}, \ and\
  \bibinfo {author} {\bibfnamefont {M.}~\bibnamefont {Mitra}},\ }\href@noop {}
  {\  (\bibinfo {year} {2015})},\ \Eprint {http://arxiv.org/abs/1512.00440}
  {arXiv:1512.00440 [hep-ph]} \BibitemShut {NoStop}%
\bibitem [{htt()}]{httpmaj}%
  \BibitemOpen
  \href@noop {} {}\bibinfo {note} {Web page,
  http://inspirehep.net/record/8251/citations}\BibitemShut {NoStop}%
\bibitem [{\citenamefont {Allahverdi}\ \emph {et~al.}(2011)\citenamefont
  {Allahverdi}, \citenamefont {Dutta},\ and\ \citenamefont
  {Mohapatra}}]{Allahverdi:2010us}%
  \BibitemOpen
  \bibfield  {author} {\bibinfo {author} {\bibfnamefont {R.}~\bibnamefont
  {Allahverdi}}, \bibinfo {author} {\bibfnamefont {B.}~\bibnamefont {Dutta}}, \
  and\ \bibinfo {author} {\bibfnamefont {R.~N.}\ \bibnamefont {Mohapatra}},\
  }\href {\doibase 10.1016/j.physletb.2010.11.006} {\bibfield  {journal}
  {\bibinfo  {journal} {Phys. Lett.}\ }\textbf {\bibinfo {volume} {B695}},\
  \bibinfo {pages} {181} (\bibinfo {year} {2011})},\ \Eprint
  {http://arxiv.org/abs/1008.1232} {arXiv:1008.1232 [hep-ph]} \BibitemShut
  {NoStop}%
\bibitem [{\citenamefont {Valle}\ and\ \citenamefont
  {Vaquera-Araujo}(2016)}]{Valle:2016kyz}%
  \BibitemOpen
  \bibfield  {author} {\bibinfo {author} {\bibfnamefont {J.~W.~F.}\
  \bibnamefont {Valle}}\ and\ \bibinfo {author} {\bibfnamefont {C.~A.}\
  \bibnamefont {Vaquera-Araujo}},\ }\href {\doibase
  10.1016/j.physletb.2016.02.031} {\bibfield  {journal} {\bibinfo  {journal}
  {Phys. Lett.}\ }\textbf {\bibinfo {volume} {B755}},\ \bibinfo {pages} {363}
  (\bibinfo {year} {2016})},\ \Eprint {http://arxiv.org/abs/1601.05237}
  {arXiv:1601.05237 [hep-ph]} \BibitemShut {NoStop}%
\bibitem [{\citenamefont {Yanagida}(1980)}]{Yanagida:1980xy}%
  \BibitemOpen
  \bibfield  {author} {\bibinfo {author} {\bibfnamefont {T.}~\bibnamefont
  {Yanagida}},\ }\href {\doibase 10.1143/PTP.64.1103} {\bibfield  {journal}
  {\bibinfo  {journal} {Prog. Theor. Phys.}\ }\textbf {\bibinfo {volume}
  {64}},\ \bibinfo {pages} {1103} (\bibinfo {year} {1980})}\BibitemShut
  {NoStop}%
\bibitem [{\citenamefont {Mohapatra}\ and\ \citenamefont
  {Senjanovic}(1981)}]{Mohapatra:1980yp}%
  \BibitemOpen
  \bibfield  {author} {\bibinfo {author} {\bibfnamefont {R.~N.}\ \bibnamefont
  {Mohapatra}}\ and\ \bibinfo {author} {\bibfnamefont {G.}~\bibnamefont
  {Senjanovic}},\ }\href {\doibase 10.1103/PhysRevD.23.165} {\bibfield
  {journal} {\bibinfo  {journal} {Phys.Rev.}\ }\textbf {\bibinfo {volume}
  {D23}},\ \bibinfo {pages} {165} (\bibinfo {year} {1981})}\BibitemShut
  {NoStop}%
\bibitem [{\citenamefont {Magg}\ and\ \citenamefont
  {Wetterich}(1980)}]{Magg:1980ut}%
  \BibitemOpen
  \bibfield  {author} {\bibinfo {author} {\bibfnamefont {M.}~\bibnamefont
  {Magg}}\ and\ \bibinfo {author} {\bibfnamefont {C.}~\bibnamefont
  {Wetterich}},\ }\href {\doibase 10.1016/0370-2693(80)90825-4} {\bibfield
  {journal} {\bibinfo  {journal} {Phys. Lett.}\ }\textbf {\bibinfo {volume}
  {B94}},\ \bibinfo {pages} {61} (\bibinfo {year} {1980})}\BibitemShut
  {NoStop}%
\bibitem [{\citenamefont {Foot}\ \emph {et~al.}(1989)\citenamefont {Foot},
  \citenamefont {Lew}, \citenamefont {He},\ and\ \citenamefont
  {Joshi}}]{Foot:1988aq}%
  \BibitemOpen
  \bibfield  {author} {\bibinfo {author} {\bibfnamefont {R.}~\bibnamefont
  {Foot}}, \bibinfo {author} {\bibfnamefont {H.}~\bibnamefont {Lew}}, \bibinfo
  {author} {\bibfnamefont {X.~G.}\ \bibnamefont {He}}, \ and\ \bibinfo {author}
  {\bibfnamefont {G.~C.}\ \bibnamefont {Joshi}},\ }\href {\doibase
  10.1007/BF01415558} {\bibfield  {journal} {\bibinfo  {journal} {Z. Phys.}\
  }\textbf {\bibinfo {volume} {C44}},\ \bibinfo {pages} {441} (\bibinfo {year}
  {1989})}\BibitemShut {NoStop}%
\bibitem [{\citenamefont {Franco}(2015)}]{Franco:2015pva}%
  \BibitemOpen
  \bibfield  {author} {\bibinfo {author} {\bibfnamefont {E.~T.}\ \bibnamefont
  {Franco}},\ }\href {\doibase 10.1103/PhysRevD.92.113010} {\bibfield
  {journal} {\bibinfo  {journal} {Phys. Rev.}\ }\textbf {\bibinfo {volume}
  {D92}},\ \bibinfo {pages} {113010} (\bibinfo {year} {2015})},\ \Eprint
  {http://arxiv.org/abs/1510.06240} {arXiv:1510.06240 [hep-ph]} \BibitemShut
  {NoStop}%
\bibitem [{\citenamefont {Gavela}\ \emph {et~al.}(2009)\citenamefont {Gavela},
  \citenamefont {Hambye}, \citenamefont {Hernandez},\ and\ \citenamefont
  {Hernandez}}]{Gavela:2009cd}%
  \BibitemOpen
  \bibfield  {author} {\bibinfo {author} {\bibfnamefont {M.~B.}\ \bibnamefont
  {Gavela}}, \bibinfo {author} {\bibfnamefont {T.}~\bibnamefont {Hambye}},
  \bibinfo {author} {\bibfnamefont {D.}~\bibnamefont {Hernandez}}, \ and\
  \bibinfo {author} {\bibfnamefont {P.}~\bibnamefont {Hernandez}},\ }\href
  {\doibase 10.1088/1126-6708/2009/09/038} {\bibfield  {journal} {\bibinfo
  {journal} {JHEP}\ }\textbf {\bibinfo {volume} {09}},\ \bibinfo {pages} {038}
  (\bibinfo {year} {2009})},\ \Eprint {http://arxiv.org/abs/0906.1461}
  {arXiv:0906.1461 [hep-ph]} \BibitemShut {NoStop}%
\bibitem [{\citenamefont {Dev}\ and\ \citenamefont
  {Pilaftsis}(2012)}]{Dev:2012sg}%
  \BibitemOpen
  \bibfield  {author} {\bibinfo {author} {\bibfnamefont {P.~S.~B.}\
  \bibnamefont {Dev}}\ and\ \bibinfo {author} {\bibfnamefont {A.}~\bibnamefont
  {Pilaftsis}},\ }\href {\doibase 10.1103/PhysRevD.86.113001} {\bibfield
  {journal} {\bibinfo  {journal} {Phys. Rev.}\ }\textbf {\bibinfo {volume}
  {D86}},\ \bibinfo {pages} {113001} (\bibinfo {year} {2012})},\ \Eprint
  {http://arxiv.org/abs/1209.4051} {arXiv:1209.4051 [hep-ph]} \BibitemShut
  {NoStop}%
\bibitem [{\citenamefont {Bhupal~Dev}\ and\ \citenamefont
  {Pilaftsis}(2013)}]{Dev:2012bd}%
  \BibitemOpen
  \bibfield  {author} {\bibinfo {author} {\bibfnamefont {P.~S.}\ \bibnamefont
  {Bhupal~Dev}}\ and\ \bibinfo {author} {\bibfnamefont {A.}~\bibnamefont
  {Pilaftsis}},\ }\href {\doibase 10.1103/PhysRevD.87.053007} {\bibfield
  {journal} {\bibinfo  {journal} {Phys. Rev.}\ }\textbf {\bibinfo {volume}
  {D87}},\ \bibinfo {pages} {053007} (\bibinfo {year} {2013})},\ \Eprint
  {http://arxiv.org/abs/1212.3808} {arXiv:1212.3808 [hep-ph]} \BibitemShut
  {NoStop}%
\bibitem [{\citenamefont {Awasthi}\ \emph {et~al.}(2013)\citenamefont
  {Awasthi}, \citenamefont {Parida},\ and\ \citenamefont
  {Patra}}]{Awasthi:2013ff}%
  \BibitemOpen
  \bibfield  {author} {\bibinfo {author} {\bibfnamefont {R.~L.}\ \bibnamefont
  {Awasthi}}, \bibinfo {author} {\bibfnamefont {M.~K.}\ \bibnamefont {Parida}},
  \ and\ \bibinfo {author} {\bibfnamefont {S.}~\bibnamefont {Patra}},\ }\href
  {\doibase 10.1007/JHEP08(2013)122} {\bibfield  {journal} {\bibinfo  {journal}
  {JHEP}\ }\textbf {\bibinfo {volume} {08}},\ \bibinfo {pages} {122} (\bibinfo
  {year} {2013})},\ \Eprint {http://arxiv.org/abs/1302.0672} {arXiv:1302.0672
  [hep-ph]} \BibitemShut {NoStop}%
\bibitem [{\citenamefont {del Aguila}\ \emph
  {et~al.}(2007{\natexlab{b}})\citenamefont {del Aguila}, \citenamefont
  {Aguilar-Saavedra},\ and\ \citenamefont {Pittau}}]{delAguila:2007em}%
  \BibitemOpen
  \bibfield  {author} {\bibinfo {author} {\bibfnamefont {F.}~\bibnamefont {del
  Aguila}}, \bibinfo {author} {\bibfnamefont {J.~A.}\ \bibnamefont
  {Aguilar-Saavedra}}, \ and\ \bibinfo {author} {\bibfnamefont
  {R.}~\bibnamefont {Pittau}},\ }\href {\doibase 10.1088/1126-6708/2007/10/047}
  {\bibfield  {journal} {\bibinfo  {journal} {JHEP}\ }\textbf {\bibinfo
  {volume} {10}},\ \bibinfo {pages} {047} (\bibinfo {year}
  {2007}{\natexlab{b}})},\ \Eprint {http://arxiv.org/abs/hep-ph/0703261}
  {arXiv:hep-ph/0703261 [hep-ph]} \BibitemShut {NoStop}%
\end{thebibliography}
%

\end{document}